\documentclass[aps,prl,reprint,superscriptaddress]{revtex4-2}

\usepackage{graphicx}
\usepackage{physics}
\usepackage{xcolor}
\usepackage{float}
\usepackage[section]{placeins}
\usepackage{chapterbib}

\usepackage{comment}

\begin{document}

\title{A light-induced Weyl semiconductor-to-metal transition mediated by Peierls instability}

\author{Honglie Ning}
\thanks{These authors contributed equally to this work}
\affiliation{Institute for Quantum Information and Matter, California Institute of Technology, Pasadena, CA 91125}
\affiliation{Department of Physics, California Institute of Technology, Pasadena, CA 91125}

\author{Omar Mehio}
\thanks{These authors contributed equally to this work}
\affiliation{Institute for Quantum Information and Matter, California Institute of Technology, Pasadena, CA 91125}
\affiliation{Department of Physics, California Institute of Technology, Pasadena, CA 91125}

\author{Chao Lian}
\thanks{These authors contributed equally to this work}
\affiliation{Department of Chemical and Environmental Engineering, Materials Science and Engineering Program, and Department of Physics and Astronomy, University of California, Riverside, CA 92521}

\author{Xinwei Li}
\affiliation{Institute for Quantum Information and Matter, California Institute of Technology, Pasadena, CA 91125}
\affiliation{Department of Physics, California Institute of Technology, Pasadena, CA 91125}

\author{Eli Zoghlin}
\affiliation{Materials Department, University of California, Santa Barbara, CA 93106}

\author{Preston Zhou}
\affiliation{Department of Physics, California Institute of Technology, Pasadena, CA 91125}

\author{Bryan Cheng}
\affiliation{Department of Physics, California Institute of Technology, Pasadena, CA 91125}

\author{Stephen D. Wilson}
\affiliation{Materials Department, University of California, Santa Barbara, CA 93106}

\author{Bryan M. Wong}
\affiliation{Department of Chemical and Environmental Engineering, Materials Science and Engineering Program, and Department of Physics and Astronomy, University of California, Riverside, CA 92521}

\author{David Hsieh}
\email[Author to whom the correspondence should be addressed: ]{dhsieh@caltech.edu}
\affiliation{Institute for Quantum Information and Matter, California Institute of Technology, Pasadena, CA 91125}
\affiliation{Department of Physics, California Institute of Technology, Pasadena, CA 91125}

\begin{abstract}
Elemental tellurium is a strongly spin-orbit coupled Peierls-distorted semiconductor whose band structure features topologically protected Weyl nodes. Using time-dependent density functional theory calculations, we show that impulsive optical excitation can be used to transiently control the amplitude of the Peierls distortion, realizing a mechanism to switch tellurium between three-states: Weyl semiconductor, Weyl metal and non-Weyl metal. Further, we present experimental evidence of this inverse-Peierls distortion using time-resolved optical second harmonic generation measurements. These results provide a pathway to multifunctional ultrafast Weyl devices and introduce Peierls systems as viable hosts of light-induced topological transitions.
\end{abstract}

\maketitle

Weyl nodes are topologically stable crossing points between non-degenerate bands in a crystal, which impart unconventional properties including ultrahigh charge mobility and chiral magneto-transport \cite{FelserARCMP2017}. The possibility to create or annihilate Weyl nodes \textit{in situ} using ultrashort light pulses has been broadly explored theoretically \cite{RubioNCOMMS2017, PatrickLeePRB2016, ZWangPRL2016, SentefNCOMMS2017, MengNCOMMS2017, Ebihara2016} and was recently demonstrated experimentally in Dirac and type-II Weyl semimetal materials via impulsively driven lattice symmetry changes \cite{WJGPRX2020, WJGNMAT2021, LindenbergNature2019,WNLPRX2019}. However, efforts have so far focused on binary switching between semi-metallic states with and without Weyl nodes. 

In this Letter, we first use density functional theory (DFT) calculations to show a three-state switch from Weyl semiconductor to Weyl metal to non-Weyl metal in chiral Peierls-distorted tellurium crystals as a function of the chiral chain radius. By performing time-dependent DFT calculations, we then demonstrate that these states can be transiently stabilized via a light-induced inverse-Peierls distortion. Predicted signatures of the inverse-Peierls distortion are experimentally reproduced using time-resolved optical second harmonic generation. 

\begin{figure*}[t]
\includegraphics[width=6.65 in]{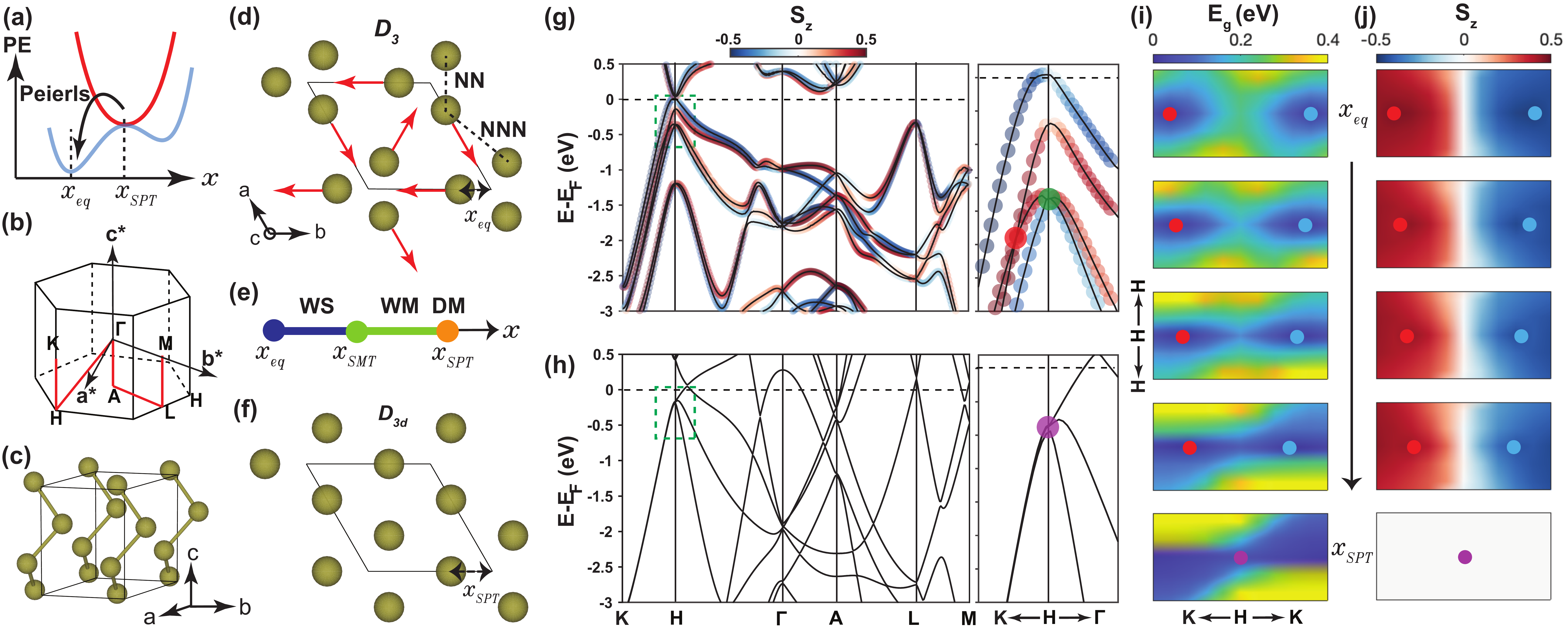}
\label{Fig1}
\caption{Band evolution of Te across the Peierls transition. (a) Schematic evolution of the potential energy (PE) surface across a Peierls transition. (b) Brillouin zone and (c) unit cell of bulk Te. (d) Equilibrium structure of Te ($D\textsubscript{3}$ point group). Black rhombus denotes the unit cell and red arrows point along the $A_1$ phonon coordinate $x$. (e) Phase diagram as a function of $x$ determined from DFT calculations (WS: Weyl semiconductor; WM: Weyl metal; DM: Dirac metal). (f) Peierls un-distorted structure of Te ($D\textsubscript{3d}$ point group). (g) DFT calculated band structure of Te for the equilibrium and (h) Peierls un-distorted structures. The color scale represents the spin polarization along $z$. Zoom-ins on the window spanning 0.05 eV to -0.7 eV (dashed rectangle) are displayed on the right. Red, green and purple circles mark WN1, the Kramers-Weyl node at $H$, and the Dirac node respectively. (i) Momentum space maps of the energy gap between the two bands forming WN1 (red circle) and WN1' (blue circle) and (j) the $z$-component of spin polarization for the upper band at select $x$ values of 0.269, 0.290, 0.305, 0.319 and 0.333 (top to bottom). The horizontal and vertical ranges of each panel span -0.1$c^*$ to 0.1$c^*$ and -0.05$b^*$ to 0.05$b^*$ relative to $H$, respectively, where $b^*$ and $c^*$ are the reciprocal lattice vectors defined in panel (b).}
\end{figure*}

The Peierls instability is a spontaneous symmetry-lowering lattice deformation that lifts the degeneracy of electronic states at the Fermi level in order to reduce the overall system energy [Fig. 1(a)]. Elemental tellurium (Te) is a prototypical Peierls-distorted system, which crystallizes in a non-centrosymmetric trigonal structure composed of chiral chains of Te atoms oriented along the $c$-axis, with space group $P3\textsubscript{1}21$ or $P3\textsubscript{2}21$ depending on the chain chirality [Fig. 1(c) depicts $P3\textsubscript{1}21$ structure]. Each atom has two intra-chain nearest-neighbors (NNs) and four inter-chain next-nearest-neighbors (NNNs). This structure can be regarded as arising from the Peierls distortion of an achiral centrosymmetric rhombohedral structure (space group $R\overline{3}m$) in which the NN and NNN distances are equal [Fig. 1(d),(f)] \cite{TangneyPRB2002}. The structural evolution is parameterized by the chain radius $x$, expressed in units of the lattice constant or inter-chain distance $a$, which is a displacement along the $A_1$ phonon coordinate. In the equilibrium phase $x_{eq}$ = 0.269 and at the structural phase transition (SPT) into the rhombohedral phase $x_{SPT}$  = 0.333, amounting to a difference of around 0.26 \AA.

To understand the evolution of the electronic structure with $x$, we performed fully relativistic DFT calculations \cite{SM}. In the equilibrium phase ($x_{eq}$ = 0.269), we find that Te is a semiconductor with non-degenerate bands harboring Weyl nodes (WNs) below the Fermi level along the $H$-$K$ line in the Brillouin zone [Fig. 1(g)], as well as Kramers-WNs \cite{Chang2018} at the time-reversal invariant momenta $\Gamma$, $M$ and $A$. These results are consistent with previous reports \cite{TakahashiPRB2017, KondoPRL2020, CrepaldiPRL2020, Zhang2020, MiyakePRL2015} classifying Te as a Weyl semiconductor (WS). To study the evolution of band topology in detail, we focus on a characteristic WN pair (WN1 and WN1') near the Fermi level along $K$-$H$-$K$. Since WNs arise from a two-band crossing, their locations in momentum space can be identified by mapping the energy gap between the two bands [Fig. 1(i)]. This can be corroborated by additionally mapping the the $z-$component of spin polarization for the upper band, which is expected to reach a maximum amplitude with opposite signs at WN1 and WN1' \cite{KondoPRL2020, CrepaldiPRL2020}.

Upon increasing $x$, Te first undergoes a semiconductor-to-metal transition (SMT) at $x_{SMT} \approx 0.283$ [Fig. 1(e)] due to the sinking of the conduction bands at $A$ \cite{SM}. At this stage, WN1 and WN1' remain well separated, thus realizing an intermediate Weyl metal (WM). As $x$ further increases, WN1 and WN1' approach each other and the energy gap along $H$-$K$ continues to shrink, with little change in the spin texture [Figs. 1(i),(j)]. Finally, on reaching the rhombohedral phase ($x_{SPT}$  = 0.333), band degeneracy is restored and a merger of WN1 and WN1' into a Dirac node accompanies a closing of the direct gap near $H$ [Figs. 1(h)-(j)], giving rise to a Dirac metal (DM). Tuning $x$ therefore provides a mechanism for simultaneous WN, spin texture, and band gap control.

Previous DFT studies have shown that trigonal Te can be driven into topological Weyl semimetal, three-dimensional topological insulator or WM phases by applying either hydrostatic pressure, shear or uniaxial strains to alter its structure \cite{KioussisPRL2013,MiyakePRL2015,Xue2018}. However, no equilibrium pathway to directly invert the Peierls distortion by tuning $x$, either through thermal or mechanical deformation \cite{UykurPRL2020, IwasaPNAS2019}, is known to exist.

Impulsive excitation by an intense laser pulse offers a potential out-of-equilibrium pathway to induce an inverse Peierls transition. By optically de-populating states near the Fermi level, the energy increase due to the lattice distortion is no longer balanced by the energy decrease due to the lifting of band degeneracy. This causes a sudden change in the potential energy surface of the lattice, generating a restoring force that drives coherent atomic motion reversing the Peierls distortion. For underdamped motion, a totally symmetric $A_{1(g)}$ Raman active mode is expected to be coherently launched through this displacive excitation mechanism \cite{DresselhausPRB1992, SM} [Fig. 1(a)]. Light-induced inverse Peierls transitions are accessible in a variety of systems including A7-structured semimetals \cite{NelsonPRX2018}, VO$_2$ \cite{WallNCOMM2012} and charge density wave materials \cite{WallPRL2012,JohnsonPRL2014,BeaudNMAT2014}. However, this mechanism has so far not been explored for ultrafast control of band topology.  

To study the possibility of a light-induced inverse Peierls distortion in Te, we carried out time-dependent (TD) DFT calculations \cite{SM} to simulate the real-time lattice dynamics following impulsive optical excitation. Our method provides a fully \textit{ab initio} description of the electronic, phononic and photonic degrees of freedom on equal footing. Using an advanced evolutionary algorithm, the velocity-gauge formalism, and symmetry-reduced momentum space sampling, we efficiently calculate the periodic Te system up to an unprecedented 3 ps, providing comprehensive information about not only the fast electronic response but also the long-time structural dynamics. The pump pulse is chosen to have a Gaussian profile of 100 fs width, a linear polarization with electric field perpendicular to the $c$-axis, and a photon energy centered at 1 eV, which is above the 0.3 eV band gap of Te. Otherwise, there are no adjustable parameters. For all pump fluences sampled, we resolved atomic motion exclusively along the $A_1$ phonon coordinate. Specifically, pumping excites sinusoidal displacement oscillations in time ($t$) about a new value of $x$ that is shifted higher than $x_{eq}$ [Fig. 2(a)]. Since the lifetime of photo-carriers deduced from our simulations well exceeds our sampled time window of several picoseconds \cite{SM}, this new position ($x_0$) is metastable. With increasing fluence ($F$), both the magnitude of the oscillations and $x_0$ increase. At sufficiently high fluence $x_0$ is able to reach $x_{SPT}$, signifying complete reversal of the Peierls distortion. An effective potential energy (PE)$_{\textrm{eff}}$ can be determined by evaluating $E_{total}[x(t),t]-E_{kin}[x(t),t]$, which is the difference between the total energy - including both lattice and electronic degrees of freedom - and ionic kinetic energy [Fig. 2(b)]. To better visualize the dynamics of the inverse Peierls transition, we plot the spatio-temporal trajectory of (PE)$_{\textrm{eff}}$ for select fluences in Figure 2(c). For a fixed fluence, we observe that (PE)$_{\textrm{eff}}$ generally increases with time because $E_{total}[x(t),t]$ is fixed while $E_{kin}[x(t),t]$ is gradually damped out. But within each oscillation period there are local minima marking metastable $x$ positions. In the low fluence regime, the trajectories are parabolic and share a single minimum displaced slightly towards $x_{SPT}$ from $x_{eq}$. As fluence increases, this minimum monotonically shifts to larger $x$ and, near a critical value $F_c \approx$ 2 mJ/cm$^2$ (0.03 V/\AA~ peak field), the trajectories become flattened and highly non-parabolic. Above $F_c$, parabolic trajectories are restored about a new minimum fixed at $x_{SPT}$. The fluence dependence of the curvature and local minima of (PE)$_{\textrm{eff}}$ are signatures of a dynamical phase transition across $F_c$ that can be quantitatively tested experimentally.

\begin{figure}
\includegraphics[width=3.375in]{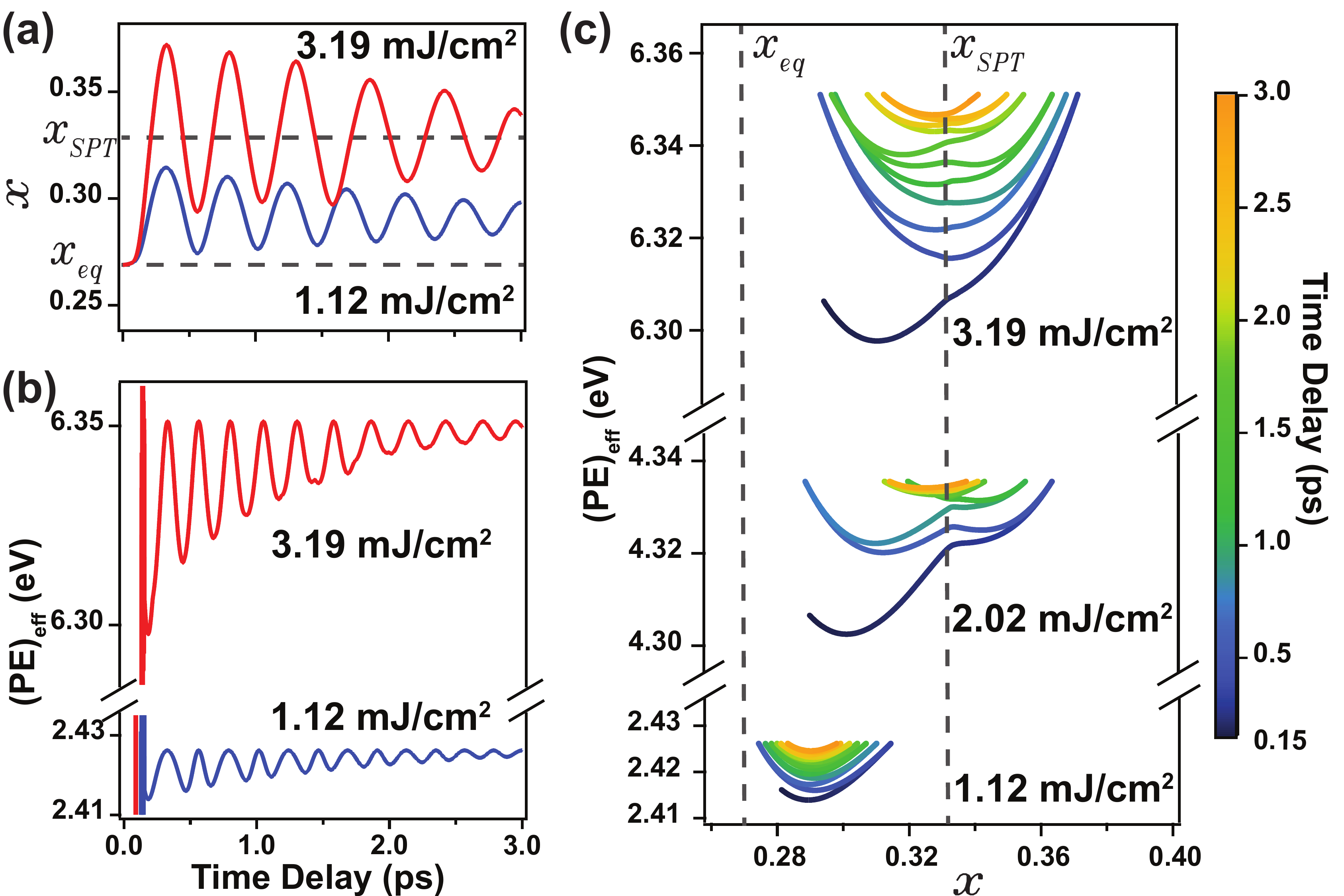}
\label{Fig2}
\caption{Light-induced lattice and effective potential energy dynamics. TDDFT calculated temporal evolution of (a) $x$ and (b) the effective potential energy (see text) following impulsive optical excitation at a fluence below (blue) and above (red) $F_c$. (c) Spatio-temporal trajectories of the effective potential energy at fluences below, near and above $F_c$. Color scales with the time delay.}
\end{figure}

Our TDDFT simulations are consistent with several previously reported experiments. Time-resolved pump-probe x-ray diffraction measurements on Te showed that the lattice undergoes transient deformation predominantly along the $A_1$ coordinate \cite{JohnsonPRL2009}, characterized by periodic oscillations about a position positively offset from $x_{eq}$. However, low fluences were used for this experiment, which only induced changes in $x$ of order 0.01. Time-resolved optical reflectivity measurements demonstrated displacive excitation of coherent $A_1$ phonon oscillations, which undergo red-shifting and chirping with increasing fluence \cite{KurzPRL1995, HunschePRL1995, SoodPRB2010}, as well as an anomalous blue-shift at higher fluence possibly due to overshooting a high symmetry point, although no explicit claim of an inverse Peierls distortion was made \cite{NelsonPRB2018}. Transient broadband optical spectroscopy measurements on Te thin films revealed slow photo-carrier recombination times ranging from tens to hundreds of picoseconds, possibly bottlenecked by weak inter-valley scattering \cite{XXFPRB2019,Smithncomm2020}, giving rise a metastable excited state. Finally, evidence of a semiconductor-to-metal transition \cite{Kim2003} was revealed by time-resolved ellipsometry measurements, which may be related to our predicted sinking of the $A$ point conduction band at $x_{SMT}$. However, no ultrafast inverse Peierls transition in Te has been experimentally reported to date. 

Under an adiabatic approximation in which the electronic response time is much shorter than the characteristic phonon period, the time-dependent band structure as a function of $x(t)$ can be captured by static DFT calculations as a function of $x$ \cite{LindenbergNature2019}. Therefore, experimental verification of the light-induced structural changes calculated using TDDFT serves as indirect evidence of the predicted three-state switch between a WS, WM and DM.

To quantitatively test our TDDFT predictions, we performed time-resolved optical second harmonic generation rotational anisotropy (SHG-RA) measurements [Fig. 3(e) inset] \cite{HsiehRSI2014,SM}. Since the leading order electric-dipole contribution to SHG directly couples to the inversion odd structural order parameter of Te \cite{GGYPRB2019}, this technique is simultaneously sensitive to the metastable $x$ coordinate and the $A_1$ phonon properties under identical experimental conditions. Static SHG-RA patterns were measured using 1.5 eV incident probe light on Te single crystals polished with the $c$-axis parallel to the surface plane. By acquiring patterns in both parallel (S\textsubscript{in}-S\textsubscript{out}) and crossed (S\textsubscript{in}-P\textsubscript{out}) polarization channels, we verified that the entire signal is attributable to a bulk electric-dipole SHG susceptibility tensor respecting $D_3$ symmetry \cite{SM}. 

\begin{figure*}
\includegraphics[width=6.75 in]{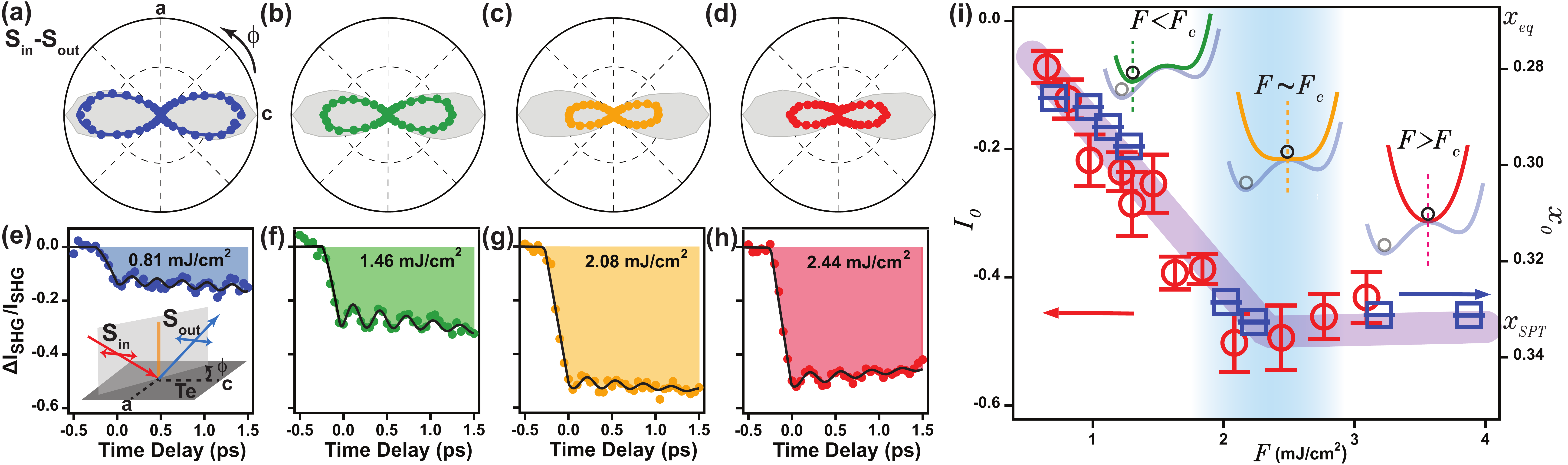}
\label{Fig3}
\caption{Fluence dependence of the metastable structure. (a)-(d) Instantaneous SHG-RA patterns measured at $t$ = 50 fs for different absorbed pump fluence values. The static ($t <$ 0) pattern is overlaid and shaded gray. (e)-(h) Normalized differential SHG intensity transients acquired at the angle of maximum intensity ($\phi$ = 0$^{\circ}$) in the S\textsubscript{in}-S\textsubscript{out} pattern for each fluence. Solid lines are fits to an exponential decaying oscillation plus a constant offset (see text). A weak linear background was introduced in some traces to account for laser power drift. The inset in panel (e) shows a schematic of the SHG-RA setup. The scattering plane angle $\phi$ is measured with respect to the crystallographic $c$-axis. Red, blue and orange lines represent the incident, reflected SHG and pump beams, respectively. (i) Pump fluence dependence of the measured SHG offset term (red circles) and TDDFT calculated metastable $x$ position (blue squares). The blue shaded bar indicates the critical fluence regime. Insets show schematics of the transient potential energy surface in the low, critical and high fluence regimes, illustrating the shift in the metastable $x$ position from equilibrium (gray curves).}
\end{figure*}

Figures 3(a)-(d) show instantaneous SHG-RA patterns in the S\textsubscript{in}-S\textsubscript{out} channel measured immediately after exciting with a 1 eV pump pulse of 100 fs duration - matching our TDDFT parameters - for different absorbed fluence levels. Note that the absorbed fluence is lower than the applied fluence by a factor of 1 - $R$, where $R$ is the reflectance at 1 eV. The instantaneous patterns exhibit a uniform (independent of scattering plane angle $\phi$) decrease in intensity relative to the equilibrium pattern, indicating that all electric-dipole susceptibility tensor elements are suppressed by the same scale factor \cite{SM}. Since each tensor element is proportional to the structural order parameter, this confirms that pump excitation acts simply to reduce the structural order parameter and does not induce any symmetry breaking. The patterns subsequently undergo uniform oscillations about the reduced intensity value, consistent with a totally symmetric $A_1$ breathing mode \cite{SM}. By tracking the time dependence of the SHG intensity at $\phi = 0^{\circ}$ [Figs. 3(e)-(h)], we clearly resolve an intensity drop upon pump excitation on the timescale of a half cycle of the $A_1$ mode, followed by $A_1$ mode oscillations. While the oscillations are damped out after approximately 2 ps, the intensity offset persists out to at least 10 ps \cite{SM}. 

To directly compare the predicted and measured structural dynamics, we fit both $x(t)$ obtained from our TDDFT simulations [Fig. 2(a)] as well as the differential SHG transients [Figs. 3(e)-(h)] for $t > 0$ to the function $Ae^{-t/\tau}\cos{({2\pi\nu}t+\varphi)}+B$. This expression includes the phonon amplitude $A$, damping time $\tau$, phase $\varphi$, frequency $\nu$ and a constant offset of the $x$ coordinate ($B = x_0$) or SHG intensity ($B = I_0$). Focusing first on the offset term, we plot in Figure 3(i) the fitted values of $x_0$ for multiple fluences. In the weak excitation regime, $x_0$ increases monotonically with fluence, indicating a shift of potential energy minimum towards the centrosymmetric position. At a critical fluence near 2 mJ/cm$^2$, the Peierls non-distorted structure is reached and the potential energy surface becomes parabolic with a minimum at $x_{SPT}$. Further increasing the fluence alters the curvature of the parabola but leaves $x_0$ fixed at $x_{SPT}$. The fitted values of $I_0$ acquired over a similar fluence range are overlaid [Fig. 3(i)], which not only obeys a qualitatively similar trend to $x_0$ but also shows a quantitatively matching critical fluence value. These results suggest an experimental realization of a light-induced inverse Peierls transition. We note that the comparison is not exact because the SHG intensity saturates to a non-zero value. Such residual signals are commonly observed across photo-induced phase transitions \cite{GedikNPHY2020,JohnsonPRL2014,CavalleriPRB2013,WNLPRX2019}. They may be attributed to incomplete order parameter suppression within the probed volume due to quench-induced spatial domains and defects \cite{MihailovicNPHY2010}, penetration depth mismatch between pump and probe beams \cite{GedikNPHY2020, SM}, spatial non-uniformity of the pump intensity \cite{WNLPRX2019}, or higher multipole SHG radiation processes, all of which are not accounted for by our TDDFT simulations.

\begin{figure}
\includegraphics[width=3.375in]{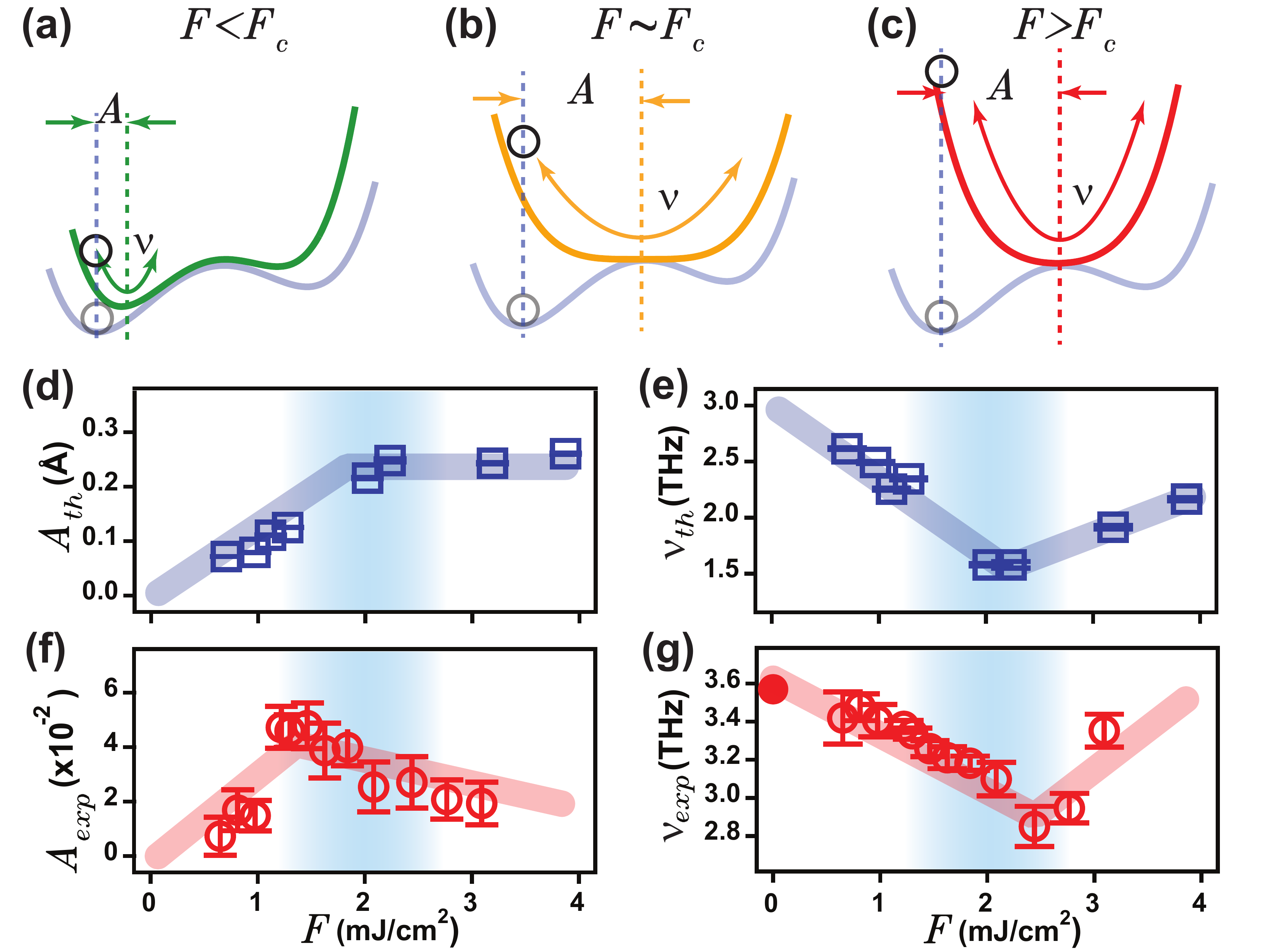}
\label{Fig4}
\caption{Fluence dependence of the coherent $A_1$ phonon dynamics. (a)-(c) Schematics of the metastable potential energy surface at select fluences. The separation between the vertical dashed lines sets the phonon amplitude $A$ and the curvature of the potential minimum sets the phonon frequency $\nu$. (d) Simulated pump fluence dependence of the phonon amplitude and (e) frequency obtained from TDDFT \cite{SM}. (f) Experimentally measured fluence dependence of the phonon amplitude and (g) frequency obtained by fitting the SHG transients in Figs. 3(e)-(h).  The solid circle at $F=0$ in panel (g) is measured with Raman scattering \cite{SM}. The shaded blue bars mark the critical regime.}
\end{figure}

The dynamics of the coherent $A_{1}$ phonon can serve as an additional diagnostic as schematically depicted in Figures 4(a)-(c). For an ultrafast inverse Peierls distortion, one expects that as the fluence increases towards $F_c$, the potential energy minimum should be displaced further away from $x_{eq}$ and its curvature should decrease as the landscape evolves from being locally parabolic to locally quartic. As illustrated in Figure 4(a), the former causes the phonon amplitude to increase while the latter causes the phonon frequency to decrease. At $F_c$, the displacement reaches its maximum value of $x_{SPT} - x_{eq}$ and so the phonon amplitude saturates [Fig. 4(b)]. On the other hand, the landscape becomes parabolic again above $F_c$ and so the curvature starts to increase with fluence, corresponding to an increasing phonon frequency [Fig. 4(c)].

Figures 4(d) and (e) show the fluence dependence of $A_{th}$ and $\nu_{th}$ extracted from fits to the TDDFT simulations. Subscripts on the phonon parameters denote theoretical (th) or experimental (exp) values. The anticipated saturation behavior of $A_{th}$ and softening and re-hardening behavior of $\nu_{th}$ are clearly borne out. Turning to the transient SHG data, we resolve coherent phonon oscillations both below and above $F_c$ [Figs. 3(e)-(h)], indicating that Te remains crystalline over our measured fluence range \cite{SM}. As shown in Figure 4(f), $A_{exp}$ increases with fluence in the weak excitation regime and then abruptly changes slope just below $F_c$, reminiscent of $A_{th}$. However, unlike $A_{th}$, $A_{exp}$ does not saturate above $F_c$ but instead exhibits a slightly downward slope. Although the origin of this discrepancy with TDDFT is unclear, a similar downward trend has been reported above other ultrafast SPTs \cite{WNLPRX2019} and may be related to cumulative heating, electronic diffusion and changes in the Raman scattering cross section of probe photons, which are not accounted for in TDDFT. Figure 4(g) shows that $\nu_{exp}$ decreases with fluence from its 3.6 THz equilibrium value in the weak excitation regime. This is quantitatively consistent with previous studies on Te \cite{KurzPRL1995,NelsonPRX2018} and indicates that TDDFT slightly underestimates the frequency. Above approximately 2 mJ/cm$^2$, there is an abrupt change in slope from negative to positive, closely following the behavior of $\nu_{th}$. 

The consistency between our TDDFT simulations and time-resolved SHG experiments across multiple observables establishes that light can be used to tune elemental Te across an inverse Peierls transition. Our TDDFT results show that this SPT should be accompanied by an ultrafast switching from WS to metastable WM and DM states. Although structural probes, while indirect, have commonly been used to infer the changes in electronic structure \cite{LindenbergNature2019,WNLPRX2019}, the topological band structure change may be directly verifiable in the future using high-resolution extreme ultraviolet time-, spin-, and angle-resolved photoemission spectroscopy. Moreover, our work showcases the effectiveness of TDDFT in predicting impulsively driven out-of-equilibrium SPTs. More generally, note that this inverse Peierls distortion may also be induced entirely through lattice degrees of freedom via ionic Raman scattering \cite{CavalleriACR2015,SM}. Therefore, our results suggest that three-dimensional Peierls systems are a vast and fertile playground for exploring the interplay of ultrafast insulator-to-metal transitions and ultrafast band topology control, two hitherto disparate areas of research. As embodied by Te, this research possibly paves the way towards multi-state-switchable and multifunctional ultrafast Weyl devices.

\begin{acknowledgements}
We thank Michael Buchhold, Alberto de la Torre, Nicholas J. Laurita, and Alon Ron for helpful discussions. We are grateful to George Rossman for assistance with and use of the Raman spectrometer. Optical spectroscopy measurements were supported by the U.S. Department of Energy under Grant No. DE SC0010533. D.H. also acknowledges funding from the David and Lucile Packard Foundation and support for instrumentation from the Institute for Quantum Information and Matter, an NSF Physics Frontiers Center (PHY-1733907). RT-TDDFT calculations by C.L. and B.M.W. were supported by the U.S. Department of Energy, Office of Science, Basic Energy Sciences, TCMP Program, under Award No. DE-SC0022209. S.D.W. and E.Z. gratefully acknowledge support via the U.C. Santa Barbara NSF Quantum Foundry funded via the Q-AMASE-i program under award DMR-1906325. 
\end{acknowledgements}


%

	\newpage
	
	\onecolumngrid	

	\begin{center}
		\textbf{\large Supplemental Material \linebreak 
A light-induced Weyl semiconductor-to-metal transition mediated by Peierls instability}
	\end{center}

\section{I. Static Density Functional Theory Simulations}
The electronic band structures of Te with lattice structures of different chiral chain radius were calculated based on the DFT first-principles \textsf{QUANTUM ESPRESSO} package using plane wave and fully relativistic norm-conserving pseudopotentials \cite{Giannozzi2009, Giannozzi2017}. Without GW correction, this method underestimates the direct bandgap at the H point but still accurately describes the valence bands, and we find an excellent match between our calculated band structures and previously reported literature \cite{CrepaldiPRL2020,KondoPRL2020}. Therefore, our results are well-suited for a qualitative check of electronic structures before and after the SPT, and the position of Weyl nodes in the valence bands should be accurate. 

To calculate the equilibrium band structure, we relieve the stress in the structure and obtain the equilibrium lattice in agreement with previous results \cite{TangneyPRL1998,TangneyPRB2002,CrepaldiPRL2020}. For the intermediate structures, we displace the Te atom along the phonon eigenvector at several values between $x_{eq}=0.269$ and $x_{SPT}=0.333$. We then obtain the electronic structure with the new lattice structures. Within the frozen-phonon approximation, these intermediate states reflect the metastable structure of the photo-induced states at different excitation levels. The results of $x_{eq}=0.269$ and $x_{SPT}=0.333$ are shown in the main text, while the band structures of several intermediate values are shown in Fig. S1. As the structure approaches the high-symmetry phase, three qualitative changes of band structure occur. First, due to the restoration of higher symmetry, the band splitting vanishes. This phenomenon is most evident along the K-H cut, where six bands become degenerate as the SPT occurs; Second, the conduction band bottom at A crosses the valence band top at H at around $x=0.283$, demonstrating a semiconductor-to-metal transition, before full metallization occurs at $x=0.319$, where the direct gap around A fully collapses. Third, due to the restoration of degeneracy, there is a dramatic decrease in the number of Weyl nodes at the H, $\Gamma$, and A points \cite{CrepaldiPRL2020}.

    \begin{figure}[h]
        \includegraphics[width=6.75in]{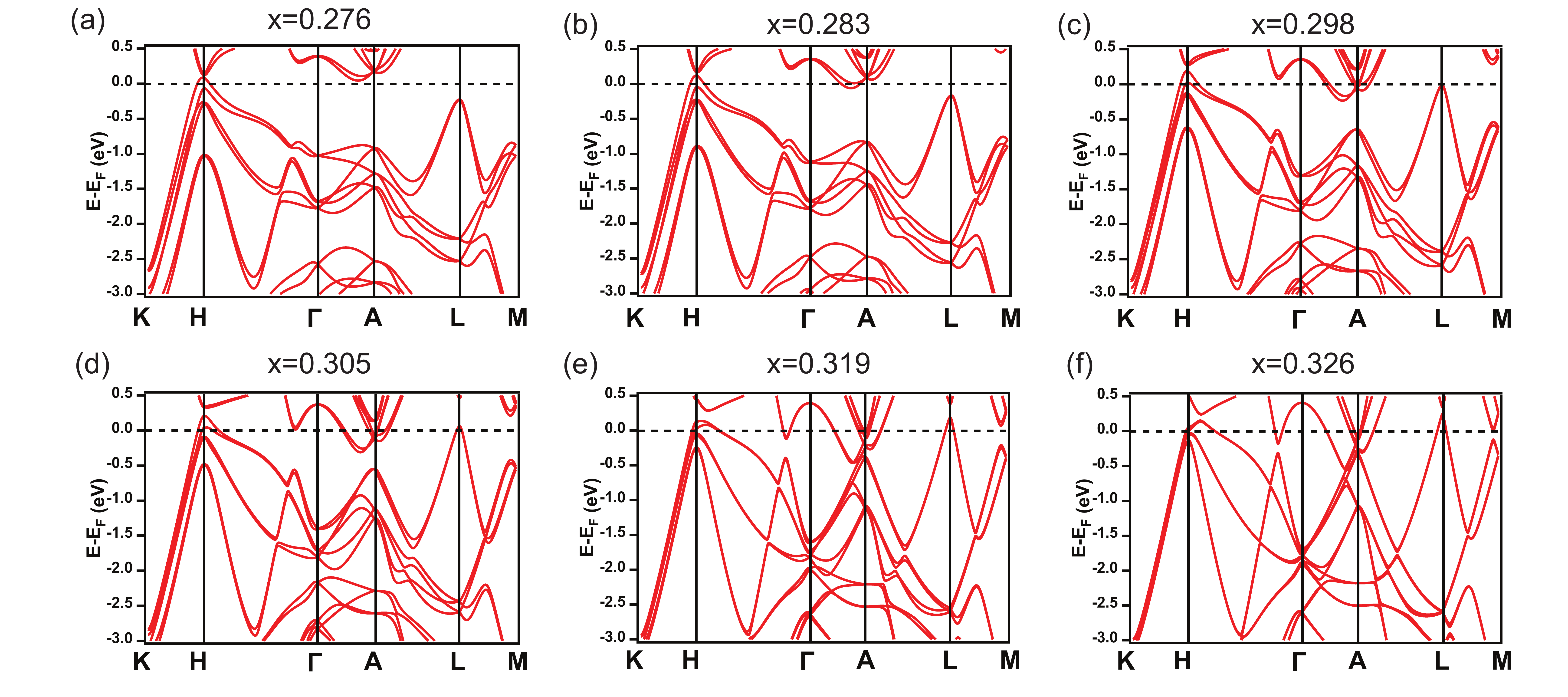}
        \label{FigS9}
        \caption{DFT calculated band structures at several intermediate coordinates between $x_{eq}=0.269$ and $x_{SPT}=0.333$.}
    \end{figure}

\newpage
\section{II. Time-Dependent Density Functional Theory Simulations}
    \begin{figure}[h]
        \includegraphics[width=5in]{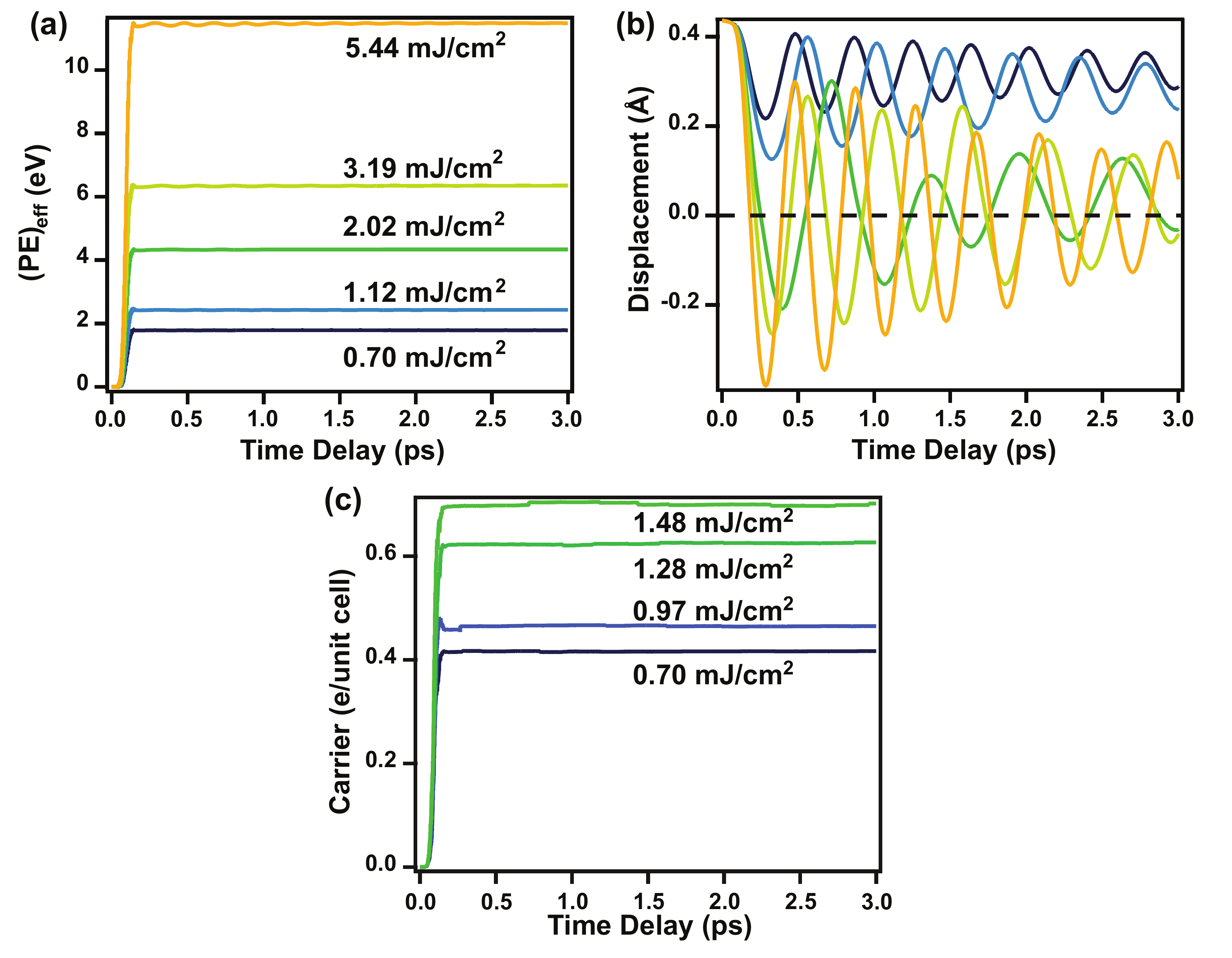}
        \label{FigS8}
        \caption{Temporal evolution of the TDDFT-calculated (a) effective potential energy , (b) atomic displacement, and (c) carrier density (electrons per unit cell) at selected absorbed fluences. The dashed line in (b) indicates $x_{SPT}$.}
    \end{figure}

We used our in-house time-dependent \textit{ab initio} package (\textsf{TDAP})~\cite{Lian2018MultiK, Lian2018AdvTheo, Lian2019CDWDyn} for real time (RT)-TDDFT calculations~\cite{Runge1984, Bertsch2000, Wang2015RTTDDFT}. The time-dependent Kohn-Sham (TDKS) equation at time $t$ in the plane wave (PW) basis $\{\mathbf{G}\}$ reads~\cite{Runge1984}:
\begin{equation}
\label{eq.TDKS}
i\hbar\frac{\partial \psi_{\gamma \mathbf{k}}(\mathbf{G},t)}{\partial t} = \mathcal{H}_{\mathbf{k}}(t) \psi_{\gamma \mathbf{k}}(\mathbf{G},t)
\end{equation}
where $\psi_{\gamma \mathbf{k}}(\mathbf{G},t)$ is the TDKS orbital, $\gamma$ denotes the band index, and $\mathbf{k}$ is the reciprocal momentum index.
$\mathcal{H}_{\mathbf{k}}(t)$ is the Hamiltonian expanded in a plane-wave basis with matrix element 
\begin{equation}
\begin{split}
\mathcal{H}_{\mathbf{k}}(\mathbf{G},\mathbf{G'},t) =& T_\mathbf{k}(\mathbf{G},\mathbf{G'},t) + V(\mathbf{G},\mathbf{G'},t) \\
=& \frac{\hbar^2}{2m}|\mathbf{k}+\mathbf{G}+\mathbf{A}(t)|^2 \delta_{\mathbf{G},\mathbf{G'}} + V(\mathbf{G},\mathbf{G'},t)
\end{split}
\end{equation} 
where $T_\mathbf{k}(\mathbf{G},\mathbf{G'}) = \frac{\hbar^2}{2m}|\mathbf{k}+\mathbf{G}+\mathbf{A}(t)|^2 \delta_{\mathbf{G},\mathbf{G'}}$ is the kinetic term. $\mathbf{A}$ is the velocity gauge potential~\cite{Bertsch2000}:
\begin{equation}
\mathbf{A}(t) = -\int_0^t \mathbf{E}(t') dt'
\end{equation}
where $\mathbf{E}$ is the electric field. $V(\mathbf{G},\mathbf{G'})$ is the potential term calculated within the corresponding module \textsf{Quantum Espresso}, including ion-electron potential, Hartree potential, and exchange-correlation potential.

Once the self-consistency in charge density evolution is satisfied, post-processing including the calculation of total energy, Hellmann-Feynman forces, and the ionic trajectory are invoked. For instance, the forces acting on the ions can be calculated through 
\begin{equation}
\mathbf{F}_{\mathbf{R}_I} = \sum_{\gamma\mathbf{k}} \braket{\psi_{\gamma\mathbf{k}}|\nabla_{\mathbf{R}_I} \mathcal{H}|\psi_{\gamma\mathbf{k}}},
\end{equation}
where $\mathbf{R}_I$ and $\mathbf{F}_{\mathbf{R}_I}$ are the position and force of the $I$th ion.

With $\mathbf{R}_I$ and $\mathbf{F}_{\mathbf{R}_I}$, we utilize the Ehrenfest theorem for evolving ions according to the equation of motion \begin{equation}
M_I \frac{d^2\mathbf{R}_I}{dt^2} = \mathbf{F}_{\mathbf{R}_I},
\end{equation}
where $M_I$ is the mass of $I$th ion. The velocity $v_I(t) = d\mathbf{R}_I/dt$ and the temperature $T(t) = \sum_I^{N_I} M_I v^2_I(t)/2N_I$ are also calculated, where $N_I$ is the total number of ions.

Also note that the Ehrenfest dynamics describe a microcanonical system where only electron-electron energy transfer and electron-phonon scattering are allowed. We neglected radiative recombination and heat radiation. Moreover, since periodic boundary conditions are utilized in the simulation, the whole crystal is homogeneously excited when the laser pulse is present and there is no driven mechanism for thermal transport. Thus, the total energy of the system is conserved. 

In both the DFT and RT-TDDFT calculations, we used the Perdew-Burke-Ernzerhof (PBE) exchange-correlation (XC) functional~\cite{Perdew1996} and norm-conserving pseudo-potential from the \textsf{PseudoDojo}~\cite{vanSetten2018} database. 
The plane-wave energy cutoff was set to 70~Ry, and the Brillouin zone was sampled using a Monkhorst-Pack scheme with an $12\times 12 \times 8$ $\mathbf{k}$-point mesh. In RT-TDDFT calculations, the electronic timestep, $\delta t$, was set to $1.94\times10^{-4}$~fs, and the ionic timestep, $\Delta t$, was $0.194$~fs. A Gaussian-type laser pulse $\label{eq:GaussianWave}
\mathbf{E}(t)=\mathbf{E}_0\cos\left(\omega t \right)\exp\left[-\frac{(t-t_0)^2}{2\sigma^2}\right]$ is utilized,
where $|\mathbf{E}_0|$ is the peak electric field, $\hbar{\omega}=1$ eV is the pump frequency, $\sigma=100$ fs is the pulse width and $t_0$ is the peak time set at 50 fs.

We show the temporal evolution of the TDDFT-calculated effective potential energy (PE)$_{\textrm{eff}}$, lattice displacement, and carrier density (electrons per unit cell) pumped with fluences across a large range in Fig. S2. As shown in Fig. S2(a), (PE)$_{\textrm{eff}}$ displays a sharp increase proportional to the absorbed fluence within the pump pulse duration followed by an oscillation due to the presence of phonons. The atomic displacement [Fig. S2(b)] shows a clear oscillation representing the $A_1$ mode at all fluences. When the fluence is higher than a critical value, the lattice oscillates around $x_{SPT}$, corresponding to the high-symmetry phase. Fits to these atomic displacement curves produce the simulation results shown in Figs. 3 and 4 in the main text. The free carrier density [Fig. S2(c)] exhibits an initial increase within the pump duration similar to (PE)$_{\textrm{eff}}$. After the initial excitation, the carrier density stays nearly constant over the sampled time window, indicating slow recombination dynamics and thus metastability.

\newpage
\section{III. Static Second Harmonic Generation}

\subsection{Sample Preparation and Characterization}
A conventional Bridgman method was used to produce high-quality single crystals of tellurium with typical dimensions of 3 mm $\times$ 3 mm $\times$ 0.5 mm. The as-grown face of the crystals were polished with 1 $\mu$m lapping film before measurement. Static Raman spectroscopy characterization was measured using a commercial Renishaw M1000 Micro Raman Spectrometer System with 514 nm laser excitation in the backscattering configuration under ambient conditions. The signal was measured with a spectral resolution of 1 cm$^{-1}$.

\subsection{Static SHG Simulation and Experiment}
Static second harmonic generation (SHG) rotational anisotropy (RA) in parallel (S\textsubscript{in}-S\textsubscript{out}) and crossed (S\textsubscript{in}-P\textsubscript{out}) polarization channels were measured using 1.5 eV probe light at an incident angle $\theta=10^{\circ}$. Te single crystals are polished with the $c$- and $a$-axes parallel to the surface plane. The experimental data in the two channels are shown in Fig. S3. Note that the intensity of the crossed polarization channel is nearly an order of magnitude smaller than for parallel polarization.

To fit the equilibrium SHG data, we calculated SHG-RA patterns in the S\textsubscript{in}-S\textsubscript{out} and S\textsubscript{in}-P\textsubscript{out}  channels under the electric-dipole approximation from a $D_{3}$ point group. We note that the $D_{3d}$ point group of the high-symmetry phase preserves inversion symmetry and thus forbids electric-dipole SHG \cite{TangneyPRL1998}. The electric-dipole SHG process is described by: $P_{i}(2\omega)\propto\chi_{ijk}(2\omega;\omega;\omega)E_{j}(\omega)E_{k}(\omega)$,
where the indices run over the Cartesian coordinates, $E(\omega)$ is the incoming electric field, $P$ is the electrical polarization induced in the sample, and $\chi$ is the second-order optical susceptibility tensor, which by Neumann's principle must respect the crystalline point group symmetry \cite{HsiehRSI2014}. Imposing the $D_{3}$ point group symmetries as well as the index permutation symmetry relevant to degenerate SHG, the original 27 independent elements of the rank-3 tensor ${\chi}$ reduce to only two independent nonzero elements: ${\chi}_{xxx}=-{\chi}_{xyy}=-{\chi}_{yyx}=-{\chi}_{yxy}$, and ${\chi}_{xyz}={\chi}_{xzy}=-{\chi}_{yxz}=-{\chi}_{yzx}$. To simulate the rotating scattering plane, we apply a basis rotation from the original tensor given above to one which has been rotated through an angle $\phi$ about the surface normal. Finally, the expression for the $\phi$-dependent SHG intensity is given by: 
\begin{equation}
    I(2\omega,\phi) \propto \left| \hat{e}_{i}(2\omega) {\chi}_{ijk}(\phi) \hat{e}_{j}(\omega) \hat{e}_{k}(\omega)\right|^{2} \times I(\omega)^{2}
\end{equation}
where $I(\omega)$ is the intensity of the incoming light. The unit vectors $\hat{e}$ represent the polarization of the incoming and outgoing light, which are selected to be either S or P polarized. The $I(2w,\phi)$ expressions for the polished surface of Te are thus given by:
\begin{equation}
    \begin{aligned}
        I_{SP}(2\omega, \phi) &\propto \left[-2\chi_{xyz}\sin(\theta)\sin(\phi)\cos(\phi) + \chi_{xxx}\cos(\theta)\cos(\phi)\sin^{2}(\phi) \right]^{2}\\
        I_{SS}(2\omega, \phi) &\propto \chi_{xxx}^{2}\sin^{6}(\phi)
    \end{aligned}
\end{equation}

These formulas provide a good fit to the experimental data as shown in Fig. S3. The ratio between $\chi_{xxx}$ and $\chi_{xyz}$ obtained from fitting is 1:0.05, in agreement with a recent theoretical SHG simulation \cite{GGYPRB2019}. 

\begin{figure}[h]
\includegraphics[width=3.5in]{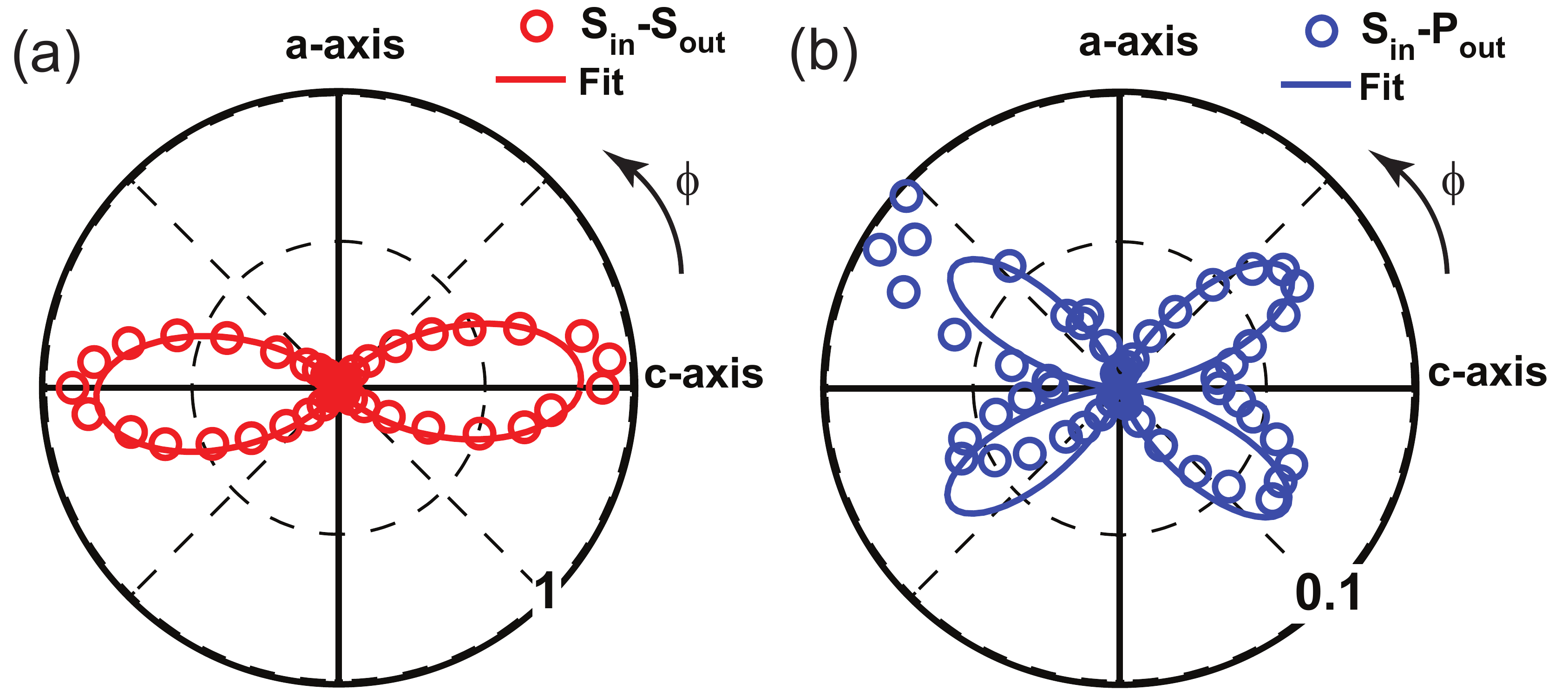}
\label{FigS5}
\caption{Static SHG-RA patterns in the (a) S\textsubscript{in}-S\textsubscript{out} and (b) S\textsubscript{in}-P\textsubscript{out} channels. The intensity is normalized to the peak intensity in the S\textsubscript{in}-S\textsubscript{out} channel.}
\end{figure}

\newpage
\section{IV. Time-Resolved Second Harmonic Generation}

\subsection{Experimental Setup}
Time-resolved SHG-RA measurements were performed using a rotating scattering plane-based technique \cite{HsiehRSI2014}. Ultrafast optical pulses (40 fs duration, 800 nm center wavelength) were generated by an amplified Ti:sapphire laser operating at a 1 kHz repetition rate. Part of the laser output was used as the probe beam, while another part was used to seed an optical parametric amplifier to generate the pump beam at 1200 nm. For the probe beam we used linear polarization, a spot size (FWHM) of 40 $\mu$m, an angle of incidence of $10^\circ$ and an absorbed fluence of 1.1 mJ/cm\textsuperscript{2}. For the pump beam we used circular polarization, a spot size (FWHM) of 50 $\mu$m and an angle of incidence of $0^\circ$. Transient SHG data were collected using a time step of 50 fs. The absorbed fluence values presented in the main text are related to the applied fluence values by a factor $1-R$, where $R$ = 0.54 (0.56) is the reflectivity at 1 eV (1.55 eV) \cite{DrewsPR1969}. Static and time-resolved SHG-RA patterns were collected within a 5 min exposure time on an electron multiplying CCD camera for each time delay. All measurements were conducted at room temperature.

\subsection{Demonstration of $A_{1}$ Phonon Symmetry}
    
A coherent $A_1$ phonon is symmetry preserving and can thus only alter the magnitude but not the symmetry of the SHG-RA patterns. In Fig. S4 we show the absolute and normalized transient SHG intensity at different fixed $\phi$ values. As seen in Fig. S4(b), all curves collapse onto one another indicating a uniform modulation of all SHG tensor elements, consistent with a symmetry preserving $A_1$ mode. 

\begin{figure}[h]
    \includegraphics[width=6.75in]{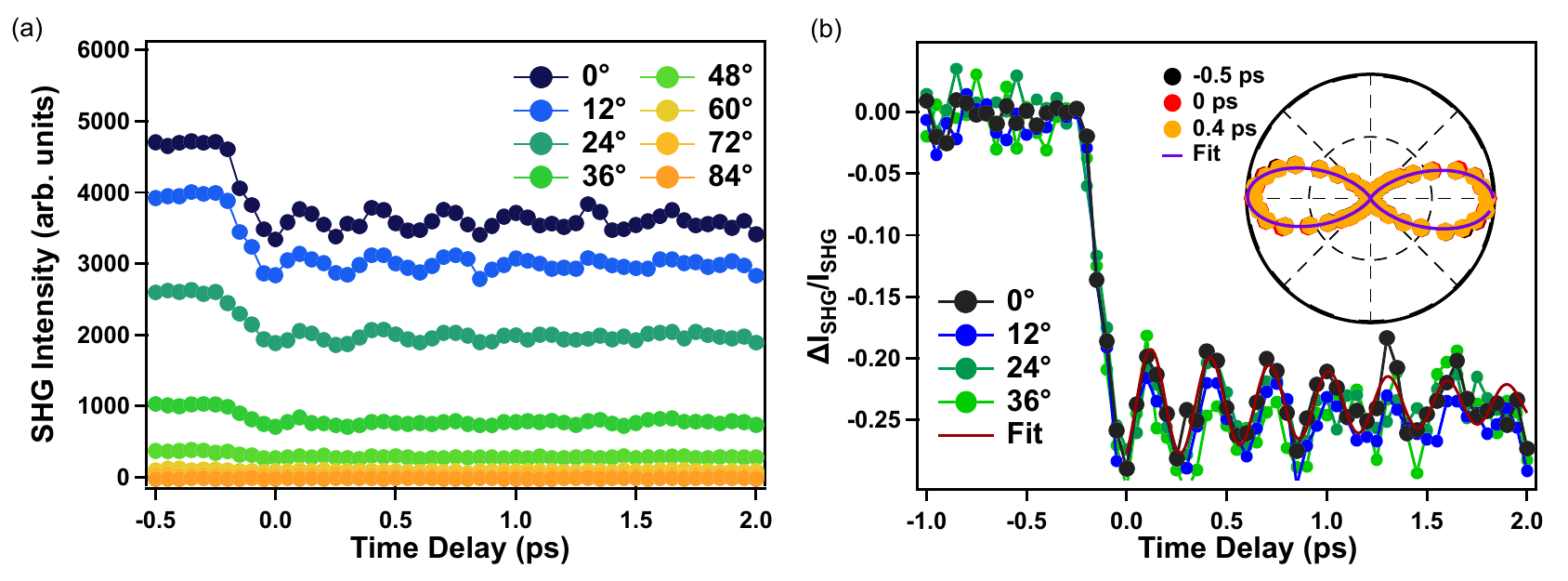}
    \label{FigS2}
    \caption{(a) Temporal evolution of SHG-RA intensity at select angles in the S\textsubscript{in}-S\textsubscript{out} channel with an absorbed pump fluence of around 1.22 mJ/cm\textsuperscript{2}. (b) The same data as in (a) but with each trace shown as a differential change in the SHG intensity normalized to its equilibrium value. The data for each angle is well-fit by a single decaying sinusoidal function, shown as a solid maroon curve. The inset shows scaled SHG-RA patterns at $t$ = -0.5 ps, 0 ps and 0.4 ps, which can all be fit with the same curve.}
\end{figure}

\newpage   
\subsection{Long Timescale Dynamics}
In Fig. S5 we plot differential SHG transients above and below the critical fluence out to 10 ps to show the longer timescale dynamics. These data were taken with a larger time step and so the phonon is not clearly resolved. Both traces show no measurable recovery towards equilibrium within 10 ps, consistent with metastable behavior.
    
\begin{figure}[h]
    \includegraphics[width=3.375in]{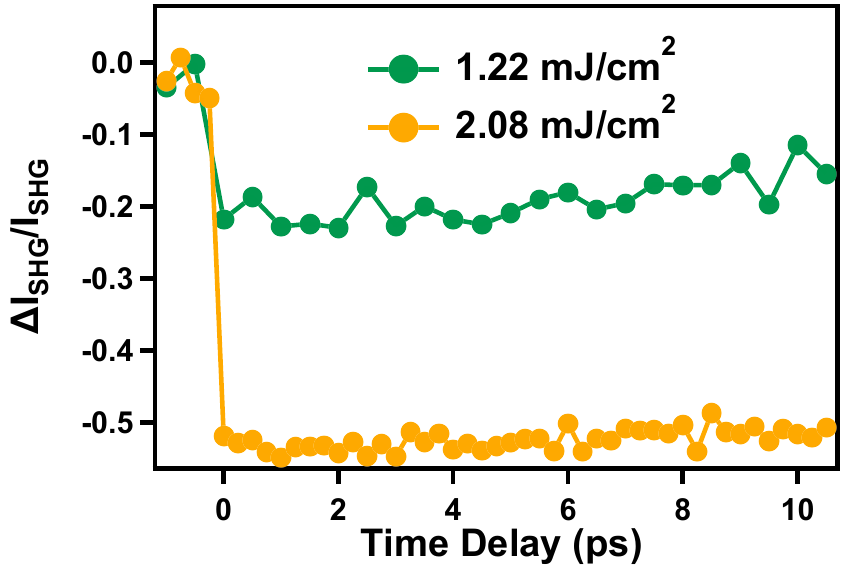}
    \label{FigS3}
    \caption{Temporal evolution of SHG intensity at two characteristic fluences for a fixed $\phi$ in the S\textsubscript{in}-S\textsubscript{out} channel.}
\end{figure}
\vspace{1cm} 

\subsection{Comparison of SHG Dynamics under Different Polarization Geometries}
Fig. S6 shows normalized differential SHG transients in the S\textsubscript{in}-P\textsubscript{out} and S\textsubscript{in}-S\textsubscript{out} channels acquired using the same absorbed pump fluence of 1.22 mJ/cm\textsuperscript{2}. Since these two channels are sensitive to different tensor elements (Section III), the fact that the curves overlap further confirms that all the SHG susceptibility tensor elements are suppressed by the same scale factor.
    
\begin{figure}[h]
    \includegraphics[width=3.375in]{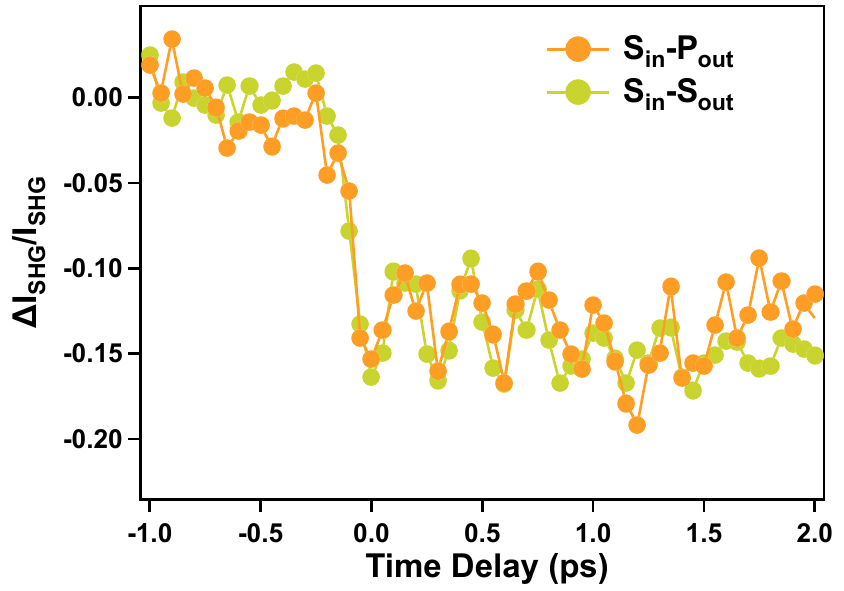}
    \label{FigS4}
    \caption{$\Delta I_{SHG}/I_{SHG}$ curves measured at the $\phi$ angle where the intensity is maximum in the S\textsubscript{in}-P\textsubscript{out} and S\textsubscript{in}-S\textsubscript{out} channels. }
\end{figure}

\newpage
\section{V. Eliminating Alternative Interpretations of SHG Dynamics}
\subsection{Thermal Heating}
To estimate the temperature increase induced by a single laser pulse, we use the equation $\Delta T = \frac{(1-R)F}{C\rho\delta}$, where $R$ is the reflectivity of the sample, $\rho$ is the sample density in g/cm$^{3}$, $C$ is the heat capacity in mJ (g K)$^{-1}$, and $\delta$ is the optical penetration depth of the pump at 1 eV. For Te, we use $R$ = 0.54, $\rho$ = 6.24 g/cm$^{3}$ and $\delta$ = 76 nm for 1 eV light \cite{DrewsPR1969}. We approximate $C$ to be constant at 202 mJ (g K)$^{-1}$ across the entire $T+\Delta T$ range, which is valid because $C$ varies by less than 5$\%$ between 300 K and 500 K \cite{PYZNCOMM2016}. Using these values, and a critical absorbed fluence $(1-R)F_c$ = 2 mJ/cm$^{2}$, we are left with a total temperature increase of around 190 K. Cumulative heating due to the laser pulses causes a static temperature increase that is estimated to be $\Delta T = \sqrt{\ln{2}/\pi}\frac{P}{l\kappa}$, where $l$ = 50 $\mu$m is the spot size (FWHM) of the Gaussian beam, $P$ is the laser power (100 $\mu$W at $F_c$), and $\kappa$ is the thermal conductivity (1.6 Wm$^{-1}$K$^{-1}$ at 300 K \cite{PYZNCOMM2016}). This gives rise to a temperature increase of 0.6 K, which is negligible compared to instantaneous temperature increase.

According to the equilibrium phase diagram of Te, there is no structural phase transition under ambient pressure until liquefaction at around 700 K \cite{HejnyPRL2003,HejnyPRB2004}. Under pressure, the rhombohedral centrosymmetric phase of Te (dubbed as Te-VI) only exists above room temperature at high pressures of order 20 GPa (compressing the lattice volume by 15\%) \cite{HejnyPRB2006}. Therefore we can rule out a thermally driven transition into the $D_{3d}$ phase.

Although both the chiral chain radius and lattice constant $a$ are known to increase with temperature, their ratio ($x$) changes by only 0.1$\%$ between 300 K and 500 K \cite{DeshpandePhysica1965}. This subtle change is unable to generate the observed large SHG intensity drop in our experiments. Moreover, this change is monotonic up to the melting point, which cannot explain our observed nonlinearity in the SHG and phonon dynamics across the critical fluence.

\vspace{1cm} 
\subsection{Laser Melting and Amorphization}    

A photo-thermal induced melting transition was previously reported in single crystalline Te based on the loss of optical anisotropy upon pumping with 100 fs long 1.5 eV pulses \cite{LindeJETP2002}. This transition was observed above approximately 15 mJ/cm$^2$ and occurs over a several picosecond timescale that decreases with fluence. In contrast, our experiments are performed at much lower fluence and the timescale over which our SHG signal drops is less than 0.2 ps and almost fluence independent. 

A reversible transition between crystalline and amorphous phases can also be realized in Te via optical illumination. We rule out this optically induced order-to-disorder transitions for two reasons. First, the amorphous phase of Te exhibits a broad mode centered at 4.7 THz in its spontaneous Raman spectrum, which is absent in the FFT spectrum of our signal \cite{BrodskyPSS1972}. Second, a recent single-shot transient reflectivity measurement on a polycrystalline Te thin film using 60 fs long 1.5 eV pulses reported laser-induced amorphization above a threshold fluence of around 6 mJ/cm$^2$ \cite{NelsonPRB2018}. Preceding this transition, inflections were observed in the fluence dependence of both the $A_1$ phonon amplitude and frequency, reminiscent of our reported TDDFT and SHG results. The authors speculated that this feature could indicate coherent phonon overshoot of a high symmetry point, in line with our picture of an inverse Peierls transition. 

\newpage
\section{VI. Time-Dependent Landau Theory Simulations}

To simulate the effects of a penetration depth mismatch between the pump and probe light on the transient SHG response, we used a phenomenological time-dependent Landau theory (TDLT) model \cite{JohnsonPRL2014} based on a dynamical double-well potential of the form:
\begin{equation}
    V(x,t)=\frac{a}{2}[\eta(t)-1]x^2+\frac{1}{4}x^4+kx,
\end{equation}
where $\eta(t)$ denotes the fluence- and time-dependent pump-induced change of the potential that onsets at time zero, and $k$ is a small slope added to prevent the system from overshooting to the other valley. In the un-pumped case, the system resides in the lower minimum of the two valleys separated by a Peierls barrier. The nonzero displacement $x$ denotes the Peierls distortion. The excitation of light increases $\eta$ from 0 (equilibrium value), leading to a quench of the order parameter. When $\eta>1$ the double well potential transforms into a parabola. 

The dynamical equation of the Peierls distortion $x$ following light excitation can be thus expressed as:
\begin{equation}
    \frac{1}{\omega_{ph}^2}\frac{\partial^2x}{{\partial}t^2}+\frac{2\gamma_{ph}}{\omega_{ph}}\frac{{\partial}x}{{\partial}t}+\frac{{\partial}V(x,t)}{{\partial}x}=0,
\end{equation}
where $\omega_{ph}$ is the phonon frequency and $\gamma_{ph}$ is the phonon lifetime. We adopt $a=0.55$, $k=0.01$, $\omega_{ph}=3.6$ THz, and $\gamma_{ph}=1$ THz to match the experimental phonon frequency and lifetime, leaving $\eta$ as the only free parameter.

Due to the pump-probe penetration depth mismatch, different depths $z$ below the sample surface experience a different level of quenching of the potential. The reported pump and probe penetration depths are $\delta_{pu}=$76 nm and $\delta_{pr}=$38 nm, respectively, an approximate two-fold difference \cite{DrewsPR1969}. We assume the sample is composed of thin layers with thickness $d=$ 1 nm and an exponential $z$-dependence $\eta(z,t)=\eta(t)\exp(-z/\delta_{pu})$ arising from an exponentially decaying photo-carrier distribution. We then solve $x(z,t)$ for each layer. The depth integrated value of $x$ is evaluated by summing over all layers, with each layer weighted by the probe penetration depth as $x_{int}(t)=\sum^\infty_{z=0}\exp(-z/\delta_{pr})x(z,t)$. The relative SHG intensity change is then defined as ${\Delta}I_{SHG}/I_{SHG}=[x_{int}(t)-x_{int}(0)]/x_{int}(0)$ [Fig.S7(a)]

    \begin{figure}[h]
        \includegraphics[width=6.7 in]{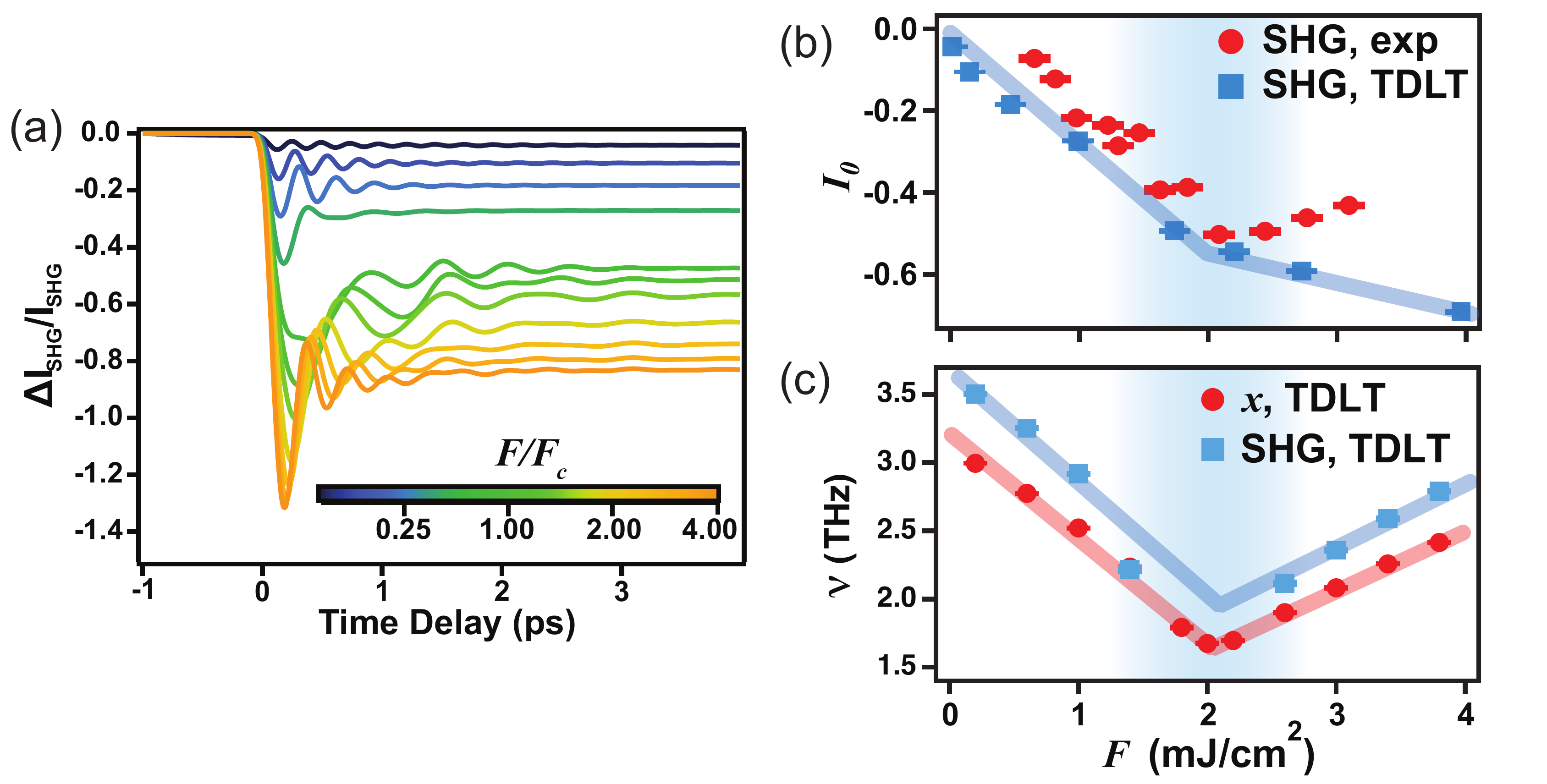}
        \label{FigS10}
        \caption{(a) TDLT simulation of relative SHG change as a function of time at different pump fluences. (b) TDLT simulation of SHG background intensity as a function of pump fluence with the experimental results overlaid. (c) Phonon frequencies as a function of fluence obtained from fitting to TDLT simulated ${\Delta}I_{SHG}/I_{SHG}(t)$ and $x(z=0,t)$. Both frequencies are scaled by the same factor to match the TDDFT results.}
    \end{figure}

By fitting the simulated ${\Delta}I_{SHG}/I_{SHG}(t)$ curves with the same formula used for the experimental ${\Delta}I_{SHG}/I_{SHG}$ curves, namely a decaying sinusoidal function plus a residual background $I_0$, we obtain the simulated $I_0$ as a function of pump fluence [Fig.S7(b)]. A decrease in slope is most pronounced near $F_c$, suggesting that the penetration depth mismatch may be partially responsible for the observed saturation-like behavior in the SHG intensity at $F_c$. The difference between simulation and experiment, which is most obvious above 2.2 mJ/cm$^2$, may be due to other factors such as spatial non-uniformity of the pump intensity, quench-induced spatial domains and defects, and higher-order multipole SHG radiation processes that were not considered in our simulation.

We want to note here that one should not interpret Fig. S7(b) as being a more refined version of Fig. 3(i). The theoretical curves shown in Fig. 3(i) and Fig. S7(b) are calculated using completely different methods. The former is an \textit{ab initio} TDDFT simulation that does not consider penetration depth mismatch, while the latter is a time-dependent Landau theory simulation, which is much cruder but does allow for consideration of penetration depth mismatch.

The penetration depth mismatch can also induce a disparity in phonon frequency between TDDFT simulations and SHG experiments, which can be quantitatively simulated by our TDLT model. We extracted the fluence dependence of the phonon frequency in two different ways, by fitting $x(z=0,t)$ and by fitting ${\Delta}I_{SHG}/I_{SHG}(t)$. We find that the former always has a lower value than the latter [Fig. S7(c)]. The reason is straightforward: the deeper the probed region lies below the surface, the lower the effective pump fluence, and thus the weaker the laser-induced phonon frequency softening. Since our TDDFT calculation directly simulates $x(z=0,t)$ while our SHG experiment measures ${\Delta}I_{SHG}/I_{SHG}(t)$, the latter should always exhibit a higher phonon frequency than the former. As a corollary, one would expect that Te thin-films should exhibit a larger frequency softening than the bulk crystals for the same applied pump fluence, which is indeed the case reported in previous studies of thin-films \cite{NelsonPRB2018} and bulk crystals \cite{KurzAPA1996,MazurPRB2007,SoodPRB2010}. 

Although our TDLT calculations provide some physical reasoning for the quantitative discrepancy between the theoretically predicted and experimentally measured phonon frequencies, a quantitative comparison between theory and experiment is challenging for the following reasons. First, DFT calculations typically underestimate phonon frequencies \cite{TangneyPRB2002}. Second, there is considerable variation of the reported phonon softening between polycrystalline and single-crystalline samples \cite{KurzAPA1996,MazurPRB2007,SoodPRB2010}. Third, many factors not considered in our TDDFT calculations, such as mismatch of pump-probe spot sizes and carrier diffusion, can all affect the measured phonon softening effects. Therefore, in the main text, we refrain from demanding a quantitative match between the calculated and measured phonon frequencies. 

\newpage
\section{VII. Elaboration on Different Displacive Excitation Mechanisms}

Tellurium is a prototypical Peierls-distorted material where optical carrier excitation can induce a sudden potential energy minimum shift. Since the lattice cannot adiabatically follow the prompt change of potential energy and remains at its equilibrium value, an effective displacive force will be imparted to the lattice and a coherent oscillation of the fully symmetric $A_{1(g)}$ mode will be initiated \cite{DresselhausPRB1992}. The key ingredients of this mechanism - dubbed displacive excitation of coherent phonons (DECP) - are optical-absorption-induced carrier excitation and electron-phonon coupling. DECP-launched phonons have been widely observed in a variety of systems including A7-structured semimetals \cite{NelsonPRX2018,NelsonPRB2018}, VO$_2$ \cite{WallNCOMM2012} and charge density wave materials \cite{WallPRL2012,JohnsonPRL2014,BeaudNMAT2014}. 

The general equation of motion of a DECP-launched phonon $Q_R$ can be expressed as a driven damped harmonic oscillator with the driving force proportional to the excited carrier density $n(t)$ \cite{DresselhausPRB1992}:
\begin{equation}
\begin{split}
    & \frac{dn(t)}{dt}=F(t)-\beta n(t), \\
    \frac{d^2Q_R(t)}{dt^2}+&2\gamma_R\frac{dQ_R(t)}{dt}+\omega_R^2Q_R(t)\propto n(t),
\end{split}    
\end{equation}
where $\beta$ and $\gamma_{R}$ are the damping rate of excited carrier density and phonon, respectively, and $\omega_R$ is the Raman-active phonon frequency. If one assumes $\beta=0$ and an instantaneous optically induced force $F(t)=I_0 \delta(t)$, where $\delta(t)$ is Dirac $\delta-$function, the solution can be expressed as \citep{MisochkoPRB2016}:
\begin{equation}\label{eq9}
    Q_R(t)\propto Im(\epsilon)I_0H(t)\{1-e^{-\gamma_Rt}[\cos(\sqrt{\omega_R^2-\gamma_R^2} t)+\frac{\gamma_R}{\sqrt{\omega_R^2-\gamma_R^2}}\sin(\sqrt{\omega_R^2-\gamma_R^2}t)]\},
\end{equation}
where $\epsilon$ is the dielectric constant of the material and $H(t)$ is the Heaviside step function. This formula clearly shows that a damped sinusoidal oscillation will be launched upon optical excitation at an offset from zero, and that the amplitude of the mode depends on the absorption determined by $Im(\epsilon)$.

Another mechanism dubbed ionic Raman scattering (IRS) can also generate a similar potential shift or quenching. It is posited that resonant excitation of an infrared-active phonon $Q_{IR}$ can mediate Raman phonon generation $Q_{R}$, a process that relies on lattice anharmonicity in lieu of electron phonon coupling \cite{CavalleriACR2015,CavalleriNPHY2011}. This so-called ``nonlinear phononics" mechanism provides a unique pathway to optically generate Raman active phonons without creating charge excitations. Remarkable progress in IRS based materials properties engineering has been demonstrated in the past decade, including structural transitions \cite{SpaldinPRL2017}, insulator-to-metal transitions \citep{CavalleriNature2007}, light-induced superconductivity \citep{CavalleriScience2011,CavalleriNature2014,CavalleriNature2016}, magnetic switching \citep{RondinelliPRB2018,CavalleriNPHY2016,CavalleriScience2019}, modulation of magnetic interactions \citep{CavigliaNMAT2021}, and mapping of interatomic anharmonic potential energy surfaces \citep{CavalleriNature2018}.

Without loss of generality, the lowest order coupling between $Q_{IR}$ and $Q_{R}$ in a centrosymmetric system is $aQ_{IR}^2Q_R^2$, where $a$ is the coupling constant. If we consider an impulsive optical stimulus $F(t)=I_0\delta(t)$ that directly couples to $Q_{IR}$, we can write out the equations of motion for both phonon coordinates as:
\begin{equation}
\begin{split}
        \frac{d^2Q_{IR}(t)}{dt^2}+2\gamma_{IR}\frac{dQ_{IR}(t)}{dt}+\omega_{IR}^2Q_{IR}(t) &=F(t) + 2aQ_{IR}(t)Q_{R}(t), \\
        \frac{d^2Q_{R}(t)}{dt^2}+2\gamma_{R}\frac{dQ_{R}(t)}{dt}+\omega_{R}^2Q_{R}(t) &=aQ_{IR}^2(t),
\end{split}
\end{equation}
where $\gamma_{IR,R}$ and $\omega_{IR,R}$ are the damping rate and frequency of the IR or Raman mode, respectively. If one assumes $\gamma_{IR}=0$ and $\omega_{IR}\gg\omega_{R}$, the solution of $Q_R$ harbors the exactly same form as the eq.\ref{eq9}. Fig.S8 shows the simulation of $Q_R(t)$ obtained by solving the equations of motion for DECP and IRS, demonstrating that one is able to induce a lattice potential energy shift and displacively launch Raman-active phonons via either mechanism. However, we emphasize that neither of these mechanisms has so far been explored for ultrafast control of band topology. 

    \begin{figure}[h]
        \includegraphics[width=3 in]{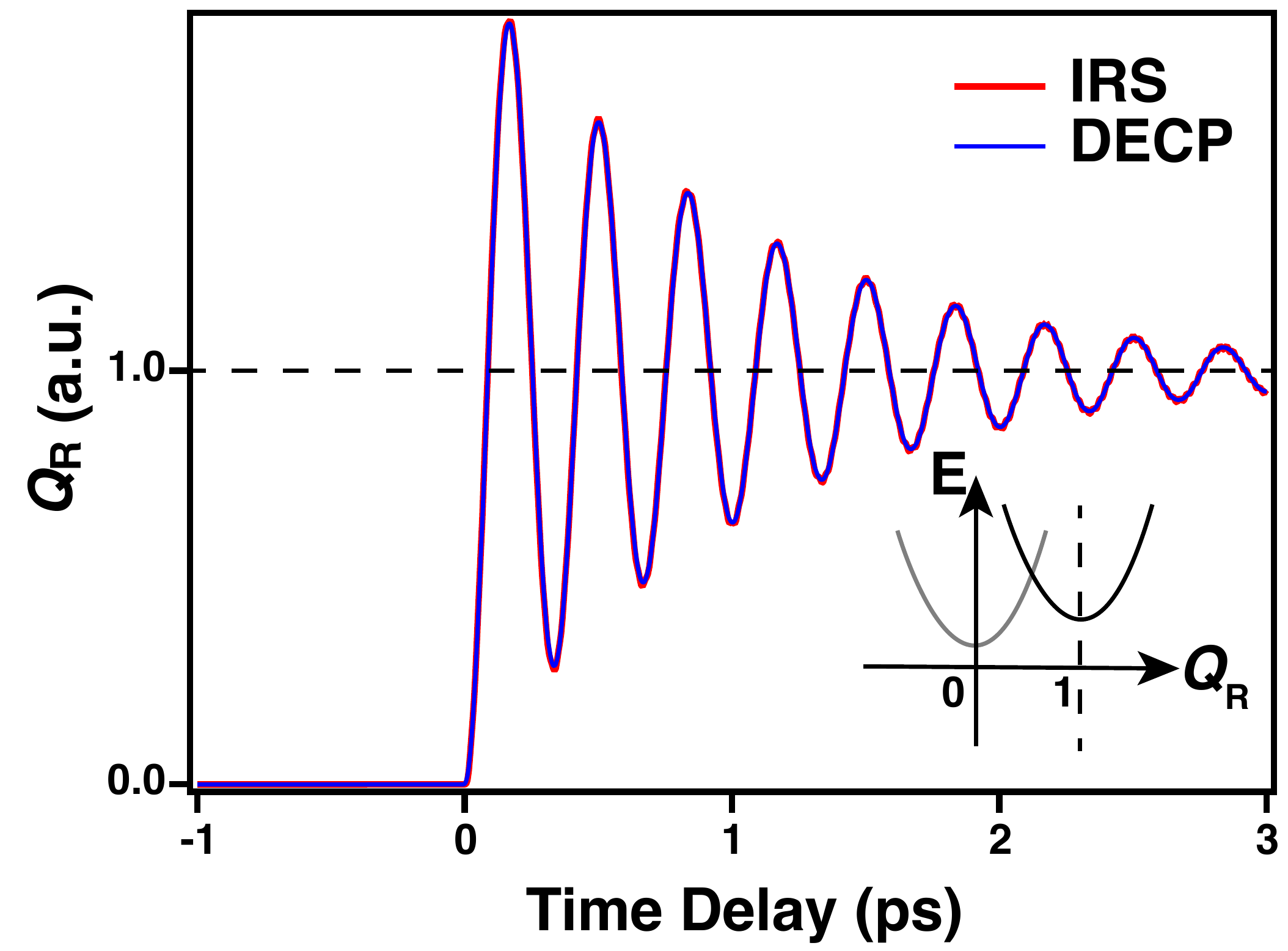}
        \label{FigS12}
        \caption{Simulation of DECP and IRS-launched Raman phonons. The two curves are vertically scaled. A set of characteristic parameter values $\gamma_{R}=1$ THz, $\omega_{R}=2\pi\cross3$ THz, $\gamma_{IR}=0$ THz, $\omega_{IR}=2\pi\cross20$ THz, $\beta=0$, $a=1$ is used. $I_0$ merely scales the curves. The inset shows a schematic of the potential energy shift.}
    \end{figure}

Despite the aforementioned similarities, DECP and IRS are in general considered as distinct mechanisms for the following reasons: first, the former depends on charge excitations while the latter does not; second, only $A_{1(g)}$ Raman-active phonons can be generated by DECP, while other Raman-active phonons can in principle be launched by IRS as long as the symmetry requirements of the infrared-active mode are satisfied; third, since DECP requires making charge excitations, pump light resonant with inter-band transitions in the near infrared or visible range is typically used. On the other hand, since IRS requires resonantly launching an infrared-active phonon, phase-stable mid-infrared light is typically employed.


\begin{thebibliography}{42}%
\makeatletter
\providecommand \@ifxundefined [1]{%
 \@ifx{#1\undefined}
}%
\providecommand \@ifnum [1]{%
 \ifnum #1\expandafter \@firstoftwo
 \else \expandafter \@secondoftwo
 \fi
}%
\providecommand \@ifx [1]{%
 \ifx #1\expandafter \@firstoftwo
 \else \expandafter \@secondoftwo
 \fi
}%
\providecommand \natexlab [1]{#1}%
\providecommand \enquote  [1]{``#1''}%
\providecommand \bibnamefont  [1]{#1}%
\providecommand \bibfnamefont [1]{#1}%
\providecommand \citenamefont [1]{#1}%
\providecommand \href@noop [0]{\@secondoftwo}%
\providecommand \href [0]{\begingroup \@sanitize@url \@href}%
\providecommand \@href[1]{\@@startlink{#1}\@@href}%
\providecommand \@@href[1]{\endgroup#1\@@endlink}%
\providecommand \@sanitize@url [0]{\catcode `\\12\catcode `\$12\catcode
  `\&12\catcode `\#12\catcode `\^12\catcode `\_12\catcode `\%12\relax}%
\providecommand \@@startlink[1]{}%
\providecommand \@@endlink[0]{}%
\providecommand \url  [0]{\begingroup\@sanitize@url \@url }%
\providecommand \@url [1]{\endgroup\@href {#1}{\urlprefix }}%
\providecommand \urlprefix  [0]{URL }%
\providecommand \Eprint [0]{\href }%
\providecommand \doibase [0]{http://dx.doi.org/}%
\providecommand \selectlanguage [0]{\@gobble}%
\providecommand \bibinfo  [0]{\@secondoftwo}%
\providecommand \bibfield  [0]{\@secondoftwo}%
\providecommand \translation [1]{[#1]}%
\providecommand \BibitemOpen [0]{}%
\providecommand \bibitemStop [0]{}%
\providecommand \bibitemNoStop [0]{.\EOS\space}%
\providecommand \EOS [0]{\spacefactor3000\relax}%
\providecommand \BibitemShut  [1]{\csname bibitem#1\endcsname}%
\let\auto@bib@innerbib\@empty
\bibitem [{\citenamefont {Yan}\ and\ \citenamefont
  {Felser}(2017)}]{FelserARCMP2017}%
  \BibitemOpen
  \bibfield  {author} {\bibinfo {author} {\bibfnamefont {B.}~\bibnamefont
  {Yan}}\ and\ \bibinfo {author} {\bibfnamefont {C.}~\bibnamefont {Felser}},\
  }\bibfield  {booktitle} {\emph {\bibinfo {booktitle} {Annual Review of
  Condensed Matter Physics}},\ }\href {\doibase
  10.1146/annurev-conmatphys-031016-025458} {\bibfield  {journal} {\bibinfo
  {journal} {Annual Review of Condensed Matter Physics}\ }\textbf {\bibinfo
  {volume} {8}},\ \bibinfo {pages} {337} (\bibinfo {year} {2017})}\BibitemShut
  {NoStop}%
\bibitem [{\citenamefont {H{\"u}bener}\ \emph {et~al.}(2017)\citenamefont
  {H{\"u}bener}, \citenamefont {Sentef}, \citenamefont {De~Giovannini},
  \citenamefont {Kemper},\ and\ \citenamefont {Rubio}}]{RubioNCOMMS2017}%
  \BibitemOpen
  \bibfield  {author} {\bibinfo {author} {\bibfnamefont {H.}~\bibnamefont
  {H{\"u}bener}}, \bibinfo {author} {\bibfnamefont {M.~A.}\ \bibnamefont
  {Sentef}}, \bibinfo {author} {\bibfnamefont {U.}~\bibnamefont
  {De~Giovannini}}, \bibinfo {author} {\bibfnamefont {A.~F.}\ \bibnamefont
  {Kemper}}, \ and\ \bibinfo {author} {\bibfnamefont {A.}~\bibnamefont
  {Rubio}},\ }\href {\doibase 10.1038/ncomms13940} {\bibfield  {journal}
  {\bibinfo  {journal} {Nature Communications}\ }\textbf {\bibinfo {volume}
  {8}},\ \bibinfo {pages} {13940} (\bibinfo {year} {2017})}\BibitemShut {NoStop}%
\bibitem [{\citenamefont {Chan}\ \emph {et~al.}(2016)\citenamefont {Chan},
  \citenamefont {Oh}, \citenamefont {Han},\ and\ \citenamefont
  {Lee}}]{PatrickLeePRB2016}%
  \BibitemOpen
  \bibfield  {author} {\bibinfo {author} {\bibfnamefont {C.-K.}\ \bibnamefont
  {Chan}}, \bibinfo {author} {\bibfnamefont {Y.-T.}\ \bibnamefont {Oh}},
  \bibinfo {author} {\bibfnamefont {J.~H.}\ \bibnamefont {Han}}, \ and\
  \bibinfo {author} {\bibfnamefont {P.~A.}\ \bibnamefont {Lee}},\ }\href
  {\doibase 10.1103/PhysRevB.94.121106} {\bibfield  {journal} {\bibinfo
  {journal} {Phys. Rev. B}\ }\textbf {\bibinfo {volume} {94}},\ \bibinfo
  {pages} {121106} (\bibinfo {year} {2016})}\BibitemShut {NoStop}%
\bibitem [{\citenamefont {Yan}\ and\ \citenamefont
  {Wang}(2016)}]{ZWangPRL2016}%
  \BibitemOpen
  \bibfield  {author} {\bibinfo {author} {\bibfnamefont {Z.}~\bibnamefont
  {Yan}}\ and\ \bibinfo {author} {\bibfnamefont {Z.}~\bibnamefont {Wang}},\
  }\href {\doibase 10.1103/PhysRevLett.117.087402} {\bibfield  {journal}
  {\bibinfo  {journal} {Phys. Rev. Lett.}\ }\textbf {\bibinfo {volume} {117}},\
  \bibinfo {pages} {087402} (\bibinfo {year} {2016})}\BibitemShut {NoStop}%
\bibitem [{\citenamefont {Topp}\ \emph {et~al.}(2018)\citenamefont {Topp},
  \citenamefont {Tancogne-Dejean}, \citenamefont {Kemper}, \citenamefont
  {Rubio},\ and\ \citenamefont {Sentef}}]{SentefNCOMMS2017}%
  \BibitemOpen
  \bibfield  {author} {\bibinfo {author} {\bibfnamefont {G.~E.}\ \bibnamefont
  {Topp}}, \bibinfo {author} {\bibfnamefont {N.}~\bibnamefont
  {Tancogne-Dejean}}, \bibinfo {author} {\bibfnamefont {A.~F.}\ \bibnamefont
  {Kemper}}, \bibinfo {author} {\bibfnamefont {A.}~\bibnamefont {Rubio}}, \
  and\ \bibinfo {author} {\bibfnamefont {M.~A.}\ \bibnamefont {Sentef}},\
  }\href {\doibase 10.1038/s41467-018-06991-8} {\bibfield  {journal} {\bibinfo
  {journal} {Nature Communications}\ }\textbf {\bibinfo {volume} {9}},\ \bibinfo {pages} {4452} (\bibinfo
  {year} {2018})}\BibitemShut {NoStop}%
\bibitem [{\citenamefont {Guan}\ \emph {et~al.}(2021)\citenamefont {Guan},
  \citenamefont {Wang}, \citenamefont {You}, \citenamefont {Sun},\ and\
  \citenamefont {Meng}}]{MengNCOMMS2017}%
  \BibitemOpen
  \bibfield  {author} {\bibinfo {author} {\bibfnamefont {M.-X.}\ \bibnamefont
  {Guan}}, \bibinfo {author} {\bibfnamefont {E.}~\bibnamefont {Wang}}, \bibinfo
  {author} {\bibfnamefont {P.-W.}\ \bibnamefont {You}}, \bibinfo {author}
  {\bibfnamefont {J.-T.}\ \bibnamefont {Sun}}, \ and\ \bibinfo {author}
  {\bibfnamefont {S.}~\bibnamefont {Meng}},\ }\href {\doibase
  10.1038/s41467-021-22056-9} {\bibfield  {journal} {\bibinfo  {journal}
  {Nature Communications}\ }\textbf {\bibinfo {volume} {12}},\ \bibinfo {pages} {1885} (\bibinfo {year}
  {2021})}\BibitemShut {NoStop}%
\bibitem [{\citenamefont {Ebihara}\ \emph {et~al.}(2016)\citenamefont
  {Ebihara}, \citenamefont {Fukushima},\ and\ \citenamefont
  {Oka}}]{Ebihara2016}%
  \BibitemOpen
  \bibfield  {author} {\bibinfo {author} {\bibfnamefont {S.}~\bibnamefont
  {Ebihara}}, \bibinfo {author} {\bibfnamefont {K.}~\bibnamefont {Fukushima}},
  \ and\ \bibinfo {author} {\bibfnamefont {T.}~\bibnamefont {Oka}},\ }\href
  {\doibase 10.1103/PhysRevB.93.155107} {\bibfield  {journal} {\bibinfo
  {journal} {Phys. Rev. B}\ }\textbf {\bibinfo {volume} {93}},\ \bibinfo
  {pages} {155107} (\bibinfo {year} {2016})}\BibitemShut {NoStop}%
\bibitem [{\citenamefont {Vaswani}\ \emph {et~al.}(2020)\citenamefont
  {Vaswani}, \citenamefont {Wang}, \citenamefont {Mudiyanselage}, \citenamefont
  {Li}, \citenamefont {Lozano}, \citenamefont {Gu}, \citenamefont {Cheng},
  \citenamefont {Song}, \citenamefont {Luo}, \citenamefont {Kim}, \citenamefont
  {Huang}, \citenamefont {Liu}, \citenamefont {Mootz}, \citenamefont {Perakis},
  \citenamefont {Yao}, \citenamefont {Ho},\ and\ \citenamefont
  {Wang}}]{WJGPRX2020}%
  \BibitemOpen
  \bibfield  {author} {\bibinfo {author} {\bibfnamefont {C.}~\bibnamefont
  {Vaswani}}, \bibinfo {author} {\bibfnamefont {L.-L.}\ \bibnamefont {Wang}},
  \bibinfo {author} {\bibfnamefont {D.~H.}\ \bibnamefont {Mudiyanselage}},
  \bibinfo {author} {\bibfnamefont {Q.}~\bibnamefont {Li}}, \bibinfo {author}
  {\bibfnamefont {P.~M.}\ \bibnamefont {Lozano}}, \bibinfo {author}
  {\bibfnamefont {G.~D.}\ \bibnamefont {Gu}}, \bibinfo {author} {\bibfnamefont
  {D.}~\bibnamefont {Cheng}}, \bibinfo {author} {\bibfnamefont
  {B.}~\bibnamefont {Song}}, \bibinfo {author} {\bibfnamefont {L.}~\bibnamefont
  {Luo}}, \bibinfo {author} {\bibfnamefont {R.~H.~J.}\ \bibnamefont {Kim}},
  \bibinfo {author} {\bibfnamefont {C.}~\bibnamefont {Huang}}, \bibinfo
  {author} {\bibfnamefont {Z.}~\bibnamefont {Liu}}, \bibinfo {author}
  {\bibfnamefont {M.}~\bibnamefont {Mootz}}, \bibinfo {author} {\bibfnamefont
  {I.~E.}\ \bibnamefont {Perakis}}, \bibinfo {author} {\bibfnamefont
  {Y.}~\bibnamefont {Yao}}, \bibinfo {author} {\bibfnamefont {K.~M.}\
  \bibnamefont {Ho}}, \ and\ \bibinfo {author} {\bibfnamefont {J.}~\bibnamefont
  {Wang}},\ }\href {\doibase 10.1103/PhysRevX.10.021013} {\bibfield  {journal}
  {\bibinfo  {journal} {Phys. Rev. X}\ }\textbf {\bibinfo {volume} {10}},\
  \bibinfo {pages} {021013} (\bibinfo {year} {2020})}\BibitemShut {NoStop}%
\bibitem [{\citenamefont {Luo}\ \emph {et~al.}(2021)\citenamefont {Luo},
  \citenamefont {Cheng}, \citenamefont {Song}, \citenamefont {Wang},
  \citenamefont {Vaswani}, \citenamefont {Lozano}, \citenamefont {Gu},
  \citenamefont {Huang}, \citenamefont {Kim}, \citenamefont {Liu},
  \citenamefont {Park}, \citenamefont {Yao}, \citenamefont {Ho}, \citenamefont
  {Perakis}, \citenamefont {Li},\ and\ \citenamefont {Wang}}]{WJGNMAT2021}%
  \BibitemOpen
  \bibfield  {author} {\bibinfo {author} {\bibfnamefont {L.}~\bibnamefont
  {Luo}}, \bibinfo {author} {\bibfnamefont {D.}~\bibnamefont {Cheng}}, \bibinfo
  {author} {\bibfnamefont {B.}~\bibnamefont {Song}}, \bibinfo {author}
  {\bibfnamefont {L.-L.}\ \bibnamefont {Wang}}, \bibinfo {author}
  {\bibfnamefont {C.}~\bibnamefont {Vaswani}}, \bibinfo {author} {\bibfnamefont
  {P.~M.}\ \bibnamefont {Lozano}}, \bibinfo {author} {\bibfnamefont
  {G.}~\bibnamefont {Gu}}, \bibinfo {author} {\bibfnamefont {C.}~\bibnamefont
  {Huang}}, \bibinfo {author} {\bibfnamefont {R.~H.~J.}\ \bibnamefont {Kim}},
  \bibinfo {author} {\bibfnamefont {Z.}~\bibnamefont {Liu}}, \bibinfo {author}
  {\bibfnamefont {J.-M.}\ \bibnamefont {Park}}, \bibinfo {author}
  {\bibfnamefont {Y.}~\bibnamefont {Yao}}, \bibinfo {author} {\bibfnamefont
  {K.}~\bibnamefont {Ho}}, \bibinfo {author} {\bibfnamefont {I.~E.}\
  \bibnamefont {Perakis}}, \bibinfo {author} {\bibfnamefont {Q.}~\bibnamefont
  {Li}}, \ and\ \bibinfo {author} {\bibfnamefont {J.}~\bibnamefont {Wang}},\
  }\href {\doibase 10.1038/s41563-020-00882-4} {\bibfield  {journal} {\bibinfo
  {journal} {Nature Materials}\ }\textbf {\bibinfo {volume} {20}},\ \bibinfo
  {pages} {329} (\bibinfo {year} {2021})}\BibitemShut {NoStop}%
\bibitem [{\citenamefont {Sie}\ \emph {et~al.}(2019)\citenamefont {Sie},
  \citenamefont {Nyby}, \citenamefont {Pemmaraju}, \citenamefont {Park},
  \citenamefont {Shen}, \citenamefont {Yang}, \citenamefont {Hoffmann},
  \citenamefont {Ofori-Okai}, \citenamefont {Li}, \citenamefont {Reid},
  \citenamefont {Weathersby}, \citenamefont {Mannebach}, \citenamefont
  {Finney}, \citenamefont {Rhodes}, \citenamefont {Chenet}, \citenamefont
  {Antony}, \citenamefont {Balicas}, \citenamefont {Hone}, \citenamefont
  {Devereaux}, \citenamefont {Heinz}, \citenamefont {Wang},\ and\ \citenamefont
  {Lindenberg}}]{LindenbergNature2019}%
  \BibitemOpen
  \bibfield  {author} {\bibinfo {author} {\bibfnamefont {E.~J.}\ \bibnamefont
  {Sie}}, \bibinfo {author} {\bibfnamefont {C.~M.}\ \bibnamefont {Nyby}},
  \bibinfo {author} {\bibfnamefont {C.~D.}\ \bibnamefont {Pemmaraju}}, \bibinfo
  {author} {\bibfnamefont {S.~J.}\ \bibnamefont {Park}}, \bibinfo {author}
  {\bibfnamefont {X.}~\bibnamefont {Shen}}, \bibinfo {author} {\bibfnamefont
  {J.}~\bibnamefont {Yang}}, \bibinfo {author} {\bibfnamefont {M.~C.}\
  \bibnamefont {Hoffmann}}, \bibinfo {author} {\bibfnamefont {B.~K.}\
  \bibnamefont {Ofori-Okai}}, \bibinfo {author} {\bibfnamefont
  {R.}~\bibnamefont {Li}}, \bibinfo {author} {\bibfnamefont {A.~H.}\
  \bibnamefont {Reid}}, \bibinfo {author} {\bibfnamefont {S.}~\bibnamefont
  {Weathersby}}, \bibinfo {author} {\bibfnamefont {E.}~\bibnamefont
  {Mannebach}}, \bibinfo {author} {\bibfnamefont {N.}~\bibnamefont {Finney}},
  \bibinfo {author} {\bibfnamefont {D.}~\bibnamefont {Rhodes}}, \bibinfo
  {author} {\bibfnamefont {D.}~\bibnamefont {Chenet}}, \bibinfo {author}
  {\bibfnamefont {A.}~\bibnamefont {Antony}}, \bibinfo {author} {\bibfnamefont
  {L.}~\bibnamefont {Balicas}}, \bibinfo {author} {\bibfnamefont
  {J.}~\bibnamefont {Hone}}, \bibinfo {author} {\bibfnamefont {T.~P.}\
  \bibnamefont {Devereaux}}, \bibinfo {author} {\bibfnamefont {T.~F.}\
  \bibnamefont {Heinz}}, \bibinfo {author} {\bibfnamefont {X.}~\bibnamefont
  {Wang}}, \ and\ \bibinfo {author} {\bibfnamefont {A.~M.}\ \bibnamefont
  {Lindenberg}},\ }\href {\doibase 10.1038/s41586-018-0809-4} {\bibfield
  {journal} {\bibinfo  {journal} {Nature}\ }\textbf {\bibinfo {volume} {565}},\
  \bibinfo {pages} {61} (\bibinfo {year} {2019})}\BibitemShut {NoStop}%
\bibitem [{\citenamefont {Zhang}\ \emph {et~al.}(2019)\citenamefont {Zhang},
  \citenamefont {Wang}, \citenamefont {Li}, \citenamefont {Shi}, \citenamefont
  {Wu}, \citenamefont {Lin}, \citenamefont {Zhang}, \citenamefont {Liu},
  \citenamefont {Liu}, \citenamefont {Wang}, \citenamefont {Dong},\ and\
  \citenamefont {Wang}}]{WNLPRX2019}%
  \BibitemOpen
  \bibfield  {author} {\bibinfo {author} {\bibfnamefont {M.~Y.}\ \bibnamefont
  {Zhang}}, \bibinfo {author} {\bibfnamefont {Z.~X.}\ \bibnamefont {Wang}},
  \bibinfo {author} {\bibfnamefont {Y.~N.}\ \bibnamefont {Li}}, \bibinfo
  {author} {\bibfnamefont {L.~Y.}\ \bibnamefont {Shi}}, \bibinfo {author}
  {\bibfnamefont {D.}~\bibnamefont {Wu}}, \bibinfo {author} {\bibfnamefont
  {T.}~\bibnamefont {Lin}}, \bibinfo {author} {\bibfnamefont {S.~J.}\
  \bibnamefont {Zhang}}, \bibinfo {author} {\bibfnamefont {Y.~Q.}\ \bibnamefont
  {Liu}}, \bibinfo {author} {\bibfnamefont {Q.~M.}\ \bibnamefont {Liu}},
  \bibinfo {author} {\bibfnamefont {J.}~\bibnamefont {Wang}}, \bibinfo {author}
  {\bibfnamefont {T.}~\bibnamefont {Dong}}, \ and\ \bibinfo {author}
  {\bibfnamefont {N.~L.}\ \bibnamefont {Wang}},\ }\href {\doibase
  10.1103/PhysRevX.9.021036} {\bibfield  {journal} {\bibinfo  {journal} {Phys.
  Rev. X}\ }\textbf {\bibinfo {volume} {9}},\ \bibinfo {pages} {021036}
  (\bibinfo {year} {2019})}\BibitemShut {NoStop}%
\bibitem [{\citenamefont {Tangney}\ and\ \citenamefont
  {Fahy}(2002)}]{TangneyPRB2002}%
  \BibitemOpen
  \bibfield  {author} {\bibinfo {author} {\bibfnamefont {P.}~\bibnamefont
  {Tangney}}\ and\ \bibinfo {author} {\bibfnamefont {S.}~\bibnamefont {Fahy}},\
  }\href {\doibase 10.1103/PhysRevB.65.054302} {\bibfield  {journal} {\bibinfo
  {journal} {Phys. Rev. B}\ }\textbf {\bibinfo {volume} {65}},\ \bibinfo
  {pages} {054302} (\bibinfo {year} {2002})}\BibitemShut {NoStop}%
\bibitem [{SM()}]{SM}%
  \BibitemOpen
  \href@noop {} {\bibinfo  {journal} {See Supplemental Material at [URL] for
  extensive simulation and experimental details. See, also, references [44-76] therein}\ }\BibitemShut {NoStop}%
\bibitem [{\citenamefont {Chang}\ \emph {et~al.}(2018)\citenamefont {Chang},
  \citenamefont {Wieder}, \citenamefont {Schindler}, \citenamefont {Sanchez},
  \citenamefont {Belopolski}, \citenamefont {Huang}, \citenamefont {Singh},
  \citenamefont {Wu}, \citenamefont {Chang}, \citenamefont {Neupert},
  \citenamefont {Xu}, \citenamefont {Lin},\ and\ \citenamefont
  {Hasan}}]{Chang2018}%
  \BibitemOpen
\bibfield  {journal} {  }\bibfield  {author} {\bibinfo {author} {\bibfnamefont
  {G.}~\bibnamefont {Chang}}, \bibinfo {author} {\bibfnamefont {B.~J.}\
  \bibnamefont {Wieder}}, \bibinfo {author} {\bibfnamefont {F.}~\bibnamefont
  {Schindler}}, \bibinfo {author} {\bibfnamefont {D.~S.}\ \bibnamefont
  {Sanchez}}, \bibinfo {author} {\bibfnamefont {I.}~\bibnamefont {Belopolski}},
  \bibinfo {author} {\bibfnamefont {S.-M.}\ \bibnamefont {Huang}}, \bibinfo
  {author} {\bibfnamefont {B.}~\bibnamefont {Singh}}, \bibinfo {author}
  {\bibfnamefont {D.}~\bibnamefont {Wu}}, \bibinfo {author} {\bibfnamefont
  {T.-R.}\ \bibnamefont {Chang}}, \bibinfo {author} {\bibfnamefont
  {T.}~\bibnamefont {Neupert}}, \bibinfo {author} {\bibfnamefont {S.-Y.}\
  \bibnamefont {Xu}}, \bibinfo {author} {\bibfnamefont {H.}~\bibnamefont
  {Lin}}, \ and\ \bibinfo {author} {\bibfnamefont {M.~Z.}\ \bibnamefont
  {Hasan}},\ }\href {\doibase 10.1038/s41563-018-0169-3} {\bibfield  {journal}
  {\bibinfo  {journal} {Nature Materials}\ }\textbf {\bibinfo {volume} {17}},\
  \bibinfo {pages} {978} (\bibinfo {year} {2018})}\BibitemShut {NoStop}%
\bibitem [{\citenamefont {Nakayama}\ \emph {et~al.}(2017)\citenamefont
  {Nakayama}, \citenamefont {Kuno}, \citenamefont {Yamauchi}, \citenamefont
  {Souma}, \citenamefont {Sugawara}, \citenamefont {Oguchi}, \citenamefont
  {Sato},\ and\ \citenamefont {Takahashi}}]{TakahashiPRB2017}%
  \BibitemOpen
  \bibfield  {author} {\bibinfo {author} {\bibfnamefont {K.}~\bibnamefont
  {Nakayama}}, \bibinfo {author} {\bibfnamefont {M.}~\bibnamefont {Kuno}},
  \bibinfo {author} {\bibfnamefont {K.}~\bibnamefont {Yamauchi}}, \bibinfo
  {author} {\bibfnamefont {S.}~\bibnamefont {Souma}}, \bibinfo {author}
  {\bibfnamefont {K.}~\bibnamefont {Sugawara}}, \bibinfo {author}
  {\bibfnamefont {T.}~\bibnamefont {Oguchi}}, \bibinfo {author} {\bibfnamefont
  {T.}~\bibnamefont {Sato}}, \ and\ \bibinfo {author} {\bibfnamefont
  {T.}~\bibnamefont {Takahashi}},\ }\href {\doibase 10.1103/PhysRevB.95.125204}
  {\bibfield  {journal} {\bibinfo  {journal} {Phys. Rev. B}\ }\textbf {\bibinfo
  {volume} {95}},\ \bibinfo {pages} {125204} (\bibinfo {year}
  {2017})}\BibitemShut {NoStop}%
\bibitem [{\citenamefont {Sakano}\ \emph {et~al.}(2020)\citenamefont {Sakano},
  \citenamefont {Hirayama}, \citenamefont {Takahashi}, \citenamefont {Akebi},
  \citenamefont {Nakayama}, \citenamefont {Kuroda}, \citenamefont {Taguchi},
  \citenamefont {Yoshikawa}, \citenamefont {Miyamoto}, \citenamefont {Okuda},
  \citenamefont {Ono}, \citenamefont {Kumigashira}, \citenamefont {Ideue},
  \citenamefont {Iwasa}, \citenamefont {Mitsuishi}, \citenamefont {Ishizaka},
  \citenamefont {Shin}, \citenamefont {Miyake}, \citenamefont {Murakami},
  \citenamefont {Sasagawa},\ and\ \citenamefont {Kondo}}]{KondoPRL2020}%
  \BibitemOpen
  \bibfield  {author} {\bibinfo {author} {\bibfnamefont {M.}~\bibnamefont
  {Sakano}}, \bibinfo {author} {\bibfnamefont {M.}~\bibnamefont {Hirayama}},
  \bibinfo {author} {\bibfnamefont {T.}~\bibnamefont {Takahashi}}, \bibinfo
  {author} {\bibfnamefont {S.}~\bibnamefont {Akebi}}, \bibinfo {author}
  {\bibfnamefont {M.}~\bibnamefont {Nakayama}}, \bibinfo {author}
  {\bibfnamefont {K.}~\bibnamefont {Kuroda}}, \bibinfo {author} {\bibfnamefont
  {K.}~\bibnamefont {Taguchi}}, \bibinfo {author} {\bibfnamefont
  {T.}~\bibnamefont {Yoshikawa}}, \bibinfo {author} {\bibfnamefont
  {K.}~\bibnamefont {Miyamoto}}, \bibinfo {author} {\bibfnamefont
  {T.}~\bibnamefont {Okuda}}, \bibinfo {author} {\bibfnamefont
  {K.}~\bibnamefont {Ono}}, \bibinfo {author} {\bibfnamefont {H.}~\bibnamefont
  {Kumigashira}}, \bibinfo {author} {\bibfnamefont {T.}~\bibnamefont {Ideue}},
  \bibinfo {author} {\bibfnamefont {Y.}~\bibnamefont {Iwasa}}, \bibinfo
  {author} {\bibfnamefont {N.}~\bibnamefont {Mitsuishi}}, \bibinfo {author}
  {\bibfnamefont {K.}~\bibnamefont {Ishizaka}}, \bibinfo {author}
  {\bibfnamefont {S.}~\bibnamefont {Shin}}, \bibinfo {author} {\bibfnamefont
  {T.}~\bibnamefont {Miyake}}, \bibinfo {author} {\bibfnamefont
  {S.}~\bibnamefont {Murakami}}, \bibinfo {author} {\bibfnamefont
  {T.}~\bibnamefont {Sasagawa}}, \ and\ \bibinfo {author} {\bibfnamefont
  {T.}~\bibnamefont {Kondo}},\ }\href {\doibase 10.1103/PhysRevLett.124.136404}
  {\bibfield  {journal} {\bibinfo  {journal} {Phys. Rev. Lett.}\ }\textbf
  {\bibinfo {volume} {124}},\ \bibinfo {pages} {136404} (\bibinfo {year}
  {2020})}\BibitemShut {NoStop}%
\bibitem [{\citenamefont {Gatti}\ \emph {et~al.}(2020)\citenamefont {Gatti},
  \citenamefont {Gos\'albez-Mart\'{\i}nez}, \citenamefont {Tsirkin},
  \citenamefont {Fanciulli}, \citenamefont {Puppin}, \citenamefont
  {Polishchuk}, \citenamefont {Moser}, \citenamefont {Testa}, \citenamefont
  {Martino}, \citenamefont {Roth}, \citenamefont {Bugnon}, \citenamefont
  {Moreschini}, \citenamefont {Bostwick}, \citenamefont {Jozwiak},
  \citenamefont {Rotenberg}, \citenamefont {Di~Santo}, \citenamefont
  {Petaccia}, \citenamefont {Vobornik}, \citenamefont {Fujii}, \citenamefont
  {Wong}, \citenamefont {Jariwala}, \citenamefont {Atwater}, \citenamefont
  {R\o{}nnow}, \citenamefont {Chergui}, \citenamefont {Yazyev}, \citenamefont
  {Grioni},\ and\ \citenamefont {Crepaldi}}]{CrepaldiPRL2020}%
  \BibitemOpen
  \bibfield  {author} {\bibinfo {author} {\bibfnamefont {G.}~\bibnamefont
  {Gatti}}, \bibinfo {author} {\bibfnamefont {D.}~\bibnamefont
  {Gos\'albez-Mart\'{\i}nez}}, \bibinfo {author} {\bibfnamefont {S.~S.}\
  \bibnamefont {Tsirkin}}, \bibinfo {author} {\bibfnamefont {M.}~\bibnamefont
  {Fanciulli}}, \bibinfo {author} {\bibfnamefont {M.}~\bibnamefont {Puppin}},
  \bibinfo {author} {\bibfnamefont {S.}~\bibnamefont {Polishchuk}}, \bibinfo
  {author} {\bibfnamefont {S.}~\bibnamefont {Moser}}, \bibinfo {author}
  {\bibfnamefont {L.}~\bibnamefont {Testa}}, \bibinfo {author} {\bibfnamefont
  {E.}~\bibnamefont {Martino}}, \bibinfo {author} {\bibfnamefont
  {S.}~\bibnamefont {Roth}}, \bibinfo {author} {\bibfnamefont {P.}~\bibnamefont
  {Bugnon}}, \bibinfo {author} {\bibfnamefont {L.}~\bibnamefont {Moreschini}},
  \bibinfo {author} {\bibfnamefont {A.}~\bibnamefont {Bostwick}}, \bibinfo
  {author} {\bibfnamefont {C.}~\bibnamefont {Jozwiak}}, \bibinfo {author}
  {\bibfnamefont {E.}~\bibnamefont {Rotenberg}}, \bibinfo {author}
  {\bibfnamefont {G.}~\bibnamefont {Di~Santo}}, \bibinfo {author}
  {\bibfnamefont {L.}~\bibnamefont {Petaccia}}, \bibinfo {author}
  {\bibfnamefont {I.}~\bibnamefont {Vobornik}}, \bibinfo {author}
  {\bibfnamefont {J.}~\bibnamefont {Fujii}}, \bibinfo {author} {\bibfnamefont
  {J.}~\bibnamefont {Wong}}, \bibinfo {author} {\bibfnamefont {D.}~\bibnamefont
  {Jariwala}}, \bibinfo {author} {\bibfnamefont {H.~A.}\ \bibnamefont
  {Atwater}}, \bibinfo {author} {\bibfnamefont {H.~M.}\ \bibnamefont
  {R\o{}nnow}}, \bibinfo {author} {\bibfnamefont {M.}~\bibnamefont {Chergui}},
  \bibinfo {author} {\bibfnamefont {O.~V.}\ \bibnamefont {Yazyev}}, \bibinfo
  {author} {\bibfnamefont {M.}~\bibnamefont {Grioni}}, \ and\ \bibinfo {author}
  {\bibfnamefont {A.}~\bibnamefont {Crepaldi}},\ }\href {\doibase
  10.1103/PhysRevLett.125.216402} {\bibfield  {journal} {\bibinfo  {journal}
  {Phys. Rev. Lett.}\ }\textbf {\bibinfo {volume} {125}},\ \bibinfo {pages}
  {216402} (\bibinfo {year} {2020})}\BibitemShut {NoStop}%
\bibitem [{\citenamefont {Zhang}\ \emph {et~al.}(2020)\citenamefont {Zhang},
  \citenamefont {Zhao}, \citenamefont {Li}, \citenamefont {Wang}, \citenamefont
  {Xie}, \citenamefont {Cheng}, \citenamefont {Li}, \citenamefont {Lin},
  \citenamefont {Xi}, \citenamefont {Ke}, \citenamefont {Yang}, \citenamefont
  {He}, \citenamefont {Sun}, \citenamefont {Wang}, \citenamefont {Zhang},\ and\
  \citenamefont {Zeng}}]{Zhang2020}%
  \BibitemOpen
  \bibfield  {author} {\bibinfo {author} {\bibfnamefont {N.}~\bibnamefont
  {Zhang}}, \bibinfo {author} {\bibfnamefont {G.}~\bibnamefont {Zhao}},
  \bibinfo {author} {\bibfnamefont {L.}~\bibnamefont {Li}}, \bibinfo {author}
  {\bibfnamefont {P.}~\bibnamefont {Wang}}, \bibinfo {author} {\bibfnamefont
  {L.}~\bibnamefont {Xie}}, \bibinfo {author} {\bibfnamefont {B.}~\bibnamefont
  {Cheng}}, \bibinfo {author} {\bibfnamefont {H.}~\bibnamefont {Li}}, \bibinfo
  {author} {\bibfnamefont {Z.}~\bibnamefont {Lin}}, \bibinfo {author}
  {\bibfnamefont {C.}~\bibnamefont {Xi}}, \bibinfo {author} {\bibfnamefont
  {J.}~\bibnamefont {Ke}}, \bibinfo {author} {\bibfnamefont {M.}~\bibnamefont
  {Yang}}, \bibinfo {author} {\bibfnamefont {J.}~\bibnamefont {He}}, \bibinfo
  {author} {\bibfnamefont {Z.}~\bibnamefont {Sun}}, \bibinfo {author}
  {\bibfnamefont {Z.}~\bibnamefont {Wang}}, \bibinfo {author} {\bibfnamefont
  {Z.}~\bibnamefont {Zhang}}, \ and\ \bibinfo {author} {\bibfnamefont
  {C.}~\bibnamefont {Zeng}},\ }\href {\doibase 10.1073/pnas.2002913117}
  {\bibfield  {journal} {\bibinfo  {journal} {Proceedings of the National
  Academy of Sciences}\ }\textbf {\bibinfo {volume} {117}},\ \bibinfo {pages}
  {11337} (\bibinfo {year} {2020})}\BibitemShut {NoStop}%
\bibitem [{\citenamefont {Hirayama}\ \emph {et~al.}(2015)\citenamefont
  {Hirayama}, \citenamefont {Okugawa}, \citenamefont {Ishibashi}, \citenamefont
  {Murakami},\ and\ \citenamefont {Miyake}}]{MiyakePRL2015}%
  \BibitemOpen
  \bibfield  {author} {\bibinfo {author} {\bibfnamefont {M.}~\bibnamefont
  {Hirayama}}, \bibinfo {author} {\bibfnamefont {R.}~\bibnamefont {Okugawa}},
  \bibinfo {author} {\bibfnamefont {S.}~\bibnamefont {Ishibashi}}, \bibinfo
  {author} {\bibfnamefont {S.}~\bibnamefont {Murakami}}, \ and\ \bibinfo
  {author} {\bibfnamefont {T.}~\bibnamefont {Miyake}},\ }\href {\doibase
  10.1103/PhysRevLett.114.206401} {\bibfield  {journal} {\bibinfo  {journal}
  {Phys. Rev. Lett.}\ }\textbf {\bibinfo {volume} {114}},\ \bibinfo {pages}
  {206401} (\bibinfo {year} {2015})}\BibitemShut {NoStop}%
\bibitem [{\citenamefont {Agapito}\ \emph {et~al.}(2013)\citenamefont
  {Agapito}, \citenamefont {Kioussis}, \citenamefont {Goddard},\ and\
  \citenamefont {Ong}}]{KioussisPRL2013}%
  \BibitemOpen
  \bibfield  {author} {\bibinfo {author} {\bibfnamefont {L.~A.}\ \bibnamefont
  {Agapito}}, \bibinfo {author} {\bibfnamefont {N.}~\bibnamefont {Kioussis}},
  \bibinfo {author} {\bibfnamefont {W.~A.}\ \bibnamefont {Goddard}}, \ and\
  \bibinfo {author} {\bibfnamefont {N.~P.}\ \bibnamefont {Ong}},\ }\href
  {\doibase 10.1103/PhysRevLett.110.176401} {\bibfield  {journal} {\bibinfo
  {journal} {Phys. Rev. Lett.}\ }\textbf {\bibinfo {volume} {110}},\ \bibinfo
  {pages} {176401} (\bibinfo {year} {2013})}\BibitemShut {NoStop}%
\bibitem [{\citenamefont {Xue}\ \emph {et~al.}(2018)\citenamefont {Xue},
  \citenamefont {Feng}, \citenamefont {Liao}, \citenamefont {Chen},
  \citenamefont {Wang}, \citenamefont {Tang},\ and\ \citenamefont
  {Chen}}]{Xue2018}%
  \BibitemOpen
  \bibfield  {author} {\bibinfo {author} {\bibfnamefont {X.-X.}\ \bibnamefont
  {Xue}}, \bibinfo {author} {\bibfnamefont {Y.-X.}\ \bibnamefont {Feng}},
  \bibinfo {author} {\bibfnamefont {L.}~\bibnamefont {Liao}}, \bibinfo {author}
  {\bibfnamefont {Q.-J.}\ \bibnamefont {Chen}}, \bibinfo {author}
  {\bibfnamefont {D.}~\bibnamefont {Wang}}, \bibinfo {author} {\bibfnamefont
  {L.-M.}\ \bibnamefont {Tang}}, \ and\ \bibinfo {author} {\bibfnamefont
  {K.}~\bibnamefont {Chen}},\ }\href {\doibase 10.1088/1361-648x/aaaea1}
  {\bibfield  {journal} {\bibinfo  {journal} {Journal of Physics: Condensed
  Matter}\ }\textbf {\bibinfo {volume} {30}},\ \bibinfo {pages} {125001}
  (\bibinfo {year} {2018})}\BibitemShut {NoStop}%
\bibitem [{\citenamefont {Rodriguez}\ \emph {et~al.}(2020)\citenamefont
  {Rodriguez}, \citenamefont {Tsirlin}, \citenamefont {Biesner}, \citenamefont
  {Ueno}, \citenamefont {Takahashi}, \citenamefont {Kobayashi}, \citenamefont
  {Dressel},\ and\ \citenamefont {Uykur}}]{UykurPRL2020}%
  \BibitemOpen
  \bibfield  {author} {\bibinfo {author} {\bibfnamefont {D.}~\bibnamefont
  {Rodriguez}}, \bibinfo {author} {\bibfnamefont {A.~A.}\ \bibnamefont
  {Tsirlin}}, \bibinfo {author} {\bibfnamefont {T.}~\bibnamefont {Biesner}},
  \bibinfo {author} {\bibfnamefont {T.}~\bibnamefont {Ueno}}, \bibinfo {author}
  {\bibfnamefont {T.}~\bibnamefont {Takahashi}}, \bibinfo {author}
  {\bibfnamefont {K.}~\bibnamefont {Kobayashi}}, \bibinfo {author}
  {\bibfnamefont {M.}~\bibnamefont {Dressel}}, \ and\ \bibinfo {author}
  {\bibfnamefont {E.}~\bibnamefont {Uykur}},\ }\href {\doibase
  10.1103/PhysRevLett.124.136402} {\bibfield  {journal} {\bibinfo  {journal}
  {Phys. Rev. Lett.}\ }\textbf {\bibinfo {volume} {124}},\ \bibinfo {pages}
  {136402} (\bibinfo {year} {2020})}\BibitemShut {NoStop}%
\bibitem [{\citenamefont {Ideue}\ \emph {et~al.}(2019)\citenamefont {Ideue},
  \citenamefont {Hirayama}, \citenamefont {Taiko}, \citenamefont {Takahashi},
  \citenamefont {Murase}, \citenamefont {Miyake}, \citenamefont {Murakami},
  \citenamefont {Sasagawa},\ and\ \citenamefont {Iwasa}}]{IwasaPNAS2019}%
  \BibitemOpen
  \bibfield  {author} {\bibinfo {author} {\bibfnamefont {T.}~\bibnamefont
  {Ideue}}, \bibinfo {author} {\bibfnamefont {M.}~\bibnamefont {Hirayama}},
  \bibinfo {author} {\bibfnamefont {H.}~\bibnamefont {Taiko}}, \bibinfo
  {author} {\bibfnamefont {T.}~\bibnamefont {Takahashi}}, \bibinfo {author}
  {\bibfnamefont {M.}~\bibnamefont {Murase}}, \bibinfo {author} {\bibfnamefont
  {T.}~\bibnamefont {Miyake}}, \bibinfo {author} {\bibfnamefont
  {S.}~\bibnamefont {Murakami}}, \bibinfo {author} {\bibfnamefont
  {T.}~\bibnamefont {Sasagawa}}, \ and\ \bibinfo {author} {\bibfnamefont
  {Y.}~\bibnamefont {Iwasa}},\ }\href {\doibase 10.1073/pnas.1905524116}
  {\bibfield  {journal} {\bibinfo  {journal} {Proceedings of the National
  Academy of Sciences}\ }\textbf {\bibinfo {volume} {116}},\ \bibinfo {pages}
  {25530} (\bibinfo {year} {2019})} \BibitemShut {NoStop}%
\bibitem [{\citenamefont {Zeiger}\ \emph {et~al.}(1992)\citenamefont {Zeiger},
  \citenamefont {Vidal}, \citenamefont {Cheng}, \citenamefont {Ippen},
  \citenamefont {Dresselhaus},\ and\ \citenamefont
  {Dresselhaus}}]{DresselhausPRB1992}%
  \BibitemOpen
  \bibfield  {author} {\bibinfo {author} {\bibfnamefont {H.~J.}\ \bibnamefont
  {Zeiger}}, \bibinfo {author} {\bibfnamefont {J.}~\bibnamefont {Vidal}},
  \bibinfo {author} {\bibfnamefont {T.~K.}\ \bibnamefont {Cheng}}, \bibinfo
  {author} {\bibfnamefont {E.~P.}\ \bibnamefont {Ippen}}, \bibinfo {author}
  {\bibfnamefont {G.}~\bibnamefont {Dresselhaus}}, \ and\ \bibinfo {author}
  {\bibfnamefont {M.~S.}\ \bibnamefont {Dresselhaus}},\ }\href {\doibase
  10.1103/PhysRevB.45.768} {\bibfield  {journal} {\bibinfo  {journal} {Phys.
  Rev. B}\ }\textbf {\bibinfo {volume} {45}},\ \bibinfo {pages} {768} (\bibinfo
  {year} {1992})}\BibitemShut {NoStop}%
\bibitem [{\citenamefont {F{\"o}rst}\ \emph {et~al.}(2015)\citenamefont
  {F{\"o}rst}, \citenamefont {Mankowsky},\ and\ \citenamefont
  {Cavalleri}}]{CavalleriACR2015}%
  \BibitemOpen
  \bibfield  {author} {\bibinfo {author} {\bibfnamefont {M.}~\bibnamefont
  {F{\"o}rst}}, \bibinfo {author} {\bibfnamefont {R.}~\bibnamefont
  {Mankowsky}}, \ and\ \bibinfo {author} {\bibfnamefont {A.}~\bibnamefont
  {Cavalleri}},\ }\href {\doibase 10.1021/ar500391x} {\bibfield  {journal}
  {\bibinfo  {journal} {Accounts of Chemical Research}\ }\textbf {\bibinfo
  {volume} {48}},\ \bibinfo {pages} {380} (\bibinfo {year} {2015})}\BibitemShut
  {NoStop}%
\bibitem [{\citenamefont {Teitelbaum}\ \emph {et~al.}(2018)\citenamefont
  {Teitelbaum}, \citenamefont {Shin}, \citenamefont {Wolfson}, \citenamefont
  {Cheng}, \citenamefont {Porter}, \citenamefont {Kandyla},\ and\ \citenamefont
  {Nelson}}]{NelsonPRX2018}%
  \BibitemOpen
  \bibfield  {author} {\bibinfo {author} {\bibfnamefont {S.~W.}\ \bibnamefont
  {Teitelbaum}}, \bibinfo {author} {\bibfnamefont {T.}~\bibnamefont {Shin}},
  \bibinfo {author} {\bibfnamefont {J.~W.}\ \bibnamefont {Wolfson}}, \bibinfo
  {author} {\bibfnamefont {Y.-H.}\ \bibnamefont {Cheng}}, \bibinfo {author}
  {\bibfnamefont {I.~J.}\ \bibnamefont {Porter}}, \bibinfo {author}
  {\bibfnamefont {M.}~\bibnamefont {Kandyla}}, \ and\ \bibinfo {author}
  {\bibfnamefont {K.~A.}\ \bibnamefont {Nelson}},\ }\href {\doibase
  10.1103/PhysRevX.8.031081} {\bibfield  {journal} {\bibinfo  {journal} {Phys.
  Rev. X}\ }\textbf {\bibinfo {volume} {8}},\ \bibinfo {pages} {031081}
  (\bibinfo {year} {2018})}\BibitemShut {NoStop}%
\bibitem [{\citenamefont {Wall}\ \emph
  {et~al.}(2012{\natexlab{a}})\citenamefont {Wall}, \citenamefont {Wegkamp},
  \citenamefont {Foglia}, \citenamefont {Appavoo}, \citenamefont {Nag},
  \citenamefont {Haglund}, \citenamefont {St{\"a}hler},\ and\ \citenamefont
  {Wolf}}]{WallNCOMM2012}%
  \BibitemOpen
  \bibfield  {author} {\bibinfo {author} {\bibfnamefont {S.}~\bibnamefont
  {Wall}}, \bibinfo {author} {\bibfnamefont {D.}~\bibnamefont {Wegkamp}},
  \bibinfo {author} {\bibfnamefont {L.}~\bibnamefont {Foglia}}, \bibinfo
  {author} {\bibfnamefont {K.}~\bibnamefont {Appavoo}}, \bibinfo {author}
  {\bibfnamefont {J.}~\bibnamefont {Nag}}, \bibinfo {author} {\bibfnamefont
  {R.~F.}\ \bibnamefont {Haglund}}, \bibinfo {author} {\bibfnamefont
  {J.}~\bibnamefont {St{\"a}hler}}, \ and\ \bibinfo {author} {\bibfnamefont
  {M.}~\bibnamefont {Wolf}},\ }\href {\doibase 10.1038/ncomms1719} {\bibfield
  {journal} {\bibinfo  {journal} {Nature Communications}\ }\textbf {\bibinfo
  {volume} {3}},\ \bibinfo {pages} {721} (\bibinfo {year}
  {2012}{\natexlab{a}})}\BibitemShut {NoStop}%
\bibitem [{\citenamefont {Wall}\ \emph
  {et~al.}(2012{\natexlab{b}})\citenamefont {Wall}, \citenamefont {Krenzer},
  \citenamefont {Wippermann}, \citenamefont {Sanna}, \citenamefont {Klasing},
  \citenamefont {Hanisch-Blicharski}, \citenamefont {Kammler}, \citenamefont
  {Schmidt},\ and\ \citenamefont {Horn-von Hoegen}}]{WallPRL2012}%
  \BibitemOpen
  \bibfield  {author} {\bibinfo {author} {\bibfnamefont {S.}~\bibnamefont
  {Wall}}, \bibinfo {author} {\bibfnamefont {B.}~\bibnamefont {Krenzer}},
  \bibinfo {author} {\bibfnamefont {S.}~\bibnamefont {Wippermann}}, \bibinfo
  {author} {\bibfnamefont {S.}~\bibnamefont {Sanna}}, \bibinfo {author}
  {\bibfnamefont {F.}~\bibnamefont {Klasing}}, \bibinfo {author} {\bibfnamefont
  {A.}~\bibnamefont {Hanisch-Blicharski}}, \bibinfo {author} {\bibfnamefont
  {M.}~\bibnamefont {Kammler}}, \bibinfo {author} {\bibfnamefont {W.~G.}\
  \bibnamefont {Schmidt}}, \ and\ \bibinfo {author} {\bibfnamefont
  {M.}~\bibnamefont {Horn-von Hoegen}},\ }\href {\doibase
  10.1103/PhysRevLett.109.186101} {\bibfield  {journal} {\bibinfo  {journal}
  {Phys. Rev. Lett.}\ }\textbf {\bibinfo {volume} {109}},\ \bibinfo {pages}
  {186101} (\bibinfo {year} {2012}{\natexlab{b}})}\BibitemShut {NoStop}%
\bibitem [{\citenamefont {Huber}\ \emph {et~al.}(2014)\citenamefont {Huber},
  \citenamefont {Mariager}, \citenamefont {Ferrer}, \citenamefont {Sch\"afer},
  \citenamefont {Johnson}, \citenamefont {Gr\"ubel}, \citenamefont {L\"ubcke},
  \citenamefont {Huber}, \citenamefont {Kubacka}, \citenamefont {Dornes},
  \citenamefont {Laulhe}, \citenamefont {Ravy}, \citenamefont {Ingold},
  \citenamefont {Beaud}, \citenamefont {Demsar},\ and\ \citenamefont
  {Johnson}}]{JohnsonPRL2014}%
  \BibitemOpen
  \bibfield  {author} {\bibinfo {author} {\bibfnamefont {T.}~\bibnamefont
  {Huber}}, \bibinfo {author} {\bibfnamefont {S.~O.}\ \bibnamefont {Mariager}},
  \bibinfo {author} {\bibfnamefont {A.}~\bibnamefont {Ferrer}}, \bibinfo
  {author} {\bibfnamefont {H.}~\bibnamefont {Sch\"afer}}, \bibinfo {author}
  {\bibfnamefont {J.~A.}\ \bibnamefont {Johnson}}, \bibinfo {author}
  {\bibfnamefont {S.}~\bibnamefont {Gr\"ubel}}, \bibinfo {author}
  {\bibfnamefont {A.}~\bibnamefont {L\"ubcke}}, \bibinfo {author}
  {\bibfnamefont {L.}~\bibnamefont {Huber}}, \bibinfo {author} {\bibfnamefont
  {T.}~\bibnamefont {Kubacka}}, \bibinfo {author} {\bibfnamefont
  {C.}~\bibnamefont {Dornes}}, \bibinfo {author} {\bibfnamefont
  {C.}~\bibnamefont {Laulhe}}, \bibinfo {author} {\bibfnamefont
  {S.}~\bibnamefont {Ravy}}, \bibinfo {author} {\bibfnamefont {G.}~\bibnamefont
  {Ingold}}, \bibinfo {author} {\bibfnamefont {P.}~\bibnamefont {Beaud}},
  \bibinfo {author} {\bibfnamefont {J.}~\bibnamefont {Demsar}}, \ and\ \bibinfo
  {author} {\bibfnamefont {S.~L.}\ \bibnamefont {Johnson}},\ }\href {\doibase
  10.1103/PhysRevLett.113.026401} {\bibfield  {journal} {\bibinfo  {journal}
  {Phys. Rev. Lett.}\ }\textbf {\bibinfo {volume} {113}},\ \bibinfo {pages}
  {026401} (\bibinfo {year} {2014})}\BibitemShut {NoStop}%
\bibitem [{\citenamefont {Beaud}\ \emph {et~al.}(2014)\citenamefont {Beaud},
  \citenamefont {Caviezel}, \citenamefont {Mariager}, \citenamefont {Rettig},
  \citenamefont {Ingold}, \citenamefont {Dornes}, \citenamefont {Huang},
  \citenamefont {Johnson}, \citenamefont {Radovic}, \citenamefont {Huber},
  \citenamefont {Kubacka}, \citenamefont {Ferrer}, \citenamefont {Lemke},
  \citenamefont {Chollet}, \citenamefont {Zhu}, \citenamefont {Glownia},
  \citenamefont {Sikorski}, \citenamefont {Robert}, \citenamefont {Wadati},
  \citenamefont {Nakamura}, \citenamefont {Kawasaki}, \citenamefont {Tokura},
  \citenamefont {Johnson},\ and\ \citenamefont {Staub}}]{BeaudNMAT2014}%
  \BibitemOpen
  \bibfield  {author} {\bibinfo {author} {\bibfnamefont {P.}~\bibnamefont
  {Beaud}}, \bibinfo {author} {\bibfnamefont {A.}~\bibnamefont {Caviezel}},
  \bibinfo {author} {\bibfnamefont {S.~O.}\ \bibnamefont {Mariager}}, \bibinfo
  {author} {\bibfnamefont {L.}~\bibnamefont {Rettig}}, \bibinfo {author}
  {\bibfnamefont {G.}~\bibnamefont {Ingold}}, \bibinfo {author} {\bibfnamefont
  {C.}~\bibnamefont {Dornes}}, \bibinfo {author} {\bibfnamefont {S.-W.}\
  \bibnamefont {Huang}}, \bibinfo {author} {\bibfnamefont {J.~A.}\ \bibnamefont
  {Johnson}}, \bibinfo {author} {\bibfnamefont {M.}~\bibnamefont {Radovic}},
  \bibinfo {author} {\bibfnamefont {T.}~\bibnamefont {Huber}}, \bibinfo
  {author} {\bibfnamefont {T.}~\bibnamefont {Kubacka}}, \bibinfo {author}
  {\bibfnamefont {A.}~\bibnamefont {Ferrer}}, \bibinfo {author} {\bibfnamefont
  {H.~T.}\ \bibnamefont {Lemke}}, \bibinfo {author} {\bibfnamefont
  {M.}~\bibnamefont {Chollet}}, \bibinfo {author} {\bibfnamefont
  {D.}~\bibnamefont {Zhu}}, \bibinfo {author} {\bibfnamefont {J.~M.}\
  \bibnamefont {Glownia}}, \bibinfo {author} {\bibfnamefont {M.}~\bibnamefont
  {Sikorski}}, \bibinfo {author} {\bibfnamefont {A.}~\bibnamefont {Robert}},
  \bibinfo {author} {\bibfnamefont {H.}~\bibnamefont {Wadati}}, \bibinfo
  {author} {\bibfnamefont {M.}~\bibnamefont {Nakamura}}, \bibinfo {author}
  {\bibfnamefont {M.}~\bibnamefont {Kawasaki}}, \bibinfo {author}
  {\bibfnamefont {Y.}~\bibnamefont {Tokura}}, \bibinfo {author} {\bibfnamefont
  {S.~L.}\ \bibnamefont {Johnson}}, \ and\ \bibinfo {author} {\bibfnamefont
  {U.}~\bibnamefont {Staub}},\ }\href {\doibase 10.1038/nmat4046} {\bibfield
  {journal} {\bibinfo  {journal} {Nature Materials}\ }\textbf {\bibinfo
  {volume} {13}},\ \bibinfo {pages} {923} (\bibinfo {year} {2014})}\BibitemShut
  {NoStop}%
\bibitem [{\citenamefont {Johnson}\ \emph {et~al.}(2009)\citenamefont
  {Johnson}, \citenamefont {Vorobeva}, \citenamefont {Beaud}, \citenamefont
  {Milne},\ and\ \citenamefont {Ingold}}]{JohnsonPRL2009}%
  \BibitemOpen
  \bibfield  {author} {\bibinfo {author} {\bibfnamefont {S.~L.}\ \bibnamefont
  {Johnson}}, \bibinfo {author} {\bibfnamefont {E.}~\bibnamefont {Vorobeva}},
  \bibinfo {author} {\bibfnamefont {P.}~\bibnamefont {Beaud}}, \bibinfo
  {author} {\bibfnamefont {C.~J.}\ \bibnamefont {Milne}}, \ and\ \bibinfo
  {author} {\bibfnamefont {G.}~\bibnamefont {Ingold}},\ }\href {\doibase
  10.1103/PhysRevLett.103.205501} {\bibfield  {journal} {\bibinfo  {journal}
  {Phys. Rev. Lett.}\ }\textbf {\bibinfo {volume} {103}},\ \bibinfo {pages}
  {205501} (\bibinfo {year} {2009})}\BibitemShut {NoStop}%
\bibitem [{\citenamefont {Dekorsy}\ \emph {et~al.}(1995)\citenamefont
  {Dekorsy}, \citenamefont {Auer}, \citenamefont {Waschke}, \citenamefont
  {Bakker}, \citenamefont {Roskos}, \citenamefont {Kurz}, \citenamefont
  {Wagner},\ and\ \citenamefont {Grosse}}]{KurzPRL1995}%
  \BibitemOpen
  \bibfield  {author} {\bibinfo {author} {\bibfnamefont {T.}~\bibnamefont
  {Dekorsy}}, \bibinfo {author} {\bibfnamefont {H.}~\bibnamefont {Auer}},
  \bibinfo {author} {\bibfnamefont {C.}~\bibnamefont {Waschke}}, \bibinfo
  {author} {\bibfnamefont {H.~J.}\ \bibnamefont {Bakker}}, \bibinfo {author}
  {\bibfnamefont {H.~G.}\ \bibnamefont {Roskos}}, \bibinfo {author}
  {\bibfnamefont {H.}~\bibnamefont {Kurz}}, \bibinfo {author} {\bibfnamefont
  {V.}~\bibnamefont {Wagner}}, \ and\ \bibinfo {author} {\bibfnamefont
  {P.}~\bibnamefont {Grosse}},\ }\href {\doibase 10.1103/PhysRevLett.74.738}
  {\bibfield  {journal} {\bibinfo  {journal} {Phys. Rev. Lett.}\ }\textbf
  {\bibinfo {volume} {74}},\ \bibinfo {pages} {738} (\bibinfo {year}
  {1995})}\BibitemShut {NoStop}%
\bibitem [{\citenamefont {Hunsche}\ \emph {et~al.}(1995)\citenamefont
  {Hunsche}, \citenamefont {Wienecke}, \citenamefont {Dekorsy},\ and\
  \citenamefont {Kurz}}]{HunschePRL1995}%
  \BibitemOpen
  \bibfield  {author} {\bibinfo {author} {\bibfnamefont {S.}~\bibnamefont
  {Hunsche}}, \bibinfo {author} {\bibfnamefont {K.}~\bibnamefont {Wienecke}},
  \bibinfo {author} {\bibfnamefont {T.}~\bibnamefont {Dekorsy}}, \ and\
  \bibinfo {author} {\bibfnamefont {H.}~\bibnamefont {Kurz}},\ }\href {\doibase
  10.1103/PhysRevLett.75.1815} {\bibfield  {journal} {\bibinfo  {journal}
  {Phys. Rev. Lett.}\ }\textbf {\bibinfo {volume} {75}},\ \bibinfo {pages}
  {1815} (\bibinfo {year} {1995})}\BibitemShut {NoStop}%
\bibitem [{\citenamefont {Kamaraju}\ \emph {et~al.}(2010)\citenamefont
  {Kamaraju}, \citenamefont {Kumar}, \citenamefont {Anija},\ and\ \citenamefont
  {Sood}}]{SoodPRB2010}%
  \BibitemOpen
  \bibfield  {author} {\bibinfo {author} {\bibfnamefont {N.}~\bibnamefont
  {Kamaraju}}, \bibinfo {author} {\bibfnamefont {S.}~\bibnamefont {Kumar}},
  \bibinfo {author} {\bibfnamefont {M.}~\bibnamefont {Anija}}, \ and\ \bibinfo
  {author} {\bibfnamefont {A.~K.}\ \bibnamefont {Sood}},\ }\href {\doibase
  10.1103/PhysRevB.82.195202} {\bibfield  {journal} {\bibinfo  {journal} {Phys.
  Rev. B}\ }\textbf {\bibinfo {volume} {82}},\ \bibinfo {pages} {195202}
  (\bibinfo {year} {2010})}\BibitemShut {NoStop}%
\bibitem [{\citenamefont {Cheng}\ \emph {et~al.}(2018)\citenamefont {Cheng},
  \citenamefont {Teitelbaum}, \citenamefont {Gao},\ and\ \citenamefont
  {Nelson}}]{NelsonPRB2018}%
  \BibitemOpen
  \bibfield  {author} {\bibinfo {author} {\bibfnamefont {Y.-H.}\ \bibnamefont
  {Cheng}}, \bibinfo {author} {\bibfnamefont {S.~W.}\ \bibnamefont
  {Teitelbaum}}, \bibinfo {author} {\bibfnamefont {F.~Y.}\ \bibnamefont {Gao}},
  \ and\ \bibinfo {author} {\bibfnamefont {K.~A.}\ \bibnamefont {Nelson}},\
  }\href {\doibase 10.1103/PhysRevB.98.134112} {\bibfield  {journal} {\bibinfo
  {journal} {Phys. Rev. B}\ }\textbf {\bibinfo {volume} {98}},\ \bibinfo
  {pages} {134112} (\bibinfo {year} {2018})}\BibitemShut {NoStop}%
\bibitem [{\citenamefont {Iyer}\ \emph {et~al.}(2019)\citenamefont {Iyer},
  \citenamefont {Segovia}, \citenamefont {Wang}, \citenamefont {Wu},
  \citenamefont {Ye},\ and\ \citenamefont {Xu}}]{XXFPRB2019}%
  \BibitemOpen
  \bibfield  {author} {\bibinfo {author} {\bibfnamefont {V.}~\bibnamefont
  {Iyer}}, \bibinfo {author} {\bibfnamefont {M.}~\bibnamefont {Segovia}},
  \bibinfo {author} {\bibfnamefont {Y.}~\bibnamefont {Wang}}, \bibinfo {author}
  {\bibfnamefont {W.}~\bibnamefont {Wu}}, \bibinfo {author} {\bibfnamefont
  {P.}~\bibnamefont {Ye}}, \ and\ \bibinfo {author} {\bibfnamefont
  {X.}~\bibnamefont {Xu}},\ }\href {\doibase 10.1103/PhysRevB.100.075436}
  {\bibfield  {journal} {\bibinfo  {journal} {Phys. Rev. B}\ }\textbf {\bibinfo
  {volume} {100}},\ \bibinfo {pages} {075436} (\bibinfo {year}
  {2019})}\BibitemShut {NoStop}%
\bibitem [{\citenamefont {Jnawali}\ \emph {et~al.}(2020)\citenamefont
  {Jnawali}, \citenamefont {Xiang}, \citenamefont {Linser}, \citenamefont
  {Shojaei}, \citenamefont {Wang}, \citenamefont {Qiu}, \citenamefont {Lian},
  \citenamefont {Wong}, \citenamefont {Wu}, \citenamefont {Ye}, \citenamefont
  {Leng}, \citenamefont {Jackson},\ and\ \citenamefont
  {Smith}}]{Smithncomm2020}%
  \BibitemOpen
  \bibfield  {author} {\bibinfo {author} {\bibfnamefont {G.}~\bibnamefont
  {Jnawali}}, \bibinfo {author} {\bibfnamefont {Y.}~\bibnamefont {Xiang}},
  \bibinfo {author} {\bibfnamefont {S.~M.}\ \bibnamefont {Linser}}, \bibinfo
  {author} {\bibfnamefont {I.~A.}\ \bibnamefont {Shojaei}}, \bibinfo {author}
  {\bibfnamefont {R.}~\bibnamefont {Wang}}, \bibinfo {author} {\bibfnamefont
  {G.}~\bibnamefont {Qiu}}, \bibinfo {author} {\bibfnamefont {C.}~\bibnamefont
  {Lian}}, \bibinfo {author} {\bibfnamefont {B.~M.}\ \bibnamefont {Wong}},
  \bibinfo {author} {\bibfnamefont {W.}~\bibnamefont {Wu}}, \bibinfo {author}
  {\bibfnamefont {P.~D.}\ \bibnamefont {Ye}}, \bibinfo {author} {\bibfnamefont
  {Y.}~\bibnamefont {Leng}}, \bibinfo {author} {\bibfnamefont {H.~E.}\
  \bibnamefont {Jackson}}, \ and\ \bibinfo {author} {\bibfnamefont {L.~M.}\
  \bibnamefont {Smith}},\ }\href {\doibase 10.1038/s41467-020-17766-5}
  {\bibfield  {journal} {\bibinfo  {journal} {Nature Communications}\ }\textbf
  {\bibinfo {volume} {11}},\ \bibinfo {pages} {3991} (\bibinfo {year}
  {2020})}\BibitemShut {NoStop}%
\bibitem [{\citenamefont {Kim}\ \emph {et~al.}(2003)\citenamefont {Kim},
  \citenamefont {Roeser},\ and\ \citenamefont {Mazur}}]{Kim2003}%
  \BibitemOpen
  \bibfield  {author} {\bibinfo {author} {\bibfnamefont {A.~M.-T.}\
  \bibnamefont {Kim}}, \bibinfo {author} {\bibfnamefont {C.~A.~D.}\
  \bibnamefont {Roeser}}, \ and\ \bibinfo {author} {\bibfnamefont
  {E.}~\bibnamefont {Mazur}},\ }\href {\doibase 10.1103/PhysRevB.68.012301}
  {\bibfield  {journal} {\bibinfo  {journal} {Phys. Rev. B}\ }\textbf {\bibinfo
  {volume} {68}},\ \bibinfo {pages} {012301} (\bibinfo {year}
  {2003})}\BibitemShut {NoStop}%
\bibitem [{\citenamefont {Torchinsky}\ \emph {et~al.}(2014)\citenamefont
  {Torchinsky}, \citenamefont {Chu}, \citenamefont {Qi}, \citenamefont {Cao},\
  and\ \citenamefont {Hsieh}}]{HsiehRSI2014}%
  \BibitemOpen
  \bibfield  {author} {\bibinfo {author} {\bibfnamefont {D.~H.}\ \bibnamefont
  {Torchinsky}}, \bibinfo {author} {\bibfnamefont {H.}~\bibnamefont {Chu}},
  \bibinfo {author} {\bibfnamefont {T.}~\bibnamefont {Qi}}, \bibinfo {author}
  {\bibfnamefont {G.}~\bibnamefont {Cao}}, \ and\ \bibinfo {author}
  {\bibfnamefont {D.}~\bibnamefont {Hsieh}},\ }\href {\doibase
  10.1063/1.4891417} {\bibfield  {journal} {\bibinfo  {journal} {Review of
  Scientific Instruments}\ }\textbf {\bibinfo {volume} {85}},\ \bibinfo {pages}
  {083102} (\bibinfo {year} {2014})} \BibitemShut {NoStop}%
\bibitem [{\citenamefont {Cheng}\ \emph {et~al.}(2019)\citenamefont {Cheng},
  \citenamefont {Wu}, \citenamefont {Zhu},\ and\ \citenamefont
  {Guo}}]{GGYPRB2019}%
  \BibitemOpen
  \bibfield  {author} {\bibinfo {author} {\bibfnamefont {M.}~\bibnamefont
  {Cheng}}, \bibinfo {author} {\bibfnamefont {S.}~\bibnamefont {Wu}}, \bibinfo
  {author} {\bibfnamefont {Z.-Z.}\ \bibnamefont {Zhu}}, \ and\ \bibinfo
  {author} {\bibfnamefont {G.-Y.}\ \bibnamefont {Guo}},\ }\href {\doibase
  10.1103/PhysRevB.100.035202} {\bibfield  {journal} {\bibinfo  {journal}
  {Phys. Rev. B}\ }\textbf {\bibinfo {volume} {100}},\ \bibinfo {pages}
  {035202} (\bibinfo {year} {2019})}\BibitemShut {NoStop}%
\bibitem [{\citenamefont {Kogar}\ \emph {et~al.}(2020)\citenamefont {Kogar},
  \citenamefont {Zong}, \citenamefont {Dolgirev}, \citenamefont {Shen},
  \citenamefont {Straquadine}, \citenamefont {Bie}, \citenamefont {Wang},
  \citenamefont {Rohwer}, \citenamefont {Tung}, \citenamefont {Yang},
  \citenamefont {Li}, \citenamefont {Yang}, \citenamefont {Weathersby},
  \citenamefont {Park}, \citenamefont {Kozina}, \citenamefont {Sie},
  \citenamefont {Wen}, \citenamefont {Jarillo-Herrero}, \citenamefont {Fisher},
  \citenamefont {Wang},\ and\ \citenamefont {Gedik}}]{GedikNPHY2020}%
  \BibitemOpen
  \bibfield  {author} {\bibinfo {author} {\bibfnamefont {A.}~\bibnamefont
  {Kogar}}, \bibinfo {author} {\bibfnamefont {A.}~\bibnamefont {Zong}},
  \bibinfo {author} {\bibfnamefont {P.~E.}\ \bibnamefont {Dolgirev}}, \bibinfo
  {author} {\bibfnamefont {X.}~\bibnamefont {Shen}}, \bibinfo {author}
  {\bibfnamefont {J.}~\bibnamefont {Straquadine}}, \bibinfo {author}
  {\bibfnamefont {Y.-Q.}\ \bibnamefont {Bie}}, \bibinfo {author} {\bibfnamefont
  {X.}~\bibnamefont {Wang}}, \bibinfo {author} {\bibfnamefont {T.}~\bibnamefont
  {Rohwer}}, \bibinfo {author} {\bibfnamefont {I.-C.}\ \bibnamefont {Tung}},
  \bibinfo {author} {\bibfnamefont {Y.}~\bibnamefont {Yang}}, \bibinfo {author}
  {\bibfnamefont {R.}~\bibnamefont {Li}}, \bibinfo {author} {\bibfnamefont
  {J.}~\bibnamefont {Yang}}, \bibinfo {author} {\bibfnamefont {S.}~\bibnamefont
  {Weathersby}}, \bibinfo {author} {\bibfnamefont {S.}~\bibnamefont {Park}},
  \bibinfo {author} {\bibfnamefont {M.~E.}\ \bibnamefont {Kozina}}, \bibinfo
  {author} {\bibfnamefont {E.~J.}\ \bibnamefont {Sie}}, \bibinfo {author}
  {\bibfnamefont {H.}~\bibnamefont {Wen}}, \bibinfo {author} {\bibfnamefont
  {P.}~\bibnamefont {Jarillo-Herrero}}, \bibinfo {author} {\bibfnamefont
  {I.~R.}\ \bibnamefont {Fisher}}, \bibinfo {author} {\bibfnamefont
  {X.}~\bibnamefont {Wang}}, \ and\ \bibinfo {author} {\bibfnamefont
  {N.}~\bibnamefont {Gedik}},\ }\href {\doibase 10.1038/s41567-019-0705-3}
  {\bibfield  {journal} {\bibinfo  {journal} {Nature Physics}\ }\textbf
  {\bibinfo {volume} {16}},\ \bibinfo {pages} {159} (\bibinfo {year}
  {2020})}\BibitemShut {NoStop}%
\bibitem [{\citenamefont {Caviglia}\ \emph {et~al.}(2013)\citenamefont
  {Caviglia}, \citenamefont {F\"orst}, \citenamefont {Scherwitzl},
  \citenamefont {Khanna}, \citenamefont {Bromberger}, \citenamefont
  {Mankowsky}, \citenamefont {Singla}, \citenamefont {Chuang}, \citenamefont
  {Lee}, \citenamefont {Krupin}, \citenamefont {Schlotter}, \citenamefont
  {Turner}, \citenamefont {Dakovski}, \citenamefont {Minitti}, \citenamefont
  {Robinson}, \citenamefont {Scagnoli}, \citenamefont {Wilkins}, \citenamefont
  {Cavill}, \citenamefont {Gibert}, \citenamefont {Gariglio}, \citenamefont
  {Zubko}, \citenamefont {Triscone}, \citenamefont {Hill}, \citenamefont
  {Dhesi},\ and\ \citenamefont {Cavalleri}}]{CavalleriPRB2013}%
  \BibitemOpen
  \bibfield  {author} {\bibinfo {author} {\bibfnamefont {A.~D.}\ \bibnamefont
  {Caviglia}}, \bibinfo {author} {\bibfnamefont {M.}~\bibnamefont {F\"orst}},
  \bibinfo {author} {\bibfnamefont {R.}~\bibnamefont {Scherwitzl}}, \bibinfo
  {author} {\bibfnamefont {V.}~\bibnamefont {Khanna}}, \bibinfo {author}
  {\bibfnamefont {H.}~\bibnamefont {Bromberger}}, \bibinfo {author}
  {\bibfnamefont {R.}~\bibnamefont {Mankowsky}}, \bibinfo {author}
  {\bibfnamefont {R.}~\bibnamefont {Singla}}, \bibinfo {author} {\bibfnamefont
  {Y.-D.}\ \bibnamefont {Chuang}}, \bibinfo {author} {\bibfnamefont {W.~S.}\
  \bibnamefont {Lee}}, \bibinfo {author} {\bibfnamefont {O.}~\bibnamefont
  {Krupin}}, \bibinfo {author} {\bibfnamefont {W.~F.}\ \bibnamefont
  {Schlotter}}, \bibinfo {author} {\bibfnamefont {J.~J.}\ \bibnamefont
  {Turner}}, \bibinfo {author} {\bibfnamefont {G.~L.}\ \bibnamefont
  {Dakovski}}, \bibinfo {author} {\bibfnamefont {M.~P.}\ \bibnamefont
  {Minitti}}, \bibinfo {author} {\bibfnamefont {J.}~\bibnamefont {Robinson}},
  \bibinfo {author} {\bibfnamefont {V.}~\bibnamefont {Scagnoli}}, \bibinfo
  {author} {\bibfnamefont {S.~B.}\ \bibnamefont {Wilkins}}, \bibinfo {author}
  {\bibfnamefont {S.~A.}\ \bibnamefont {Cavill}}, \bibinfo {author}
  {\bibfnamefont {M.}~\bibnamefont {Gibert}}, \bibinfo {author} {\bibfnamefont
  {S.}~\bibnamefont {Gariglio}}, \bibinfo {author} {\bibfnamefont
  {P.}~\bibnamefont {Zubko}}, \bibinfo {author} {\bibfnamefont {J.-M.}\
  \bibnamefont {Triscone}}, \bibinfo {author} {\bibfnamefont {J.~P.}\
  \bibnamefont {Hill}}, \bibinfo {author} {\bibfnamefont {S.~S.}\ \bibnamefont
  {Dhesi}}, \ and\ \bibinfo {author} {\bibfnamefont {A.}~\bibnamefont
  {Cavalleri}},\ }\href {\doibase 10.1103/PhysRevB.88.220401} {\bibfield
  {journal} {\bibinfo  {journal} {Phys. Rev. B}\ }\textbf {\bibinfo {volume}
  {88}},\ \bibinfo {pages} {220401} (\bibinfo {year} {2013})}\BibitemShut
  {NoStop}%
\bibitem [{\citenamefont {Yusupov}\ \emph {et~al.}(2010)\citenamefont
  {Yusupov}, \citenamefont {Mertelj}, \citenamefont {Kabanov}, \citenamefont
  {Brazovskii}, \citenamefont {Kusar}, \citenamefont {Chu}, \citenamefont
  {Fisher},\ and\ \citenamefont {Mihailovic}}]{MihailovicNPHY2010}%
  \BibitemOpen
  \bibfield  {author} {\bibinfo {author} {\bibfnamefont {R.}~\bibnamefont
  {Yusupov}}, \bibinfo {author} {\bibfnamefont {T.}~\bibnamefont {Mertelj}},
  \bibinfo {author} {\bibfnamefont {V.~V.}\ \bibnamefont {Kabanov}}, \bibinfo
  {author} {\bibfnamefont {S.}~\bibnamefont {Brazovskii}}, \bibinfo {author}
  {\bibfnamefont {P.}~\bibnamefont {Kusar}}, \bibinfo {author} {\bibfnamefont
  {J.-H.}\ \bibnamefont {Chu}}, \bibinfo {author} {\bibfnamefont {I.~R.}\
  \bibnamefont {Fisher}}, \ and\ \bibinfo {author} {\bibfnamefont
  {D.}~\bibnamefont {Mihailovic}},\ }\href {\doibase 10.1038/nphys1738}
  {\bibfield  {journal} {\bibinfo  {journal} {Nature Physics}\ }\textbf
  {\bibinfo {volume} {6}},\ \bibinfo {pages} {681} (\bibinfo {year}
  {2010})}\BibitemShut {NoStop}%
\bibitem [{\citenamefont {Giannozzi}\ \emph {et~al.}(2009)\citenamefont
  {Giannozzi}, \citenamefont {Baroni}, \citenamefont {Bonini}, \citenamefont
  {Calandra}, \citenamefont {Car}, \citenamefont {Cavazzoni}, \citenamefont
  {Ceresoli}, \citenamefont {Chiarotti}, \citenamefont {Cococcioni},
  \citenamefont {Dabo}, \citenamefont {Corso}, \citenamefont {de~Gironcoli},
  \citenamefont {Fabris}, \citenamefont {Fratesi}, \citenamefont {Gebauer},
  \citenamefont {Gerstmann}, \citenamefont {Gougoussis}, \citenamefont
  {Kokalj}, \citenamefont {Lazzeri}, \citenamefont {Martin-Samos},
  \citenamefont {Marzari}, \citenamefont {Mauri}, \citenamefont {Mazzarello},
  \citenamefont {Paolini}, \citenamefont {Pasquarello}, \citenamefont
  {Paulatto}, \citenamefont {Sbraccia}, \citenamefont {Scandolo}, \citenamefont
  {Sclauzero}, \citenamefont {Seitsonen}, \citenamefont {Smogunov},
  \citenamefont {Umari},\ and\ \citenamefont {Wentzcovitch}}]{Giannozzi2009}%
  \BibitemOpen
  \bibfield  {author} {\bibinfo {author} {\bibfnamefont {P.}~\bibnamefont
  {Giannozzi}}, \bibinfo {author} {\bibfnamefont {S.}~\bibnamefont {Baroni}},
  \bibinfo {author} {\bibfnamefont {N.}~\bibnamefont {Bonini}}, \bibinfo
  {author} {\bibfnamefont {M.}~\bibnamefont {Calandra}}, \bibinfo {author}
  {\bibfnamefont {R.}~\bibnamefont {Car}}, \bibinfo {author} {\bibfnamefont
  {C.}~\bibnamefont {Cavazzoni}}, \bibinfo {author} {\bibfnamefont
  {D.}~\bibnamefont {Ceresoli}}, \bibinfo {author} {\bibfnamefont {G.~L.}\
  \bibnamefont {Chiarotti}}, \bibinfo {author} {\bibfnamefont {M.}~\bibnamefont
  {Cococcioni}}, \bibinfo {author} {\bibfnamefont {I.}~\bibnamefont {Dabo}},
  \bibinfo {author} {\bibfnamefont {A.~D.}\ \bibnamefont {Corso}}, \bibinfo
  {author} {\bibfnamefont {S.}~\bibnamefont {de~Gironcoli}}, \bibinfo {author}
  {\bibfnamefont {S.}~\bibnamefont {Fabris}}, \bibinfo {author} {\bibfnamefont
  {G.}~\bibnamefont {Fratesi}}, \bibinfo {author} {\bibfnamefont
  {R.}~\bibnamefont {Gebauer}}, \bibinfo {author} {\bibfnamefont
  {U.}~\bibnamefont {Gerstmann}}, \bibinfo {author} {\bibfnamefont
  {C.}~\bibnamefont {Gougoussis}}, \bibinfo {author} {\bibfnamefont
  {A.}~\bibnamefont {Kokalj}}, \bibinfo {author} {\bibfnamefont
  {M.}~\bibnamefont {Lazzeri}}, \bibinfo {author} {\bibfnamefont
  {L.}~\bibnamefont {Martin-Samos}}, \bibinfo {author} {\bibfnamefont
  {N.}~\bibnamefont {Marzari}}, \bibinfo {author} {\bibfnamefont
  {F.}~\bibnamefont {Mauri}}, \bibinfo {author} {\bibfnamefont
  {R.}~\bibnamefont {Mazzarello}}, \bibinfo {author} {\bibfnamefont
  {S.}~\bibnamefont {Paolini}}, \bibinfo {author} {\bibfnamefont
  {A.}~\bibnamefont {Pasquarello}}, \bibinfo {author} {\bibfnamefont
  {L.}~\bibnamefont {Paulatto}}, \bibinfo {author} {\bibfnamefont
  {C.}~\bibnamefont {Sbraccia}}, \bibinfo {author} {\bibfnamefont
  {S.}~\bibnamefont {Scandolo}}, \bibinfo {author} {\bibfnamefont
  {G.}~\bibnamefont {Sclauzero}}, \bibinfo {author} {\bibfnamefont {A.~P.}\
  \bibnamefont {Seitsonen}}, \bibinfo {author} {\bibfnamefont {A.}~\bibnamefont
  {Smogunov}}, \bibinfo {author} {\bibfnamefont {P.}~\bibnamefont {Umari}}, \
  and\ \bibinfo {author} {\bibfnamefont {R.~M.}\ \bibnamefont {Wentzcovitch}},\
  }\href {\doibase 10.1088/0953-8984/21/39/395502} {\bibfield  {journal}
  {\bibinfo  {journal} {Journal of Physics: Condensed Matter}\ }\textbf
  {\bibinfo {volume} {21}},\ \bibinfo {pages} {395502} (\bibinfo {year}
  {2009})}\BibitemShut {NoStop}%
\bibitem [{\citenamefont {Giannozzi}\ \emph {et~al.}(2017)\citenamefont
  {Giannozzi}, \citenamefont {Andreussi}, \citenamefont {Brumme}, \citenamefont
  {Bunau}, \citenamefont {Nardelli}, \citenamefont {Calandra}, \citenamefont
  {Car}, \citenamefont {Cavazzoni}, \citenamefont {Ceresoli}, \citenamefont
  {Cococcioni}, \citenamefont {Colonna}, \citenamefont {Carnimeo},
  \citenamefont {Corso}, \citenamefont {de~Gironcoli}, \citenamefont {Delugas},
  \citenamefont {DiStasio}, \citenamefont {Ferretti}, \citenamefont {Floris},
  \citenamefont {Fratesi}, \citenamefont {Fugallo}, \citenamefont {Gebauer},
  \citenamefont {Gerstmann}, \citenamefont {Giustino}, \citenamefont {Gorni},
  \citenamefont {Jia}, \citenamefont {Kawamura}, \citenamefont {Ko},
  \citenamefont {Kokalj}, \citenamefont {K{\"u}{\c{c}}{\"u}kbenli},
  \citenamefont {Lazzeri}, \citenamefont {Marsili}, \citenamefont {Marzari},
  \citenamefont {Mauri}, \citenamefont {Nguyen}, \citenamefont {Nguyen},
  \citenamefont {de-la Roza}, \citenamefont {Paulatto}, \citenamefont
  {Ponc{\'{e}}}, \citenamefont {Rocca}, \citenamefont {Sabatini}, \citenamefont
  {Santra}, \citenamefont {Schlipf}, \citenamefont {Seitsonen}, \citenamefont
  {Smogunov}, \citenamefont {Timrov}, \citenamefont {Thonhauser}, \citenamefont
  {Umari}, \citenamefont {Vast}, \citenamefont {Wu},\ and\ \citenamefont
  {Baroni}}]{Giannozzi2017}%
  \BibitemOpen
  \bibfield  {author} {\bibinfo {author} {\bibfnamefont {P.}~\bibnamefont
  {Giannozzi}}, \bibinfo {author} {\bibfnamefont {O.}~\bibnamefont
  {Andreussi}}, \bibinfo {author} {\bibfnamefont {T.}~\bibnamefont {Brumme}},
  \bibinfo {author} {\bibfnamefont {O.}~\bibnamefont {Bunau}}, \bibinfo
  {author} {\bibfnamefont {M.~B.}\ \bibnamefont {Nardelli}}, \bibinfo {author}
  {\bibfnamefont {M.}~\bibnamefont {Calandra}}, \bibinfo {author}
  {\bibfnamefont {R.}~\bibnamefont {Car}}, \bibinfo {author} {\bibfnamefont
  {C.}~\bibnamefont {Cavazzoni}}, \bibinfo {author} {\bibfnamefont
  {D.}~\bibnamefont {Ceresoli}}, \bibinfo {author} {\bibfnamefont
  {M.}~\bibnamefont {Cococcioni}}, \bibinfo {author} {\bibfnamefont
  {N.}~\bibnamefont {Colonna}}, \bibinfo {author} {\bibfnamefont
  {I.}~\bibnamefont {Carnimeo}}, \bibinfo {author} {\bibfnamefont {A.~D.}\
  \bibnamefont {Corso}}, \bibinfo {author} {\bibfnamefont {S.}~\bibnamefont
  {de~Gironcoli}}, \bibinfo {author} {\bibfnamefont {P.}~\bibnamefont
  {Delugas}}, \bibinfo {author} {\bibfnamefont {R.~A.}\ \bibnamefont
  {DiStasio}}, \bibinfo {author} {\bibfnamefont {A.}~\bibnamefont {Ferretti}},
  \bibinfo {author} {\bibfnamefont {A.}~\bibnamefont {Floris}}, \bibinfo
  {author} {\bibfnamefont {G.}~\bibnamefont {Fratesi}}, \bibinfo {author}
  {\bibfnamefont {G.}~\bibnamefont {Fugallo}}, \bibinfo {author} {\bibfnamefont
  {R.}~\bibnamefont {Gebauer}}, \bibinfo {author} {\bibfnamefont
  {U.}~\bibnamefont {Gerstmann}}, \bibinfo {author} {\bibfnamefont
  {F.}~\bibnamefont {Giustino}}, \bibinfo {author} {\bibfnamefont
  {T.}~\bibnamefont {Gorni}}, \bibinfo {author} {\bibfnamefont
  {J.}~\bibnamefont {Jia}}, \bibinfo {author} {\bibfnamefont {M.}~\bibnamefont
  {Kawamura}}, \bibinfo {author} {\bibfnamefont {H.-Y.}\ \bibnamefont {Ko}},
  \bibinfo {author} {\bibfnamefont {A.}~\bibnamefont {Kokalj}}, \bibinfo
  {author} {\bibfnamefont {E.}~\bibnamefont {K{\"u}{\c{c}}{\"u}kbenli}},
  \bibinfo {author} {\bibfnamefont {M.}~\bibnamefont {Lazzeri}}, \bibinfo
  {author} {\bibfnamefont {M.}~\bibnamefont {Marsili}}, \bibinfo {author}
  {\bibfnamefont {N.}~\bibnamefont {Marzari}}, \bibinfo {author} {\bibfnamefont
  {F.}~\bibnamefont {Mauri}}, \bibinfo {author} {\bibfnamefont {N.~L.}\
  \bibnamefont {Nguyen}}, \bibinfo {author} {\bibfnamefont {H.-V.}\
  \bibnamefont {Nguyen}}, \bibinfo {author} {\bibfnamefont {A.~O.}\
  \bibnamefont {de-la Roza}}, \bibinfo {author} {\bibfnamefont
  {L.}~\bibnamefont {Paulatto}}, \bibinfo {author} {\bibfnamefont
  {S.}~\bibnamefont {Ponc{\'{e}}}}, \bibinfo {author} {\bibfnamefont
  {D.}~\bibnamefont {Rocca}}, \bibinfo {author} {\bibfnamefont
  {R.}~\bibnamefont {Sabatini}}, \bibinfo {author} {\bibfnamefont
  {B.}~\bibnamefont {Santra}}, \bibinfo {author} {\bibfnamefont
  {M.}~\bibnamefont {Schlipf}}, \bibinfo {author} {\bibfnamefont {A.~P.}\
  \bibnamefont {Seitsonen}}, \bibinfo {author} {\bibfnamefont {A.}~\bibnamefont
  {Smogunov}}, \bibinfo {author} {\bibfnamefont {I.}~\bibnamefont {Timrov}},
  \bibinfo {author} {\bibfnamefont {T.}~\bibnamefont {Thonhauser}}, \bibinfo
  {author} {\bibfnamefont {P.}~\bibnamefont {Umari}}, \bibinfo {author}
  {\bibfnamefont {N.}~\bibnamefont {Vast}}, \bibinfo {author} {\bibfnamefont
  {X.}~\bibnamefont {Wu}}, \ and\ \bibinfo {author} {\bibfnamefont
  {S.}~\bibnamefont {Baroni}},\ }\href {\doibase 10.1088/1361-648X/aa8f79}
  {\bibfield  {journal} {\bibinfo  {journal} {Journal of Physics: Condensed
  Matter}\ }\textbf {\bibinfo {volume} {29}},\ \bibinfo {pages} {465901}
  (\bibinfo {year} {2017})}\BibitemShut {NoStop}%
\bibitem [{\citenamefont {Tangney}\ and\ \citenamefont
  {Fahy}(1999)}]{TangneyPRL1998}%
  \BibitemOpen
  \bibfield  {author} {\bibinfo {author} {\bibfnamefont {P.}~\bibnamefont
  {Tangney}}\ and\ \bibinfo {author} {\bibfnamefont {S.}~\bibnamefont {Fahy}},\
  }\href {\doibase 10.1103/PhysRevLett.82.4340} {\bibfield  {journal} {\bibinfo
   {journal} {Phys. Rev. Lett.}\ }\textbf {\bibinfo {volume} {82}},\ \bibinfo
  {pages} {4340} (\bibinfo {year} {1999})}\BibitemShut {NoStop}%
\bibitem [{\citenamefont {Lian}\ \emph
  {et~al.}(2018{\natexlab{a}})\citenamefont {Lian}, \citenamefont {Hu},
  \citenamefont {Guan},\ and\ \citenamefont {Meng}}]{Lian2018MultiK}%
  \BibitemOpen
  \bibfield  {author} {\bibinfo {author} {\bibfnamefont {C.}~\bibnamefont
  {Lian}}, \bibinfo {author} {\bibfnamefont {S.-Q.}\ \bibnamefont {Hu}},
  \bibinfo {author} {\bibfnamefont {M.-X.}\ \bibnamefont {Guan}}, \ and\
  \bibinfo {author} {\bibfnamefont {S.}~\bibnamefont {Meng}},\ }\href {\doibase
  10.1063/1.5036543} {\bibfield  {journal} {\bibinfo  {journal} {The Journal of
  Chemical Physics}\ }\textbf {\bibinfo {volume} {149}},\ \bibinfo {pages}
  {154104} (\bibinfo {year} {2018}{\natexlab{a}})}\BibitemShut {NoStop}%
\bibitem [{\citenamefont {Lian}\ \emph
  {et~al.}(2018{\natexlab{b}})\citenamefont {Lian}, \citenamefont {Guan},
  \citenamefont {Hu}, \citenamefont {Zhang},\ and\ \citenamefont
  {Meng}}]{Lian2018AdvTheo}%
  \BibitemOpen
  \bibfield  {author} {\bibinfo {author} {\bibfnamefont {C.}~\bibnamefont
  {Lian}}, \bibinfo {author} {\bibfnamefont {M.}~\bibnamefont {Guan}}, \bibinfo
  {author} {\bibfnamefont {S.}~\bibnamefont {Hu}}, \bibinfo {author}
  {\bibfnamefont {J.}~\bibnamefont {Zhang}}, \ and\ \bibinfo {author}
  {\bibfnamefont {S.}~\bibnamefont {Meng}},\ }\href {\doibase
  10.1002/adts.201800055} {\bibfield  {journal} {\bibinfo  {journal} {Advanced
  Theory and Simulations}\ }\textbf {\bibinfo {volume} {1}},\ \bibinfo {pages}
  {1800055} (\bibinfo {year} {2018}{\natexlab{b}})}\BibitemShut {NoStop}%
\bibitem [{\citenamefont {Lian}\ \emph {et~al.}(2020)\citenamefont {Lian},
  \citenamefont {Zhang}, \citenamefont {Hu}, \citenamefont {Guan},\ and\
  \citenamefont {Meng}}]{Lian2019CDWDyn}%
  \BibitemOpen
  \bibfield  {author} {\bibinfo {author} {\bibfnamefont {C.}~\bibnamefont
  {Lian}}, \bibinfo {author} {\bibfnamefont {S.-J.}\ \bibnamefont {Zhang}},
  \bibinfo {author} {\bibfnamefont {S.-Q.}\ \bibnamefont {Hu}}, \bibinfo
  {author} {\bibfnamefont {M.-X.}\ \bibnamefont {Guan}}, \ and\ \bibinfo
  {author} {\bibfnamefont {S.}~\bibnamefont {Meng}},\ }\href {\doibase
  10.1038/s41467-019-13672-7} {\bibfield  {journal} {\bibinfo  {journal}
  {Nature Communications}\ }\textbf {\bibinfo {volume} {11}},\ \bibinfo {pages}
  {1} (\bibinfo {year} {2020})}\BibitemShut {NoStop}%
\bibitem [{\citenamefont {Runge}\ and\ \citenamefont
  {Gross}(1984)}]{Runge1984}%
  \BibitemOpen
  \bibfield  {author} {\bibinfo {author} {\bibfnamefont {E.}~\bibnamefont
  {Runge}}\ and\ \bibinfo {author} {\bibfnamefont {E.~K.~U.}\ \bibnamefont
  {Gross}},\ }\href {\doibase 10.1103/PhysRevLett.52.997} {\bibfield  {journal}
  {\bibinfo  {journal} {Physical Review Letters}\ }\textbf {\bibinfo {volume}
  {52}},\ \bibinfo {pages} {997} (\bibinfo {year} {1984})}\BibitemShut
  {NoStop}%
\bibitem [{\citenamefont {Bertsch}\ \emph {et~al.}(2000)\citenamefont
  {Bertsch}, \citenamefont {Iwata}, \citenamefont {Rubio},\ and\ \citenamefont
  {Yabana}}]{Bertsch2000}%
  \BibitemOpen
  \bibfield  {author} {\bibinfo {author} {\bibfnamefont {G.~F.}\ \bibnamefont
  {Bertsch}}, \bibinfo {author} {\bibfnamefont {J.-I.}\ \bibnamefont {Iwata}},
  \bibinfo {author} {\bibfnamefont {A.}~\bibnamefont {Rubio}}, \ and\ \bibinfo
  {author} {\bibfnamefont {K.}~\bibnamefont {Yabana}},\ }\href {\doibase
  10.1103/PhysRevB.62.7998} {\bibfield  {journal} {\bibinfo  {journal}
  {Physical Review B}\ }\textbf {\bibinfo {volume} {62}},\ \bibinfo {pages}
  {7998} (\bibinfo {year} {2000})}\BibitemShut {NoStop}%
\bibitem [{\citenamefont {Wang}\ \emph {et~al.}(2015)\citenamefont {Wang},
  \citenamefont {Li},\ and\ \citenamefont {Wang}}]{Wang2015RTTDDFT}%
  \BibitemOpen
  \bibfield  {author} {\bibinfo {author} {\bibfnamefont {Z.}~\bibnamefont
  {Wang}}, \bibinfo {author} {\bibfnamefont {S.-S.}\ \bibnamefont {Li}}, \ and\
  \bibinfo {author} {\bibfnamefont {L.-W.}\ \bibnamefont {Wang}},\ }\href
  {\doibase 10.1103/PhysRevLett.114.063004} {\bibfield  {journal} {\bibinfo
  {journal} {Physical Review Letters}\ }\textbf {\bibinfo {volume} {114}},\
  \bibinfo {pages} {063004} (\bibinfo {year} {2015})}\BibitemShut {NoStop}%
\bibitem [{\citenamefont {Perdew}\ \emph {et~al.}(1996)\citenamefont {Perdew},
  \citenamefont {Burke},\ and\ \citenamefont {Ernzerhof}}]{Perdew1996}%
  \BibitemOpen
  \bibfield  {author} {\bibinfo {author} {\bibfnamefont {J.~P.}\ \bibnamefont
  {Perdew}}, \bibinfo {author} {\bibfnamefont {K.}~\bibnamefont {Burke}}, \
  and\ \bibinfo {author} {\bibfnamefont {M.}~\bibnamefont {Ernzerhof}},\ }\href
  {\doibase 10.1103/PhysRevLett.77.3865} {\bibfield  {journal} {\bibinfo
  {journal} {Physical Review Letters}\ }\textbf {\bibinfo {volume} {77}},\
  \bibinfo {pages} {3865} (\bibinfo {year} {1996})}\BibitemShut {NoStop}%
\bibitem [{\citenamefont {van Setten}\ \emph {et~al.}(2018)\citenamefont {van
  Setten}, \citenamefont {Giantomassi}, \citenamefont {Bousquet}, \citenamefont
  {Verstraete}, \citenamefont {Hamann}, \citenamefont {Gonze},\ and\
  \citenamefont {Rignanese}}]{vanSetten2018}%
  \BibitemOpen
  \bibfield  {author} {\bibinfo {author} {\bibfnamefont {M.}~\bibnamefont {van
  Setten}}, \bibinfo {author} {\bibfnamefont {M.}~\bibnamefont {Giantomassi}},
  \bibinfo {author} {\bibfnamefont {E.}~\bibnamefont {Bousquet}}, \bibinfo
  {author} {\bibfnamefont {M.}~\bibnamefont {Verstraete}}, \bibinfo {author}
  {\bibfnamefont {D.}~\bibnamefont {Hamann}}, \bibinfo {author} {\bibfnamefont
  {X.}~\bibnamefont {Gonze}}, \ and\ \bibinfo {author} {\bibfnamefont {G.-M.}\
  \bibnamefont {Rignanese}},\ }\href {\doibase 10.1016/j.cpc.2018.01.012}
  {\bibfield  {journal} {\bibinfo  {journal} {Computer Physics Communications}\
  }\textbf {\bibinfo {volume} {226}},\ \bibinfo {pages} {39} (\bibinfo {year}
  {2018})}\BibitemShut {NoStop}%
\bibitem [{\citenamefont {Tutihasi}\ \emph {et~al.}(1969)\citenamefont
  {Tutihasi}, \citenamefont {Roberts}, \citenamefont {Keezer},\ and\
  \citenamefont {Drews}}]{DrewsPR1969}%
  \BibitemOpen
  \bibfield  {author} {\bibinfo {author} {\bibfnamefont {S.}~\bibnamefont
  {Tutihasi}}, \bibinfo {author} {\bibfnamefont {G.~G.}\ \bibnamefont
  {Roberts}}, \bibinfo {author} {\bibfnamefont {R.~C.}\ \bibnamefont {Keezer}},
  \ and\ \bibinfo {author} {\bibfnamefont {R.~E.}\ \bibnamefont {Drews}},\
  }\href {\doibase 10.1103/PhysRev.177.1143} {\bibfield  {journal} {\bibinfo
  {journal} {Phys. Rev.}\ }\textbf {\bibinfo {volume} {177}},\ \bibinfo {pages}
  {1143} (\bibinfo {year} {1969})}\BibitemShut {NoStop}%
\bibitem [{\citenamefont {Lin}\ \emph {et~al.}(2016)\citenamefont {Lin},
  \citenamefont {Li}, \citenamefont {Chen}, \citenamefont {Shen}, \citenamefont
  {Ge},\ and\ \citenamefont {Pei}}]{PYZNCOMM2016}%
  \BibitemOpen
  \bibfield  {author} {\bibinfo {author} {\bibfnamefont {S.}~\bibnamefont
  {Lin}}, \bibinfo {author} {\bibfnamefont {W.}~\bibnamefont {Li}}, \bibinfo
  {author} {\bibfnamefont {Z.}~\bibnamefont {Chen}}, \bibinfo {author}
  {\bibfnamefont {J.}~\bibnamefont {Shen}}, \bibinfo {author} {\bibfnamefont
  {B.}~\bibnamefont {Ge}}, \ and\ \bibinfo {author} {\bibfnamefont
  {Y.}~\bibnamefont {Pei}},\ }\href {\doibase 10.1038/ncomms10287} {\bibfield
  {journal} {\bibinfo  {journal} {Nature Communications}\ }\textbf {\bibinfo
  {volume} {7}},\ \bibinfo {pages} {10287} (\bibinfo {year}
  {2016})}\BibitemShut {NoStop}%
\bibitem [{\citenamefont {Hejny}\ and\ \citenamefont
  {McMahon}(2003)}]{HejnyPRL2003}%
  \BibitemOpen
  \bibfield  {author} {\bibinfo {author} {\bibfnamefont {C.}~\bibnamefont
  {Hejny}}\ and\ \bibinfo {author} {\bibfnamefont {M.~I.}\ \bibnamefont
  {McMahon}},\ }\href {\doibase 10.1103/PhysRevLett.91.215502} {\bibfield
  {journal} {\bibinfo  {journal} {Phys. Rev. Lett.}\ }\textbf {\bibinfo
  {volume} {91}},\ \bibinfo {pages} {215502} (\bibinfo {year}
  {2003})}\BibitemShut {NoStop}%
\bibitem [{\citenamefont {Hejny}\ and\ \citenamefont
  {McMahon}(2004)}]{HejnyPRB2004}%
  \BibitemOpen
  \bibfield  {author} {\bibinfo {author} {\bibfnamefont {C.}~\bibnamefont
  {Hejny}}\ and\ \bibinfo {author} {\bibfnamefont {M.~I.}\ \bibnamefont
  {McMahon}},\ }\href {\doibase 10.1103/PhysRevB.70.184109} {\bibfield
  {journal} {\bibinfo  {journal} {Phys. Rev. B}\ }\textbf {\bibinfo {volume}
  {70}},\ \bibinfo {pages} {184109} (\bibinfo {year} {2004})}\BibitemShut
  {NoStop}%
\bibitem [{\citenamefont {Hejny}\ \emph {et~al.}(2006)\citenamefont {Hejny},
  \citenamefont {Falconi}, \citenamefont {Lundegaard},\ and\ \citenamefont
  {McMahon}}]{HejnyPRB2006}%
  \BibitemOpen
  \bibfield  {author} {\bibinfo {author} {\bibfnamefont {C.}~\bibnamefont
  {Hejny}}, \bibinfo {author} {\bibfnamefont {S.}~\bibnamefont {Falconi}},
  \bibinfo {author} {\bibfnamefont {L.~F.}\ \bibnamefont {Lundegaard}}, \ and\
  \bibinfo {author} {\bibfnamefont {M.~I.}\ \bibnamefont {McMahon}},\ }\href
  {\doibase 10.1103/PhysRevB.74.174119} {\bibfield  {journal} {\bibinfo
  {journal} {Phys. Rev. B}\ }\textbf {\bibinfo {volume} {74}},\ \bibinfo
  {pages} {174119} (\bibinfo {year} {2006})}\BibitemShut {NoStop}%
\bibitem [{\citenamefont {Deshpande}\ and\ \citenamefont
  {Pawar}(1965)}]{DeshpandePhysica1965}%
  \BibitemOpen
  \bibfield  {author} {\bibinfo {author} {\bibfnamefont {V.~T.}\ \bibnamefont
  {Deshpande}}\ and\ \bibinfo {author} {\bibfnamefont {R.}~\bibnamefont
  {Pawar}},\ }\href {\doibase https://doi.org/10.1016/0031-8914(65)90004-2}
  {\bibfield  {journal} {\bibinfo  {journal} {Physica}\ }\textbf {\bibinfo
  {volume} {31}},\ \bibinfo {pages} {671} (\bibinfo {year} {1965})}\BibitemShut
  {NoStop}%
\bibitem [{\citenamefont {Ashitkov}\ \emph {et~al.}(2002)\citenamefont
  {Ashitkov}, \citenamefont {Agranat}, \citenamefont {Kondratenko},
  \citenamefont {Anisimov}, \citenamefont {Fortov}, \citenamefont {Temnov},
  \citenamefont {Sokolowski-Tinten}, \citenamefont {Rethfeld}, \citenamefont
  {Zhou},\ and\ \citenamefont {von~der Linde}}]{LindeJETP2002}%
  \BibitemOpen
  \bibfield  {author} {\bibinfo {author} {\bibfnamefont {S.~I.}\ \bibnamefont
  {Ashitkov}}, \bibinfo {author} {\bibfnamefont {M.~B.}\ \bibnamefont
  {Agranat}}, \bibinfo {author} {\bibfnamefont {P.~S.}\ \bibnamefont
  {Kondratenko}}, \bibinfo {author} {\bibfnamefont {S.~I.}\ \bibnamefont
  {Anisimov}}, \bibinfo {author} {\bibfnamefont {V.~E.}\ \bibnamefont
  {Fortov}}, \bibinfo {author} {\bibfnamefont {V.~V.}\ \bibnamefont {Temnov}},
  \bibinfo {author} {\bibfnamefont {K.}~\bibnamefont {Sokolowski-Tinten}},
  \bibinfo {author} {\bibfnamefont {B.}~\bibnamefont {Rethfeld}}, \bibinfo
  {author} {\bibfnamefont {P.}~\bibnamefont {Zhou}}, \ and\ \bibinfo {author}
  {\bibfnamefont {D.}~\bibnamefont {von~der Linde}},\ }\href {\doibase
  10.1134/1.1528702} {\bibfield  {journal} {\bibinfo  {journal} {Journal of
  Experimental and Theoretical Physics Letters}\ }\textbf {\bibinfo {volume}
  {76}},\ \bibinfo {pages} {461} (\bibinfo {year} {2002})}\BibitemShut
  {NoStop}%
\bibitem [{\citenamefont {Brodsky}\ \emph {et~al.}(1972)\citenamefont
  {Brodsky}, \citenamefont {Gambino}, \citenamefont {Smith~Jr.},\ and\
  \citenamefont {Yacoby}}]{BrodskyPSS1972}%
  \BibitemOpen
  \bibfield  {author} {\bibinfo {author} {\bibfnamefont {M.~H.}\ \bibnamefont
  {Brodsky}}, \bibinfo {author} {\bibfnamefont {R.~J.}\ \bibnamefont
  {Gambino}}, \bibinfo {author} {\bibfnamefont {J.~E.}\ \bibnamefont
  {Smith~Jr.}}, \ and\ \bibinfo {author} {\bibfnamefont {Y.}~\bibnamefont
  {Yacoby}},\ }\href {\doibase https://doi.org/10.1002/pssb.2220520229}
  {\bibfield  {journal} {\bibinfo  {journal} {physica status solidi (b)}\
  }\textbf {\bibinfo {volume} {52}},\ \bibinfo {pages} {609} (\bibinfo {year}
  {1972})}\BibitemShut {NoStop}%
\bibitem [{\citenamefont {Hunsche}\ \emph {et~al.}(1996)\citenamefont
  {Hunsche}, \citenamefont {Wienecke},\ and\ \citenamefont
  {Kurz}}]{KurzAPA1996}%
  \BibitemOpen
  \bibfield  {author} {\bibinfo {author} {\bibfnamefont {S.}~\bibnamefont
  {Hunsche}}, \bibinfo {author} {\bibfnamefont {K.}~\bibnamefont {Wienecke}}, \
  and\ \bibinfo {author} {\bibfnamefont {H.}~\bibnamefont {Kurz}},\ }\href
  {\doibase 10.1007/BF01567124} {\bibfield  {journal} {\bibinfo  {journal}
  {Applied Physics A}\ }\textbf {\bibinfo {volume} {62}},\ \bibinfo {pages}
  {499} (\bibinfo {year} {1996})}\BibitemShut {NoStop}%
\bibitem [{\citenamefont {Kudryashov}\ \emph {et~al.}(2007)\citenamefont
  {Kudryashov}, \citenamefont {Kandyla}, \citenamefont {Roeser},\ and\
  \citenamefont {Mazur}}]{MazurPRB2007}%
  \BibitemOpen
  \bibfield  {author} {\bibinfo {author} {\bibfnamefont {S.~I.}\ \bibnamefont
  {Kudryashov}}, \bibinfo {author} {\bibfnamefont {M.}~\bibnamefont {Kandyla}},
  \bibinfo {author} {\bibfnamefont {C.~A.~D.}\ \bibnamefont {Roeser}}, \ and\
  \bibinfo {author} {\bibfnamefont {E.}~\bibnamefont {Mazur}},\ }\href
  {\doibase 10.1103/PhysRevB.75.085207} {\bibfield  {journal} {\bibinfo
  {journal} {Phys. Rev. B}\ }\textbf {\bibinfo {volume} {75}},\ \bibinfo
  {pages} {085207} (\bibinfo {year} {2007})}\BibitemShut {NoStop}%
\bibitem [{\citenamefont {Misochko}\ and\ \citenamefont
  {Lebedev}(2016)}]{MisochkoPRB2016}%
  \BibitemOpen
  \bibfield  {author} {\bibinfo {author} {\bibfnamefont {O.~V.}\ \bibnamefont
  {Misochko}}\ and\ \bibinfo {author} {\bibfnamefont {M.~V.}\ \bibnamefont
  {Lebedev}},\ }\href {\doibase 10.1103/PhysRevB.94.184307} {\bibfield
  {journal} {\bibinfo  {journal} {Phys. Rev. B}\ }\textbf {\bibinfo {volume}
  {94}},\ \bibinfo {pages} {184307} (\bibinfo {year} {2016})}\BibitemShut
  {NoStop}%
\bibitem [{\citenamefont {F{\"o}rst}\ \emph {et~al.}(2011)\citenamefont
  {F{\"o}rst}, \citenamefont {Manzoni}, \citenamefont {Kaiser}, \citenamefont
  {Tomioka}, \citenamefont {Tokura}, \citenamefont {Merlin},\ and\
  \citenamefont {Cavalleri}}]{CavalleriNPHY2011}%
  \BibitemOpen
  \bibfield  {author} {\bibinfo {author} {\bibfnamefont {M.}~\bibnamefont
  {F{\"o}rst}}, \bibinfo {author} {\bibfnamefont {C.}~\bibnamefont {Manzoni}},
  \bibinfo {author} {\bibfnamefont {S.}~\bibnamefont {Kaiser}}, \bibinfo
  {author} {\bibfnamefont {Y.}~\bibnamefont {Tomioka}}, \bibinfo {author}
  {\bibfnamefont {Y.}~\bibnamefont {Tokura}}, \bibinfo {author} {\bibfnamefont
  {R.}~\bibnamefont {Merlin}}, \ and\ \bibinfo {author} {\bibfnamefont
  {A.}~\bibnamefont {Cavalleri}},\ }\href {\doibase 10.1038/nphys2055}
  {\bibfield  {journal} {\bibinfo  {journal} {Nature Physics}\ }\textbf
  {\bibinfo {volume} {7}},\ \bibinfo {pages} {854} (\bibinfo {year}
  {2011})}\BibitemShut {NoStop}%
\bibitem [{\citenamefont {Juraschek}\ \emph {et~al.}(2017)\citenamefont
  {Juraschek}, \citenamefont {Fechner},\ and\ \citenamefont
  {Spaldin}}]{SpaldinPRL2017}%
  \BibitemOpen
  \bibfield  {author} {\bibinfo {author} {\bibfnamefont {D.~M.}\ \bibnamefont
  {Juraschek}}, \bibinfo {author} {\bibfnamefont {M.}~\bibnamefont {Fechner}},
  \ and\ \bibinfo {author} {\bibfnamefont {N.~A.}\ \bibnamefont {Spaldin}},\
  }\href {\doibase 10.1103/PhysRevLett.118.054101} {\bibfield  {journal}
  {\bibinfo  {journal} {Phys. Rev. Lett.}\ }\textbf {\bibinfo {volume} {118}},\
  \bibinfo {pages} {054101} (\bibinfo {year} {2017})}\BibitemShut {NoStop}%
\bibitem [{\citenamefont {Rini}\ \emph {et~al.}(2007)\citenamefont {Rini},
  \citenamefont {Tobey}, \citenamefont {Dean}, \citenamefont {Itatani},
  \citenamefont {Tomioka}, \citenamefont {Tokura}, \citenamefont {Schoenlein},\
  and\ \citenamefont {Cavalleri}}]{CavalleriNature2007}%
  \BibitemOpen
  \bibfield  {author} {\bibinfo {author} {\bibfnamefont {M.}~\bibnamefont
  {Rini}}, \bibinfo {author} {\bibfnamefont {R.}~\bibnamefont {Tobey}},
  \bibinfo {author} {\bibfnamefont {N.}~\bibnamefont {Dean}}, \bibinfo {author}
  {\bibfnamefont {J.}~\bibnamefont {Itatani}}, \bibinfo {author} {\bibfnamefont
  {Y.}~\bibnamefont {Tomioka}}, \bibinfo {author} {\bibfnamefont
  {Y.}~\bibnamefont {Tokura}}, \bibinfo {author} {\bibfnamefont {R.~W.}\
  \bibnamefont {Schoenlein}}, \ and\ \bibinfo {author} {\bibfnamefont
  {A.}~\bibnamefont {Cavalleri}},\ }\href {\doibase 10.1038/nature06119}
  {\bibfield  {journal} {\bibinfo  {journal} {Nature}\ }\textbf {\bibinfo
  {volume} {449}},\ \bibinfo {pages} {72} (\bibinfo {year} {2007})}\BibitemShut
  {NoStop}%
\bibitem [{\citenamefont {Fausti}\ \emph {et~al.}(2011)\citenamefont {Fausti},
  \citenamefont {Tobey}, \citenamefont {Dean}, \citenamefont {Kaiser},
  \citenamefont {Dienst}, \citenamefont {Hoffmann}, \citenamefont {Pyon},
  \citenamefont {Takayama}, \citenamefont {Takagi},\ and\ \citenamefont
  {Cavalleri}}]{CavalleriScience2011}%
  \BibitemOpen
  \bibfield  {author} {\bibinfo {author} {\bibfnamefont {D.}~\bibnamefont
  {Fausti}}, \bibinfo {author} {\bibfnamefont {R.~I.}\ \bibnamefont {Tobey}},
  \bibinfo {author} {\bibfnamefont {N.}~\bibnamefont {Dean}}, \bibinfo {author}
  {\bibfnamefont {S.}~\bibnamefont {Kaiser}}, \bibinfo {author} {\bibfnamefont
  {A.}~\bibnamefont {Dienst}}, \bibinfo {author} {\bibfnamefont {M.~C.}\
  \bibnamefont {Hoffmann}}, \bibinfo {author} {\bibfnamefont {S.}~\bibnamefont
  {Pyon}}, \bibinfo {author} {\bibfnamefont {T.}~\bibnamefont {Takayama}},
  \bibinfo {author} {\bibfnamefont {H.}~\bibnamefont {Takagi}}, \ and\ \bibinfo
  {author} {\bibfnamefont {A.}~\bibnamefont {Cavalleri}},\ }\href {\doibase
  10.1126/science.1197294} {\bibfield  {journal} {\bibinfo  {journal}
  {Science}\ }\textbf {\bibinfo {volume} {331}},\ \bibinfo {pages} {189}
  (\bibinfo {year} {2011})},\ \Eprint
  {http://arxiv.org/abs/https://www.science.org/doi/pdf/10.1126/science.1197294}
  {} \BibitemShut
  {NoStop}%
\bibitem [{\citenamefont {Mankowsky}\ \emph {et~al.}(2014)\citenamefont
  {Mankowsky}, \citenamefont {Subedi}, \citenamefont {F{\"o}rst}, \citenamefont
  {Mariager}, \citenamefont {Chollet}, \citenamefont {Lemke}, \citenamefont
  {Robinson}, \citenamefont {Glownia}, \citenamefont {Minitti}, \citenamefont
  {Frano}, \citenamefont {Fechner}, \citenamefont {Spaldin}, \citenamefont
  {Loew}, \citenamefont {Keimer}, \citenamefont {Georges},\ and\ \citenamefont
  {Cavalleri}}]{CavalleriNature2014}%
  \BibitemOpen
  \bibfield  {author} {\bibinfo {author} {\bibfnamefont {R.}~\bibnamefont
  {Mankowsky}}, \bibinfo {author} {\bibfnamefont {A.}~\bibnamefont {Subedi}},
  \bibinfo {author} {\bibfnamefont {M.}~\bibnamefont {F{\"o}rst}}, \bibinfo
  {author} {\bibfnamefont {S.~O.}\ \bibnamefont {Mariager}}, \bibinfo {author}
  {\bibfnamefont {M.}~\bibnamefont {Chollet}}, \bibinfo {author} {\bibfnamefont
  {H.~T.}\ \bibnamefont {Lemke}}, \bibinfo {author} {\bibfnamefont {J.~S.}\
  \bibnamefont {Robinson}}, \bibinfo {author} {\bibfnamefont {J.~M.}\
  \bibnamefont {Glownia}}, \bibinfo {author} {\bibfnamefont {M.~P.}\
  \bibnamefont {Minitti}}, \bibinfo {author} {\bibfnamefont {A.}~\bibnamefont
  {Frano}}, \bibinfo {author} {\bibfnamefont {M.}~\bibnamefont {Fechner}},
  \bibinfo {author} {\bibfnamefont {N.~A.}\ \bibnamefont {Spaldin}}, \bibinfo
  {author} {\bibfnamefont {T.}~\bibnamefont {Loew}}, \bibinfo {author}
  {\bibfnamefont {B.}~\bibnamefont {Keimer}}, \bibinfo {author} {\bibfnamefont
  {A.}~\bibnamefont {Georges}}, \ and\ \bibinfo {author} {\bibfnamefont
  {A.}~\bibnamefont {Cavalleri}},\ }\href {https://doi.org/10.1038/nature13875}
  {\bibfield  {journal} {\bibinfo  {journal} {Nature}\ }\textbf {\bibinfo
  {volume} {516}},\ \bibinfo {pages} {71 EP } (\bibinfo {year}
  {2014})}\BibitemShut {NoStop}%
\bibitem [{\citenamefont {Mitrano}\ \emph {et~al.}(2016)\citenamefont
  {Mitrano}, \citenamefont {Cantaluppi}, \citenamefont {Nicoletti},
  \citenamefont {Kaiser}, \citenamefont {Perucchi}, \citenamefont {Lupi},
  \citenamefont {Di~Pietro}, \citenamefont {Pontiroli}, \citenamefont
  {Ricc{\`o}}, \citenamefont {Clark}, \citenamefont {Jaksch},\ and\
  \citenamefont {Cavalleri}}]{CavalleriNature2016}%
  \BibitemOpen
  \bibfield  {author} {\bibinfo {author} {\bibfnamefont {M.}~\bibnamefont
  {Mitrano}}, \bibinfo {author} {\bibfnamefont {A.}~\bibnamefont {Cantaluppi}},
  \bibinfo {author} {\bibfnamefont {D.}~\bibnamefont {Nicoletti}}, \bibinfo
  {author} {\bibfnamefont {S.}~\bibnamefont {Kaiser}}, \bibinfo {author}
  {\bibfnamefont {A.}~\bibnamefont {Perucchi}}, \bibinfo {author}
  {\bibfnamefont {S.}~\bibnamefont {Lupi}}, \bibinfo {author} {\bibfnamefont
  {P.}~\bibnamefont {Di~Pietro}}, \bibinfo {author} {\bibfnamefont
  {D.}~\bibnamefont {Pontiroli}}, \bibinfo {author} {\bibfnamefont
  {M.}~\bibnamefont {Ricc{\`o}}}, \bibinfo {author} {\bibfnamefont {S.~R.}\
  \bibnamefont {Clark}}, \bibinfo {author} {\bibfnamefont {D.}~\bibnamefont
  {Jaksch}}, \ and\ \bibinfo {author} {\bibfnamefont {A.}~\bibnamefont
  {Cavalleri}},\ }\href {\doibase 10.1038/nature16522} {\bibfield  {journal}
  {\bibinfo  {journal} {Nature}\ }\textbf {\bibinfo {volume} {530}},\ \bibinfo
  {pages} {461} (\bibinfo {year} {2016})}\BibitemShut {NoStop}%
\bibitem [{\citenamefont {Gu}\ and\ \citenamefont
  {Rondinelli}(2018)}]{RondinelliPRB2018}%
  \BibitemOpen
  \bibfield  {author} {\bibinfo {author} {\bibfnamefont {M.}~\bibnamefont
  {Gu}}\ and\ \bibinfo {author} {\bibfnamefont {J.~M.}\ \bibnamefont
  {Rondinelli}},\ }\href {\doibase 10.1103/PhysRevB.98.024102} {\bibfield
  {journal} {\bibinfo  {journal} {Phys. Rev. B}\ }\textbf {\bibinfo {volume}
  {98}},\ \bibinfo {pages} {024102} (\bibinfo {year} {2018})}\BibitemShut
  {NoStop}%
\bibitem [{\citenamefont {Nova}\ \emph {et~al.}(2016)\citenamefont {Nova},
  \citenamefont {Cartella}, \citenamefont {Cantaluppi}, \citenamefont
  {F{\"o}rst}, \citenamefont {Bossini}, \citenamefont {Mikhaylovskiy},
  \citenamefont {Kimel}, \citenamefont {Merlin},\ and\ \citenamefont
  {Cavalleri}}]{CavalleriNPHY2016}%
  \BibitemOpen
  \bibfield  {author} {\bibinfo {author} {\bibfnamefont {T.~F.}\ \bibnamefont
  {Nova}}, \bibinfo {author} {\bibfnamefont {A.}~\bibnamefont {Cartella}},
  \bibinfo {author} {\bibfnamefont {A.}~\bibnamefont {Cantaluppi}}, \bibinfo
  {author} {\bibfnamefont {M.}~\bibnamefont {F{\"o}rst}}, \bibinfo {author}
  {\bibfnamefont {D.}~\bibnamefont {Bossini}}, \bibinfo {author} {\bibfnamefont
  {R.~V.}\ \bibnamefont {Mikhaylovskiy}}, \bibinfo {author} {\bibfnamefont
  {A.~V.}\ \bibnamefont {Kimel}}, \bibinfo {author} {\bibfnamefont
  {R.}~\bibnamefont {Merlin}}, \ and\ \bibinfo {author} {\bibfnamefont
  {A.}~\bibnamefont {Cavalleri}},\ }\href {https://doi.org/10.1038/nphys3925}
  {\bibfield  {journal} {\bibinfo  {journal} {Nature Physics}\ }\textbf
  {\bibinfo {volume} {13}},\ \bibinfo {pages} {132 EP } (\bibinfo {year}
  {2016})}\BibitemShut {NoStop}%
\bibitem [{\citenamefont {Nova}\ \emph {et~al.}(2019)\citenamefont {Nova},
  \citenamefont {Disa}, \citenamefont {Fechner},\ and\ \citenamefont
  {Cavalleri}}]{CavalleriScience2019}%
  \BibitemOpen
  \bibfield  {author} {\bibinfo {author} {\bibfnamefont {T.~F.}\ \bibnamefont
  {Nova}}, \bibinfo {author} {\bibfnamefont {A.~S.}\ \bibnamefont {Disa}},
  \bibinfo {author} {\bibfnamefont {M.}~\bibnamefont {Fechner}}, \ and\
  \bibinfo {author} {\bibfnamefont {A.}~\bibnamefont {Cavalleri}},\ }\href
  {\doibase 10.1126/science.aaw4911} {\bibfield  {journal} {\bibinfo  {journal}
  {Science}\ }\textbf {\bibinfo {volume} {364}},\ \bibinfo {pages} {1075}
  (\bibinfo {year} {2019})},\ \Eprint
  {http://arxiv.org/abs/https://science.sciencemag.org/content/364/6445/1075.full.pdf}
  {} \BibitemShut
  {NoStop}%
\bibitem [{\citenamefont {Afanasiev}\ \emph {et~al.}(2021)\citenamefont
  {Afanasiev}, \citenamefont {Hortensius}, \citenamefont {Ivanov},
  \citenamefont {Sasani}, \citenamefont {Bousquet}, \citenamefont {Blanter},
  \citenamefont {Mikhaylovskiy}, \citenamefont {Kimel},\ and\ \citenamefont
  {Caviglia}}]{CavigliaNMAT2021}%
  \BibitemOpen
  \bibfield  {author} {\bibinfo {author} {\bibfnamefont {D.}~\bibnamefont
  {Afanasiev}}, \bibinfo {author} {\bibfnamefont {J.~R.}\ \bibnamefont
  {Hortensius}}, \bibinfo {author} {\bibfnamefont {B.~A.}\ \bibnamefont
  {Ivanov}}, \bibinfo {author} {\bibfnamefont {A.}~\bibnamefont {Sasani}},
  \bibinfo {author} {\bibfnamefont {E.}~\bibnamefont {Bousquet}}, \bibinfo
  {author} {\bibfnamefont {Y.~M.}\ \bibnamefont {Blanter}}, \bibinfo {author}
  {\bibfnamefont {R.~V.}\ \bibnamefont {Mikhaylovskiy}}, \bibinfo {author}
  {\bibfnamefont {A.~V.}\ \bibnamefont {Kimel}}, \ and\ \bibinfo {author}
  {\bibfnamefont {A.~D.}\ \bibnamefont {Caviglia}},\ }\href {\doibase
  10.1038/s41563-021-00922-7} {\bibfield  {journal} {\bibinfo  {journal}
  {Nature Materials}\ }\textbf {\bibinfo {volume} {20}},\ \bibinfo {pages}
  {607} (\bibinfo {year} {2021})}\BibitemShut {NoStop}%
\bibitem [{\citenamefont {von Hoegen}\ \emph {et~al.}(2018)\citenamefont {von
  Hoegen}, \citenamefont {Mankowsky}, \citenamefont {Fechner}, \citenamefont
  {F{\"o}rst},\ and\ \citenamefont {Cavalleri}}]{CavalleriNature2018}%
  \BibitemOpen
  \bibfield  {author} {\bibinfo {author} {\bibfnamefont {A.}~\bibnamefont {von
  Hoegen}}, \bibinfo {author} {\bibfnamefont {R.}~\bibnamefont {Mankowsky}},
  \bibinfo {author} {\bibfnamefont {M.}~\bibnamefont {Fechner}}, \bibinfo
  {author} {\bibfnamefont {M.}~\bibnamefont {F{\"o}rst}}, \ and\ \bibinfo
  {author} {\bibfnamefont {A.}~\bibnamefont {Cavalleri}},\ }\href {\doibase
  10.1038/nature25484} {\bibfield  {journal} {\bibinfo  {journal} {Nature}\
  }\textbf {\bibinfo {volume} {555}},\ \bibinfo {pages} {79} (\bibinfo {year}
  {2018})}\BibitemShut {NoStop}%

\end{thebibliography}
\end{document}